\documentclass{aa}
\usepackage{epsfig,times}
\usepackage{amssymb}
\usepackage{txfonts}
\usepackage{graphicx}
\usepackage[comma]{natbib}
\begin{document}

\title{Force-free twisted magnetospheres of neutron stars}

\author{D.~Vigan\`o\inst{1} \and J.A.~Pons\inst{1} \and J.A.~Miralles\inst{1}}
\institute{Departament de F\'{\i}sica Aplicada, Universitat d'Alacant,  
Ap. Correus 99, 03080 Alacant, Spain}
 
\abstract 
{The X-ray spectra observed in the persistent emission of magnetars are evidence for the existence of a magnetosphere. The high-energy part of the spectra is explained by resonant cyclotron upscattering of soft thermal photons in a twisted magnetosphere, which has motivated an increasing number of efforts to improve and generalize existing magnetosphere models.}
{We want to build more general configurations of twisted, force-free magnetospheres as a first step to understanding the role played by the magnetic field geometry in the observed spectra.}
{First we reviewed and extended previous analytical works to assess the viability and limitations of semi-analytical approaches. Second, we built a numerical code able to relax an initial configuration of a nonrotating magnetosphere to a force-free geometry, provided any arbitrary form of the magnetic field at the star surface. The numerical code is based on a finite-difference time-domain, divergence-free, and conservative scheme, based of the magneto-frictional method used in other scenarios.}
{We obtain new numerical configurations of twisted magnetospheres, with distributions of twist and currents that differ from previous analytical solutions. The range of global twist of the new family of solutions is similar to the existing semi-analytical models (up to some radians), but the achieved geometry may be quite different.}
{The geometry of twisted, force-free magnetospheres shows a wider variety of possibilities than previously considered. This has implications for the observed spectra and opens the possibility of implementing alternative models in simulations of radiative transfer aiming at providing spectra to be compared with observations.}

\titlerunning{Force-free twisted magnetospheres of Neutron Stars} 
\authorrunning{D.~Vigan\`o, J.A.~Pons \& J.A.~Miralles} 

\maketitle

\section{Introduction}

Anomalous X-ray pulsars (AXPs) and soft gamma-ray repeaters (SGRs) are a class of neutron stars (NSs) characterized by high X-ray quiescent luminosities, short X-ray bursts, and giant flares (for SGRs). They are believed to be {\it magnetars} \citep{duncan92, thompson96}: young ($P/2\dot{P}\sim 10^2-10^5$ yr) NSs with very strong magnetic fields $\sim 10^{14}-10^{15}$~G. Magnetars are supposedly born with a very short spin period (few milliseconds), which causes the formation of an intense inner magnetic field via dynamo processes and a subsequent rapid loss of rotational energy by means of magnetic braking. As more and better data become available in the last decade,\footnote{The updated catalog of these sources can be found at \texttt{http://www.physics.mcgill.ca/$\sim$pulsar/magnetar/main.html}} the separation between the two traditional classes (SGRs and AXPs) has become thinner, and they are no longer considered different classes. 
All SGRs and most of AXPs have shown sporadic bursts in the $X$ and $\gamma$ bands, and the recent discovery of transient magnetars has made the situation even less clear. Currently, our interpretation of the term {\it magnetars} is being reconsidered (see \cite{mereghetti08,rea11} for observational reviews). In general, magnetars are characterized by long periods $P\sim 2-12$ s and high X-ray persistent luminosities $L_X\sim 2\times10^{33}-2\times10^{35}$~erg s$^{-1}$, which point to the magnetic field decay as the main source of energy, unlike radio pulsars, fed by the loss of rotational energy.  One key ingredient in those models that try to explain the variety of magnetar phenomenology and their spectra  is the presence of toroidal fields, of the same order of magnitude or even larger than the poloidal component, both in the interior and in the external magnetosphere.
  
The object of this paper is the modeling of the magnetosphere of magnetars, which is crucial because it is one of the main ingredients in the formation of spectra. The difficulty in building fully consistent solutions of NS magnetospheres is such that, after more than 40 years, the aligned rotator model proposed by \cite{goldreich69} still remains very popular for explaining the basic features of a rotating magnetosphere. We must point out that the aligned rotator model, with all its variants, is surely a naive description of pulsars that has been strongly criticized \citep[e.g.][]{michel80}, among other reasons, simply because it cannot explain {\it why pulsars pulsate}.
A further refinement consists of the ideal MHD approximation for force-free magnetospheres, which leads to the so-called pulsar equation. This is simply the balance of the electromagnetic forces around the rotating neutron star under the assumption of axial symmetry and it neglects the pressure, inertial, and gravitational terms. The rotationally induced electric field is required to be perpendicular to the magnetic field. The latter can in general have an azimuthal component, that twists the poloidal field lines. This toroidal component has to be sustained by currents, whose shape is the source term of the pulsar equation: it is the fundamental parameter to be chosen in order to find the corresponding geometric configuration of the magnetic field.

The problem is analytically solvable only for a few simple, \textit{ad-hoc} choices of the current, such as in \cite{michel73}, but these solutions generally present unphysical features in some border regions. An alternative numerical approach by \cite{contopoulos99} is useful to construct consistent force-free models, with a smooth matching across the light cylinder. In this model the corotating region only hosts an untwisted dipolar magnetic field, and the focus is on the bundle of open twisted lines that cross the light cylinder and the resulting output power. The model was refined by \cite{gruzinov05} and extended to the time-dependent, misaligned case by means of 3D simulations by \cite{spitkovsky06} and \cite{kalapotharakos09}.

Nowadays, X-ray instruments allow us to obtain more and more information from the spectra of magnetars. An especially interesting feature in the persistent emission of all of the magnetar candidates is that their spectra can be well fitted with a thermal component ($0.4-0.7$ keV) plus a hard nonthermal tail, described by a power law with photon index $\beta\sim 3-4$ \citep{mereghetti08}. This hard component is commonly attributed to resonant Compton scattering, which becomes important in the presence of a strong current (i.e. high charge density). For this reason, much attention has been given to the modeling of currents flowing in a twisted magnetosphere and obtaining estimates of the charge carrier density and the corresponding resonant optical depth. \cite{lyutikov06} propose a semi-analytical 1D model able to describe the basic properties of radiation transfer across a twisted magnetosphere. More accurate 3D Monte Carlo simulations have been performed for the nonrelativistic \citep{fernandez07, nobili08a} and relativistic cases \citep{nobili08b}. \cite{fernandez11} discuss the implications of the outgoing polarization. The implementation in \textit{XSpec} and the systematic fits to observational data have been successfully performed by \cite{rea08} and \cite{zane09} for the 1D and 3D cases, respectively. All these very remarkable results rely on the same family of models: the self-similar twisted dipole proposed by \cite{thompson02}, extended to higher multipoles by \cite{pavan09}, which represents a semi-analytical, easy-to-implement solution to the pulsar equation. However, this model lacks generality, as it relies on a very particular choice of the form of the current. 

We note that force-free magnetosphere models rely on a purely macrophysical approach but a fully coherent microphysical description is still lacking: how are the currents and electric fields in a given model  generated and sustained? Which particles are present (electrons, ions, pairs)? Where do they come from? Which velocity distribution can be assigned to each kind of particle? In the magnetar framework, there are some excellent attempts to give a microphysical description of coronae. \cite{beloborodov07} deal with the problem considering a deviation from force-free equilibrium, which has to be introduced to explain the production of pairs, but many questions remain open. Given the complexity of the problem, one has to decide whether to focus on detailed microphysical processes of an approximate macrophysical solution or to improve the large-scale electrodynamics solutions by gradually removing some of the simplifying hypotheses.

In this article, we present our efforts to obtain some more general configurations of a twisted magnetosphere and to estimate the dependence of the spectrum on the geometric configuration. In Sect. \ref{sec_analytical}, we set the formalism needed to describe the problem and present some analytical and semi-analytical generalizations of previous works. In Sect. \ref{sec_numerical} we describe the numerical code used to build force-free configurations and the battery of tests performed to verify the code. We pay special attention to the effect of key input parameters and the convergence of the method. In Sect. \ref{sec_results} we present new results for magnetosphere models, and explore the sensitivity of the relevant output (current, charge distribution, and resonant optical depth) to different input parameters. 
In Sect. \ref{conclusions} we summarize our findings and outline future improvements.

\section{Analytical and semianalytical solutions}\label{sec_analytical}
\begin{figure}
 \centering
 \includegraphics[width=6cm]{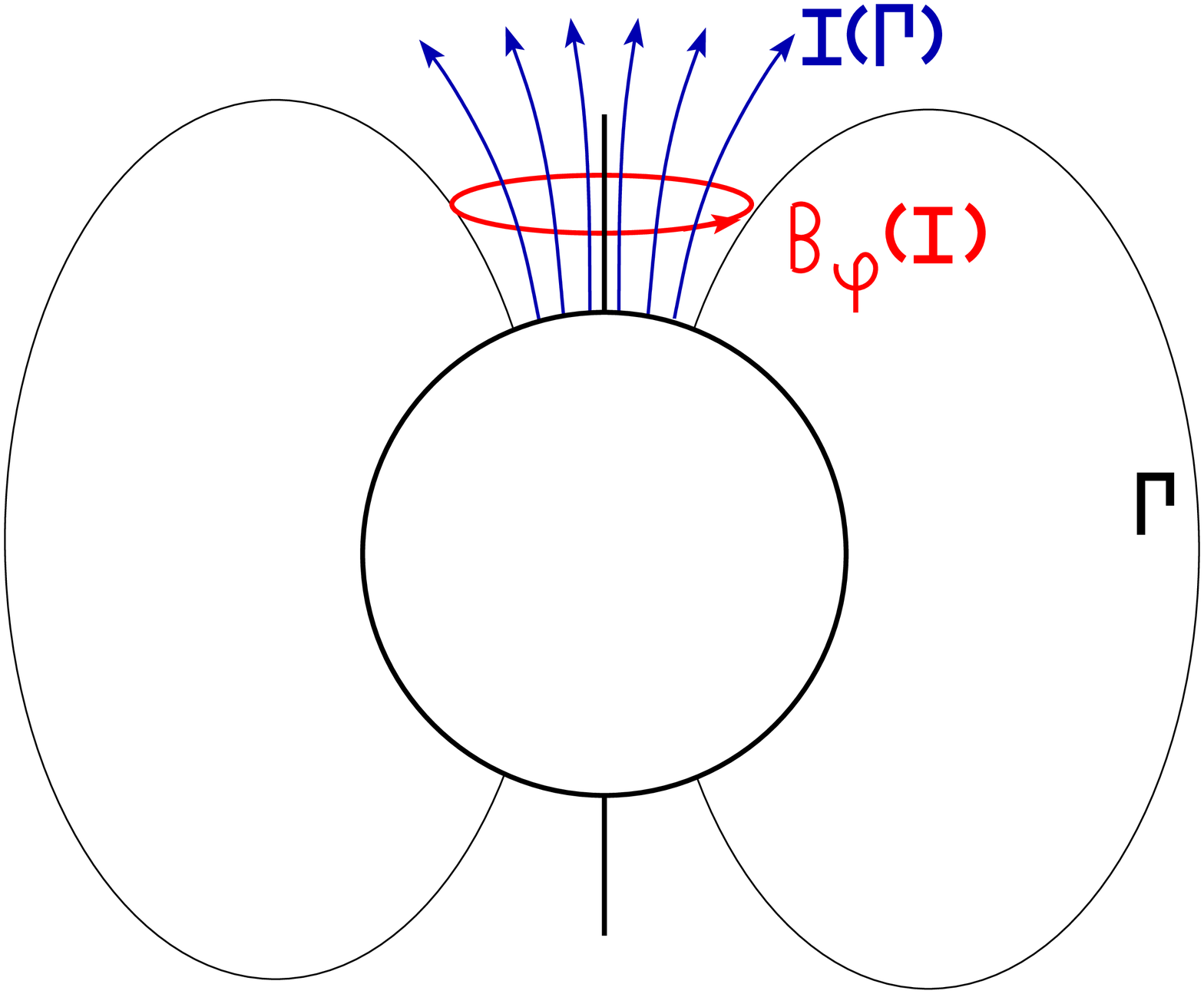}
 \caption{Relation between magnetic flux function $\Gamma$, enclosed current function $I(\Gamma)$, and toroidal field.}
 \label{axisym}
\end{figure}

\subsection{The pulsar equation}\label{sec_pulsar_eq}
We consider an NS with a magnetic field $\vec{B}$ and angular velocity $\vec{\Omega}$. Assuming that the plasma is a perfect conductor in vicinity of the star, rotation induces an electric field given by
\begin{equation}
  \vec{E}=-\frac{\vec{\Omega} \times \vec{r}}{c}\times \vec{B}.
\end{equation}
where $c$ is the velocity of light. Neglecting centrifugal (i.e. inertial), collisional, and gravitational terms compared to electromagnetic forces, the force equilibrium equation is expressed by
\begin{equation}\label{gs_gen}
  \rho_e \vec{E}+\frac{1}{c}\vec{J}\times\vec{B}=0~,
\end{equation}
where $\rho_e$ is the charge density and $\vec{J}$ the current density. Assuming axial symmetry (aligned rotator) and working in spherical coordinates ($r,\theta,\phi$), we introduce the \textit{magnetic flux function} $\Gamma(r,\theta)$, in terms of which the poloidal field is
\begin{equation}\label{Bp}
  \vec{B}_p=\frac{\vec{\nabla} \Gamma(r,\theta) \times \hat{\vec{\phi}}}{r \sin\theta},
\end{equation}
where $\hat{\vec{\phi}}$ is the unit azimuthal vector. With the choice of gauge $\Gamma=0$ on the axis, the function $\Gamma(r,\theta)$ is related to the $\phi-$component of the potential vector by
\begin{equation}\label{gamma_aphi}
  \Gamma(r,\theta)= A_\phi(r,\theta) r\sin\theta~,
\end{equation}
and it is constant along a field line ($\vec{B}_p\cdot\vec{\nabla}\Gamma=0$). Thus its value labels the axisymmetric surface $S_\Gamma$ given by the azimuthal rotation of one field line. The magnetic flux flowing within the surface $S_\Gamma$, with value $\Phi_\Gamma(\vec{B})=2\pi\Gamma$, is conserved by definition. A toroidal component (i.e. $\phi$-component in axial symmetry) of the magnetic field is present if sustained by poloidal currents. The enclosed current $I$ flowing within $S_\Gamma$ (see blue arrows in Fig. \ref{axisym} and note that the force-free condition implies that currents flow parallel to the surfaces $S_\Gamma$) has to be a function of only $\Gamma$, with the only requirement being that $I(0)=0$. Thus we have the freedom of choosing the \textit{enclosed current function} $I(\Gamma)$, and by Amp\'ere's law, the $\phi-$component of the magnetic field is related to $I$ by
\begin{equation}\label{Bt}
  {B_\phi}=\frac{2}{cr\sin\theta}I(\Gamma)~.
\end{equation}

For a given magnetosphere model, it is useful to define the twist of a field line as the azimuthal displacement between its surface footprints (for a dipolar-like configuration, one footprint in each hemisphere). The integral of the local displacement $\delta\phi(r,\theta) = B_\phi\delta\theta/B_\theta\sin\theta$ along the magnetic field line, $l_\Gamma$, between its surface footprints at magnetic colatitudes $\theta_{1,2}(\Gamma)$ is the line twist
\begin{equation}\label{def_twist}
  \Delta \phi_{tw}(\Gamma)\equiv\int_{\theta_1}^{\theta_2} \frac{B_\phi(r(\theta,\Gamma),\theta)}{B_\theta(r(\theta,\Gamma),\theta)\sin\theta}\mbox{d}\theta ~,
\end{equation}
where the dependence $r(\theta,\Gamma)$ can be found by solving the field line equations. The function $\Delta \phi_{tw}(\Gamma)$ is used to characterize a magnetosphere.

We then return to the equilibrium equation (\ref{gs_gen}). In axial symmetry, only poloidal components of the electric field are allowed. Considering the poloidal component of Eq. (\ref{gs_gen}) in cylindrical coordinates $(\rho,\phi,z)$ and denoting the radius of the light cylinder by $r_l=c/\Omega$, we obtain the so-called \textit{pulsar equation} \citep{scharlemann73, michel73}:
\begin{equation}\label{pulsar_eq}
 \left(1-\frac{\rho^2}{r_l^2}\right)(\Gamma_{zz}+\Gamma_{\rho\rho})-\frac{1}{\rho}\left(1+\frac{\rho^2}{r_l^2}\right)\Gamma_\rho=-\frac{4}{c^2}I(\Gamma)\frac{\mbox{d}I}{\mbox{d}\Gamma}~,
\end{equation}
where we indicate the partial derivatives of $\Gamma$ with a subscript. The terms $(\rho/r_l)^2$ arise from the Coulomb force $\rho_e \vec{E}$. The pulsar equation is actually the Grad-Shafranov equation for force-free fields. In general, Eq. (\ref{pulsar_eq}) has to be solved numerically. Even for trivial choices of untwisted configurations ($I(\Gamma) = 0$), the corotation velocity of the charged particles distorts the vacuum solutions, resulting in open
field lines near the light cylinder $\rho\sim O(r_l)$ \citep{michel73dis}. The only fully analytical solution including rotation and a smooth matching across the light cylinder is the split twisted monopole presented by \cite{michel73}, in which the field lines are radial and the magnetic field changes sign between the northern and southern hemispheres. A current sheet at the equator is needed to ensure the divergence-free condition. This is a wind-like solution inappropriate to describing the region with closed field lines. Later, the numerical solution by \cite{contopoulos99} for the first time provided a complete description of a magnetosphere, whose closed lines are untwisted, while the wind zone has a configuration that is very similar to the split monopole.

In the case of a nonrotating star (magnetars are slow rotators and this may be a reasonable approximation for some problems), no electric field is induced and the equilibrium equation is simply $\vec{J}\times\vec{B}=0$: the force-free condition only requires that currents flow parallel to magnetic field lines. In this limit, we can reformulate the whole problem with the simple equation
\begin{equation}\label{rotB}
\vec{\nabla}\times\vec{B}=\alpha(\Gamma)\vec{B}~,
\end{equation}
where the function $\alpha(\Gamma)$ is related to the enclosed current function introduced in Eq. (\ref{Bt}) by
\begin{equation}\label{alpha_I}
  \alpha(\Gamma)=\frac{2}{c}\frac{\mbox{d}I(\Gamma)}{\mbox{d}\Gamma}~.
\end{equation}
The pulsar equation in spherical coordinates is now expressed as
\begin{equation}\label{force-free}
\frac{\partial^2\Gamma}{\partial r^2} - \frac{\cos\theta}{\sin\theta}\frac{1}{r^2}\frac{\partial\Gamma}{\partial \theta} + \frac{1}{r^2}\frac{\partial^2\Gamma}{\partial \theta^2}=-\alpha(\Gamma)\int \alpha(\Gamma)\mbox{d}\Gamma~.
\end{equation}
Next we expand the magnetic flux function in Legendre polynomials, $P_l(\mu)$:
\begin{equation}\label{potential_gen}
  \Gamma(r,\mu)=\Gamma_0\sum_l \frac{r}{r_\star}a_l(r)(1-\mu^2)\frac{d P_l(\mu)}{d\mu},
\end{equation}
where $\mu=\cos\theta$, $a_l(r)$ is the dimensionless radial function, $r_\star$ the radius of the star, and $\Gamma_0$ the magnetic flux normalization. Denoting the strength at pole by $B_0$, we hereafter choose
\begin{equation}\label{gamma0}
  \Gamma_0=\frac{B_0r_\star^2}{2}~.
\end{equation}
This leads to the following expression for the poloidal magnetic field components:
\begin{eqnarray}\label{B_leg}
  B_r       &=&  \frac{B_0}{2}\frac{r_\star}{r}\sum_l l(l+1)P_l(\mu)a_l(r) ~, \nonumber\\
  B_\theta  &=&  - \frac{B_0}{2}\frac{r_\star}{r}\sqrt{1-\mu^2}\sum_l \frac{d P_l(\mu)}{d\mu}\frac{d (ra_l(r))}{dr}~.
\end{eqnarray}
We can obtain the governing differential equation from Eq. (\ref{force-free}):
\begin{eqnarray}\label{ode_leg_pre}
  \frac{B_0r_\star}{2}(1-\mu^2)\sum_l \frac{d P_l(\mu)}{d\mu}\left[\frac{d^2 (ra_l(r))}{dr^2}-l(l+1)\frac{a_l(r)}{r}\right]= &&\nonumber\\
  =-\alpha(\Gamma)\int \alpha(\Gamma)\mbox{d}\Gamma~. &&
\end{eqnarray}
Finally, by using the orthogonality relations of Legendre polynomials, we have
\begin{eqnarray}\label{ode_leg}
  B_0r_\star\frac{l(l+1)}{2l+1}\left[\frac{d^2 (ra_l(r))}{dr^2}-l(l+1)\frac{a_l(r)}{r}\right]= && \nonumber\\
  =-\int_{-1}^1\frac{d P_l(\mu)}{d\mu}\left[\alpha(\Gamma)\int \alpha(\Gamma)\mbox{d}\Gamma\right]\mbox{d}\mu~. && 
\end{eqnarray}
The righthand side depends on the functional form of $\alpha(\Gamma)$ and is responsible for the coupling between different multipoles. For constant $\alpha$ the equation has analytical solutions (see Sect. \ref{sec_bessel}), while the problem is more difficult for other choices. The rest of this section is devoted to reviewing some possible choices that make the problem (semi)analytically solvable in the nonrotating case, which applies to magnetars.

\subsection{Constant $\alpha$: Bessel solutions}\label{sec_bessel}
The choice of constant $\alpha=k/r_\star$, where $k$ is a dimensionless parameter, leads to decoupled equations for each single multipole, so that for each $l$, we must independently solve the equation
\begin{equation}\label{ode_alpha2}
  r^2\frac{d^2 a_l(r)}{dr^2}+2r\frac{d a_l(r)}{dr}+\left[\left(k\frac{r}{r_\star}\right)^2 - l(l+1)\right]a_l(r)=0.
\end{equation}
The analytic solutions of this equation are the spherical Bessel functions of the first and second kinds:
\begin{eqnarray}\label{spher_bessel}
  && J_l(x)=(-x)^l\left(\frac{1}{x}\frac{\partial}{\partial x}\right)^l\frac{\sin x}{x}~,\\
  && Y_l(x)=(-x)^{l+1}\left(\frac{1}{x}\frac{\partial}{\partial x}\right)^l\frac{\cos x}{x}~,
\end{eqnarray}
where $x=kr/r_\star$. The constant ratio $r_\star/k$ gives the typical scalelength of the magnetic field. The physical solutions (from which the vacuum solution $B_{p}\propto r^{-(l+1)}$ can be recovered in the limit $k\rightarrow 0$) are represented by the functions of the second kind, explicitly:
\begin{eqnarray}\label{sol_alpha2}
  && a_l(r)=c_lY_l(x)~,\nonumber\\
  && B_r=\frac{B_0}{2}\frac{r_\star}{r}\sum_l l(l+1)P_l(\mu)c_{l}Y_l(x)~,\nonumber\\
  && B_\theta=-\frac{B_0}{2}\frac{r_\star}{r}\sqrt{1-\mu^2}\sum_l \frac{d P_l(\mu)}{d\mu}c_{l}\frac{d (xY_l(x))}{d x}~,\\
  && B_\phi=k\frac{B_0}{2}{\sqrt{1-\mu^2}}\sum_l \frac{d P_l(\mu)}{d\mu}c_{l}Y_l(x) ~,\nonumber
\end{eqnarray}
where $c_l$ is the weight of the $l$-multipole.

This choice of constant $\alpha$ has been followed in several studies of the solar corona \citep{chiu77,seehafer78} or applied to the region with open field lines of the pulsar magnetosphere \citep{scharlemann73}. The $Y_l(x)$ functions are oscillatory at large distances, and the magnetic field components change sign as $r$ varies (at fixed parameter $k$). The problem with this family of analytical solutions is that, at large distances, all components (which have the same radial dependence for any $l$) decay too slowly: $B_r\rightarrow r^{-2}, B_\theta\rightarrow r^{-1}, B_\phi\rightarrow r^{-1}$. Thus this configuration cannot be a solution for the whole space, as it would imply infinite magnetic energy in an infinite volume. Also, these solutions cannot be continuously matched with vacuum, because it would require that, at the same radius $r_{out}$, $B_\phi(r_{out})=0$ and $B_r(r_{out})\neq 0$, a condition that cannot be satisfied because those two components are both proportional to the $Y_l(x)$ functions and have the same zeros.

\subsection{Self-similar models}\label{sec_selfsimilar}
\cite{low90}, followed by \cite{wolfson95} and other authors, studied a particular class of self-similar solutions to describe the opening of the solar coronal magnetic field lines due to their shear. \cite{thompson02} applied the same approach in the magnetar framework. Assuming a radial power-law form for the magnetic flux function and a radial dependence $\alpha\propto 1/r$,
\begin{eqnarray}\label{pot_tlk}
  && \Gamma=\Gamma_0 \left(\frac{r_\star}{r}\right)^pF(\mu)\label{gamma_tlk} ~,\\
  && \alpha=\frac{k_{ss}}{r}|F(\mu)|^{1/p}\label{alpha_tlk}=\frac{k_{ss}}{r_\star}\left(\frac{|\Gamma|}{\Gamma_0}\right)^{1/p} ~,
\end{eqnarray}
the enclosed current function is
\begin{equation}\label{tlk_i}
  I(\Gamma)=I_0\left|\frac{\Gamma}{\Gamma_0}\right|^{1+1/p}\label{I_tlk} ~,
\end{equation}
where
\begin{equation}\label{I0}
  I_0=k_{ss}\frac{p}{4(p+1)}cB_0r_\star
\end{equation}
is the current normalization, related to the amount of current, like the dimensionless parameter $k_{ss}$. Equation (\ref{force-free}) becomes a nonlinear second-order differential equation for the angular function $F(\mu)$:
\begin{equation}\label{ode_tlk}
  (1-\mu^2)F''(\mu)+p(p+1)F(\mu)=-k_{ss}^2\frac{p}{p+1}F(\mu)|F(\mu)|^{2/p}
\end{equation}
\noindent where primes stand for derivatives of $F(\mu)$ with respect to $\mu$. The magnetic field is given by
\begin{eqnarray}\label{bfield_tlk}
 B_r       &=& -\frac{B_0}{2}\left(\frac{r_\star}{r}\right)^{(p+2)}F'(\mu) ~,\nonumber\\
 B_\theta  &=&  \frac{B_0}{2}\left(\frac{r_\star}{r}\right)^{(p+2)}p\frac{F(\mu)}{\sqrt{1-\mu^2}} ~,\\
 B_\phi &=&  k_{ss}\frac{B_0}{2}\left(\frac{r_\star}{r}\right)^{(p+2)}\frac{p}{p+1}\frac{F(\mu)|F(\mu)|^{1/p}}{\sqrt{1-\mu^2}}~.\nonumber
\end{eqnarray}

\begin{figure}
\centering
\includegraphics[width=7.5cm]{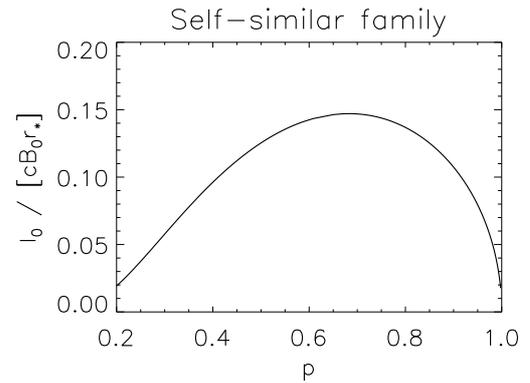}
\caption{Curve $I_0(p)$ describing the family of self-similar twisted dipoles.}
\label{fig_par_ss}
\end{figure}

The three components of the magnetic field have the same radial dependence, so we are seeking for self-similar solutions. The maximum line twist, which is the one suffered by a polar line ($\Gamma\rightarrow 0$), is called global twist, and it can be taken as the parameter that uniquely describes the $l$-pole family of self-similar solutions. Due to self-similarity, Eq. (\ref{def_twist}) for $\Gamma\rightarrow 0$ becomes very straightforward:
\begin{equation}\label{def_twist_tlk}
  \Delta \phi=\frac{2k_{ss}}{p+1}\int_0^1 \frac{|F(\mu)|^{1/p}}{1-\mu^2} \mbox{d}\mu
\end{equation}

Equation (\ref{ode_tlk}) has to be numerically solved by imposing three physical requirements: $F(1)=F(-1)=0$ (on the axis, only the radial field component is allowed), and $F'(1)=-2$ (to fix the normalization). Given a value of $k_{ss}$, one can find an infinite number of solutions characterized by an eigenvalue $p$. Each solution represents a different multipole (except the dipole for which there are two solutions). Fixing $k_{ss}=0$ (i.e. $I_0=0$, no current) implies an integer value of $p$, so that we recover the vacuum multipolar solution with $F(\mu)=(1-\mu^2)dP_p(\mu)/d\mu$ where $P_p$ is the $p^{th}$ Legendre polynomial. For each multipole, a unique relation $k_{ss}(p)$ (or $I_0(p)$ once the values of $B_0$ and $r_\star$ are fixed) defines the family of solutions. In Fig. \ref{fig_par_ss} we show the curve of parameters we obtained for the twisted dipole family. Our results agree with the dipolar solutions of \cite{thompson02} and the higher multipole solutions of \cite{pavan09}. We have used these models in the numerical code (section Sect. \ref{sec_numerical}) for testing purposes and for the sake of comparison with other numerical solutions.
\par
However, all these solutions are of limited generality because they have a defined symmetry with respect to the equator, and a linear combination of solutions for different multipoles is not a new solution, due to the nonlinear character of the problem.

\subsection{General axisymmetric solutions}\label{sec_legendre}
The righthand side depends on the functional form of $\alpha(\Gamma)$ and is responsible for the coupling between different multipoles. We have explored several possible analytical forms of $\alpha(\Gamma)$. Aiming at more general solutions, we explore several possible analytical forms of $\alpha(\Gamma)$ in Eq. (\ref{ode_leg}). In search of semianalytical solutions, we consider power laws of the form
\begin{equation}
 \alpha=\frac{k}{r_\star}\left(\frac{\Gamma}{\Gamma_0}\right)^q~,
\end{equation}
with $k$ the dimensionless parameter regulating the amount of currents, and $\Gamma_0$ given by Eq. (\ref{gamma0}).

Obviously the trivial choice $\alpha=0$ gives the vacuum (i.e. untwisted) solutions. The choice of $\alpha$ constant has already been considered in Sect. \ref{sec_bessel}. Some choices with $q<0$, such as $q=-1/2$, present mathematical problems and are definitely not physical, and high values of $q$ make the problem unsolvable with semi-analytical methods. Therefore, only a few simple choices may be useful because of their simplicity. An interesting case is $q=1/2$. Assuming that $\Gamma>0$ in the domain of integration, this particular choice leads to
\begin{equation}
B_\phi=\frac{1}{3}kB_0\frac{r_\star}{r\sin\theta}\left(\frac{\Gamma}{\Gamma_0}\right)^{3/2}~,
\end{equation}
and the following ODE for each radial function $f_l=a_l(r)r/r_\star$:
\begin{eqnarray}\label{ode_coup}
  && \frac{d^2 f_l(r)}{d r^2}-l(l+1)\frac{f_l(r)}{r^2} =-\left(\frac{k}{r_\star}\right)^2\sum_{m,n=1}^\infty f_m(r)f_n(r) g_{lmn}, 
\end{eqnarray}
where each dimensionless Gaunt factor $g_{lmn}$ involves the integral of the product of three Legendre polynomials (see Appendix \ref{app_legendre} for a detailed derivation). Numerically it is $\sim O(1)$. Therefore, the source term couples each multipole $l$ with an infinite number of multipoles $m$ and $n$. One can solve the problem by truncating the series and allowing only for a finite number of multipoles. This approach is more general than other simple cases, but it has many more degrees of freedom.

\section{Numerical solutions}\label{sec_numerical}
In the previous section we have discussed some analytical and semi-analytical solutions that share the same drawback: the arbitrary choice of the enclosed current function $I(\Gamma)$ or, equivalently, $\alpha(\Gamma)$. Some of these solutions are nonphysical, in the sense that they can neither be extended to infinity nor matched to vacuum solutions. All these limitations make the (semi-)analytical approach insufficient for general purposes, because we have no physical argument for preferring one particular form of the current over another. The alternative is to find numerical solutions of the nonlinear, force-free equations describing an NS magnetosphere. We have three main reasons for working in this direction. First, we expect these solutions to be more general and in some cases very different from the semi-analytical ones. By studying new numerical equilibrium solutions, we hope to gain insight into the form of the most realistic choices of enclosed current functions. Second, a reliable numerical code could in principle be extended, if adding the rotationally-induced electric field, to build a consistent corotating magnetosphere model from which the distribution of charge density can be computed directly. Third, the ability to build magnetospheric solutions for any prescribed form of the magnetic field at the surface is very useful for future studies of the dynamics of the magnetosphere coupled with the physics of the crust.

\subsection{The magneto-frictional method}\label{sec_magnetofrictional}
For simplicity, we begin with the two-dimensional nonrotating case, assuming that the force-free approximation is valid up to an outer radius $r_{out}$, which reduces the problem to finding solutions of Eq. (\ref{force-free}). In the so-called magneto-frictional method \citep{yang86, roumeliotis94}, one begins with an initially non-force-free configuration and defines a \textit{fictitious} velocity field proportional to the Lorentz force, $\vec{v}_f=\nu \vec{J}\times\vec{B}/B^2$, where $\vec{J}=\frac{c}{4\pi}\vec{\nabla}\times \vec{B}$.$\nu$ is a normalization constant that sets the time unit in the induction equation. Hereafter we use $\nu=1$. This results in a fictitious electric field:
\begin{eqnarray}\label{magnetofrictional}
 \vec{E}_f=-\vec{v}_f\times \vec{B}~.
\end{eqnarray}
This fictitious electric field enters into the induction equation as a frictional term that forces the solution to relax to a force-free configuration.

In the original method, \cite{roumeliotis94} write the magnetic field as
\begin{equation}\label{magnetofrictional_clebsch}
 \vec{B}=\vec{\nabla}\Gamma\times\vec{\nabla}\beta~,
\end{equation}
where $\Gamma$ is the magnetic flux function defined in Eq. (\ref{Bp}), and $\beta=\phi-\gamma(r,\theta)$, where $\gamma(r,\theta)$ is a scalar potential related with the freedom in the choice of the form of the enclosed current. The induction equation becomes a system of two advection equations for those two functions:
\begin{eqnarray}\label{Roumeqs}
 && \partial_t \Gamma+\vec{v}_f\cdot\vec{\nabla} \Gamma=0\nonumber\\
 && \partial_t \beta+\vec{v}_f\cdot\vec{\nabla} \beta=0~.
\end{eqnarray}
A static solution is achieved if and only if $\vec{v}_f=0$, since the velocity field, $\vec{\nabla}\Gamma$, and $\vec{\nabla}\beta$ are all orthogonal to the magnetic field by definition.

We apply the same idea but, instead of evolving the functions $\Gamma$ and $\beta$, we evolve the magnetic field components directly by solving the induction equation 
\begin{eqnarray}
\partial_t \vec{B} &=&  -\vec{\nabla}\times \vec{E}_f~, \label{dbdt_code}
\end{eqnarray}
where the fictitious electric field can be written as
\begin{eqnarray} 
\vec{E}_f  =\vec{J}-(\vec{J}\cdot \vec{B})\vec{B}/B^2~, \label{ef_code} 
\end{eqnarray}
where $\vec{E}_f$ is a measure of the deviation from the force-free condition, because $\vec{J}\parallel \vec{B}$ is accomplished if and only if $\vec{E}_f\equiv 0$. The main reason for solving the induction equation instead of Eqs. (\ref{Roumeqs}) is to allow for future extensions of the code by considering a real, rotationally-induced electric field. The disadvantage is that we have to be more careful when setting boundary conditions for the electric field, because we could converge to stationary solutions characterized by $\vec{\nabla}\times \vec{E}_f=0$, which are not necessarily force-free.

\subsection{Linear analysis of the magneto-frictional method}\label{sec_stability}
We now consider a background, uniform magnetic field $\vec{B}_0$ and a small perturbation $\delta \vec{B}\propto e^{i(\vec{k}\cdot\vec{r}-\omega t)}$. In the linear regime, the equations read as
\begin{eqnarray}
 && \delta \vec{J} = \frac{c}{4\pi}i\vec{k}\times \vec{\delta B}\nonumber\\
 && \delta \vec{E}_f=-\frac{1}{B_0^2}[(\delta \vec{J}\times \vec{B}_0)\times \vec{B}_0]\\
 && \frac{\partial \delta\vec{B}}{\partial t}=-i\omega \delta\vec{B}=-\vec{\nabla} \times \delta \vec{E}_f~.\nonumber
\end{eqnarray}
Explicitly, the last equation can be written as
\begin{equation}\label{dispersion}
 -i \frac{4\pi\omega}{c}\delta\vec{B} = \frac{\vec{k}\times\vec{B}_0}{B_0^2}[\delta\vec{B}\cdot (\vec{k}\times\vec{B}_0)] -k^2 \delta \vec{B}~.
\end{equation}
If the perturbed current is orthogonal to the background magnetic field (either longitudinal perturbations $\delta\vec{B}\parallel \vec{B}_0$ or transverse perturbations with $\vec{k}\parallel\vec{B}_0$), the first term on the righthand side of Eq. (\ref{dispersion}) vanishes and the dispersion relation is purely dissipative:
\begin{equation}
 \omega = -i\frac{c}{4\pi}k^2~.
\end{equation}
Any perturbation of this type will be dissipated on a timescale $\propto k^{-2}$. In contrast, for transverse perturbations with both $\delta\vec{B}$ and $\vec{k}$ orthogonal to $\vec{B}_0$, the current is parallel to the background magnetic field, and the two terms in the righthand side of Eq. (\ref{dispersion}) cancel out, so that the perturbation does not evolve (a neutral mode with $\omega=0$).
Therefore, the magneto-frictional method is designed to dissipate all induced currents nonparallel to the magnetic field but allows for stationary solutions with currents parallel to the field. Since the largest lengthscale in our problem is set by the size of the numerical domain, $\lambda_{max} \pi r_{out}$, the typical diffusion timescale on which we expect to converge to a force-free solution is $t_{dif}\propto r_{out}^2$.

\begin{figure}
\centering
\includegraphics[width=6.5cm]{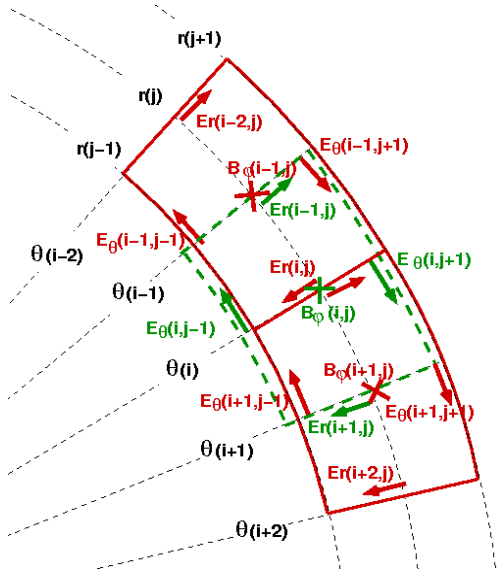}
\caption{Location of the variables in the numerical grid.}
\label{fig_grid}
\end{figure}

\subsection{The numerical method}\label{sec_method}

We work in spherical coordinates ($r,\theta,\phi$) under the assumption of axial symmetry. We employ a fully explicit finite difference time domain (FDTD) method \citep{taflove75} with a numerical grid equally spaced in $\theta$ and logarithmic in the radial direction, unless the outer radius is very close to the surface of the star (located at $r=1$), in which case we employ a linearly spaced radial grid. Our typical resolution varies between 30-200 points in the radial direction and 30-100 points in the angular direction. At each node $(\theta_i,r_j)$, we define all components of $\vec{B}^{(i,j)}$ and consider the three surfaces $S_n^{(i,j)}$ centered on this node and normal to the unit vectors $\vec{\hat{n}}=\vec{\hat{r}},\vec{\hat{\theta}},\vec{\hat{\phi}}$. The elementary areas are calculated by the following exact formulas:
\begin{eqnarray}\label{surfaces}
 && S_r^{(i,j)}      = \int_0^{2\pi} \int_{\theta_{i-1}}^{\theta_{i+1}} r_j^2 \sin\theta\,\mbox{d}\phi\,\mbox{d}\theta 	= 2\pi r_j^2(\cos\theta_{i-1}-\cos\theta_{i+1}) \\
 && S_\theta^{(i,j)} = \int_0^{2\pi} \int_{r_{j-1}}^{r_{j+1}} r\sin\theta_i \,\mbox{d}\phi\,\mbox{d}r        	= \pi(r_{j+1}^2-r_{j-1}^2)\sin\theta_i     \\
 && S_\phi^{(i,j)}   = \int_{\theta_{i-1}}^{\theta_{i+1}} \int_{r_{j-1}}^{r_{j+1}} r\,\mbox{d}\theta\,\mbox{d}r = (\theta_{i+1}-\theta_{i-1})(r^2_{j+1}-r^2_{j-1})/2~.
\end{eqnarray}
Since the the righthand side of Eq. (\ref{dbdt_code}) contains $\vec{\nabla}\times\vec{E}_f$, it is useful to apply Stokes' theorem as follows. The magnetic flux across $S_n^{(i,j)}$ is approximated by 
\begin{equation}\label{fluxes}
\Phi_n^{(i,j)}=B_n^{(i,j)}S_n^{(i,j)}~,
\end{equation}
and the magnetic fluxes are advanced in time using
\begin{equation}\label{derfluxes}
\Phi_n^{(i,j)}(t+\Delta t)=\Phi_n^{(i,j)}(t) -{\Delta}t\oint_{\partial S_n^{(i,j)}} \vec{E}_f \cdot \mbox{d}\vec{l},
\end{equation}
where the numerical circulation of the electric field along the edges of the surface $S_n$ is approximated by using the values of $\vec{E}_f$ in the middle of the edges, whose lengths are
\begin{eqnarray}\label{lengths}
 && l_r^{(i,j)}=(r_{j+1}-r_{j-1})\\
 && l_\theta^{(i,j)}=r_j(\theta_{i+1}-\theta_{i-1})\\
 && l_\phi^{(i,j)}=2\pi r_j\sin\theta_i~.
\end{eqnarray}
Explicitly, the circulation of $\vec{E}_f$ along the edge of each surface $S_n^{(i,j)}$ is
\begin{eqnarray}\label{circulations}
  \oint_{\partial S_r^{(i,j)}} \vec{E} \cdot \mbox{d}\vec{l}=&& 
E_\phi^{(i+1,j)}l_\phi^{(i+1,j)}-E_\phi^{(i-1,j)}l_\phi^{(i-1,j)}\\
  \oint_{\partial S_\theta^{(i,j)}} \vec{E} \cdot \mbox{d}\vec{l}=&& 
-E_\phi^{(i,j+1)}l_\phi^{(i,j+1)}+E_\phi^{(i,j-1)}l_\phi^{(i,j-1)}\\
  \oint_{\partial S_\phi^{(i,j)}} \vec{E} \cdot \mbox{d}\vec{l}=&& 
E_\theta^{(i,j+1)}l_\theta^{(i,j+1)}-E_\theta^{(i,j-1)}l_\theta^{(i,j-1)}\nonumber\\
&& -E_r^{(i+1,j)}l_r^{(i+1,j)}+E_r^{(i-1,j)}l_r^{(i-1,j)}~.
\end{eqnarray}
In Fig. \ref{fig_grid} we show the location of the variables needed for the time advance of $B_\phi^{(i-1,j)}$, $B_\phi^{(i+1,j)}$ (red), and $B_\phi^{(i,j)}$ (green).

We also make use of Stokes' theorem to calculate the current components at each node:
$$J_n^{(i,j)}=\frac{1}{S_n^{(i,j)}}\oint_{\partial S_n^{(i,j)}} \vec{B} \cdot \mbox{d}\vec{l}~.$$
Then, the values of $\vec{J}^{(i,j)}$ and $\vec{B}^{(i,j)}$ directly provide $\vec{E}_f^{(i,j)}$, as defined by Eq. (\ref{ef_code}). For each cell $(i,j)$ the local divergence of $\vec{B}$ is defined as the net magnetic flux flowing across the surfaces divided by the cell volume (fluxes through the toroidal surfaces $S_\phi$ cancel due to axial symmetry):
\begin{equation}\label{divb}
(\vec{\nabla}\cdot\vec{B})^{(i,j)}=[\Phi_r^{(i,j+1)}-\Phi_r^{(i,j-1)}+\Phi_\theta^{(i+1,j)}-\Phi_\theta^{(i-1,j)}]/V^{(i,j)},
\end{equation}
with
\begin{eqnarray}\label{volume}
V^{(i,j)} &=& \int_0^{2\pi} \int_{\theta_{i-1}}^{\theta_{i+1}}\int_{r_{j-1}}^{r_{j+1}} r^2\sin\theta \,\mbox{d}\phi\,\mbox{d}\theta\,\mbox{d}r \nonumber\\
	  &=& \frac{2\pi}{3}(\cos\theta_{i-1}-\cos\theta_{i+1})(r^3_{j+1}-r^3_{j-1}) ~.
\end{eqnarray}
The numerical method ensures that the local divergence is conserved to machine accuracy by construction. As a matter of fact, Eqs.~(\ref{derfluxes}) and (\ref{divb}) imply
\begin{eqnarray}
\frac{d(\vec{\nabla}\cdot\vec{B})^{(i,j)}}{dt} &=& \left[-\oint_{\partial S_r^{(i,j+1)}}\vec{E}\cdot\mbox{d}\vec{l} + \oint_{\partial S_r^{(i,j-1)}}\vec{E}\cdot\mbox{d}\vec{l}\right. \nonumber\\
 && \left.-\oint_{\partial S_\theta^{(i+1,j)}}\vec{E}\cdot\mbox{d}\vec{l} + \oint_{\partial S_\theta^{(i-1,j)}}\vec{E}\cdot\mbox{d}\vec{l} \right]~.
\end{eqnarray}
According to Eqs.~(\ref{circulations}), the last equation is written as a sum of toroidal elements $E_\phi l_\phi$. They are evaluated twice at each of the four surrounding edges (with center $\theta_{i\pm1},r_{j\pm1}$), with opposite sign, so they cancel exactly. Therefore, provided there is an initial divergence-less field, it is guaranteed that the numerical divergence defined by Eq. (\ref{divb}) remains zero to round-off level at each time step.

We must mention that we also tried a method with a staggered grid \citep{yee66}, in which each $n$-component of the magnetic field is defined only at the center of the normal surface, $S_n^{(i,j)}$, while the electric field components are defined in the middle of its delimiting edges. Methods based on staggered grid are well-suited to solving Maxwell's equations, because it provides a natural way to time-advance one field by means of the circulation of the other one. However, we are not dealing with the true Maxwell's equations, but rather with an artificial electric field, Eq. (\ref{ef_code}). Evaluating the dot product $\vec{J}\cdot \vec{B}$ requires the interpolation between two or four values of several of the six mutually displaced components. Considering the red components in Fig. \ref{fig_grid}, for instance, the calculation of $E_r^{(i+1,j)}$ also requires $B_\phi^{(i+1,j)}$, which would be not defined at that location. The unavoidable interpolation errors prevent the code from completely relax to $\vec{E}_f=0$, except in the trivial case of untwisted configurations. For this reason, we decided to work with a standard grid.

\begin{figure*}
 \centering
 \includegraphics[width=.3\textwidth,angle=270]{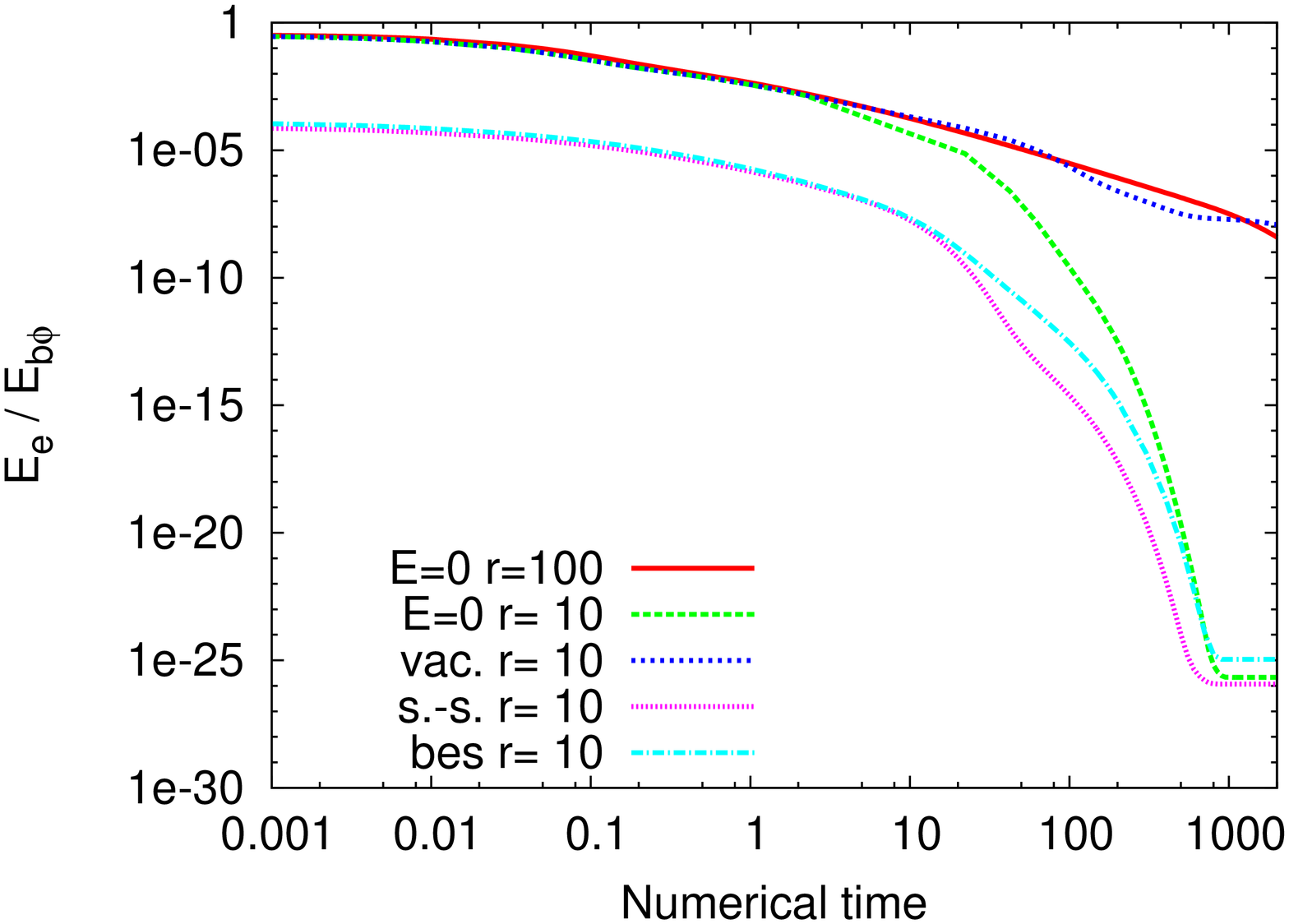}
 \includegraphics[width=.3\textwidth,angle=270]{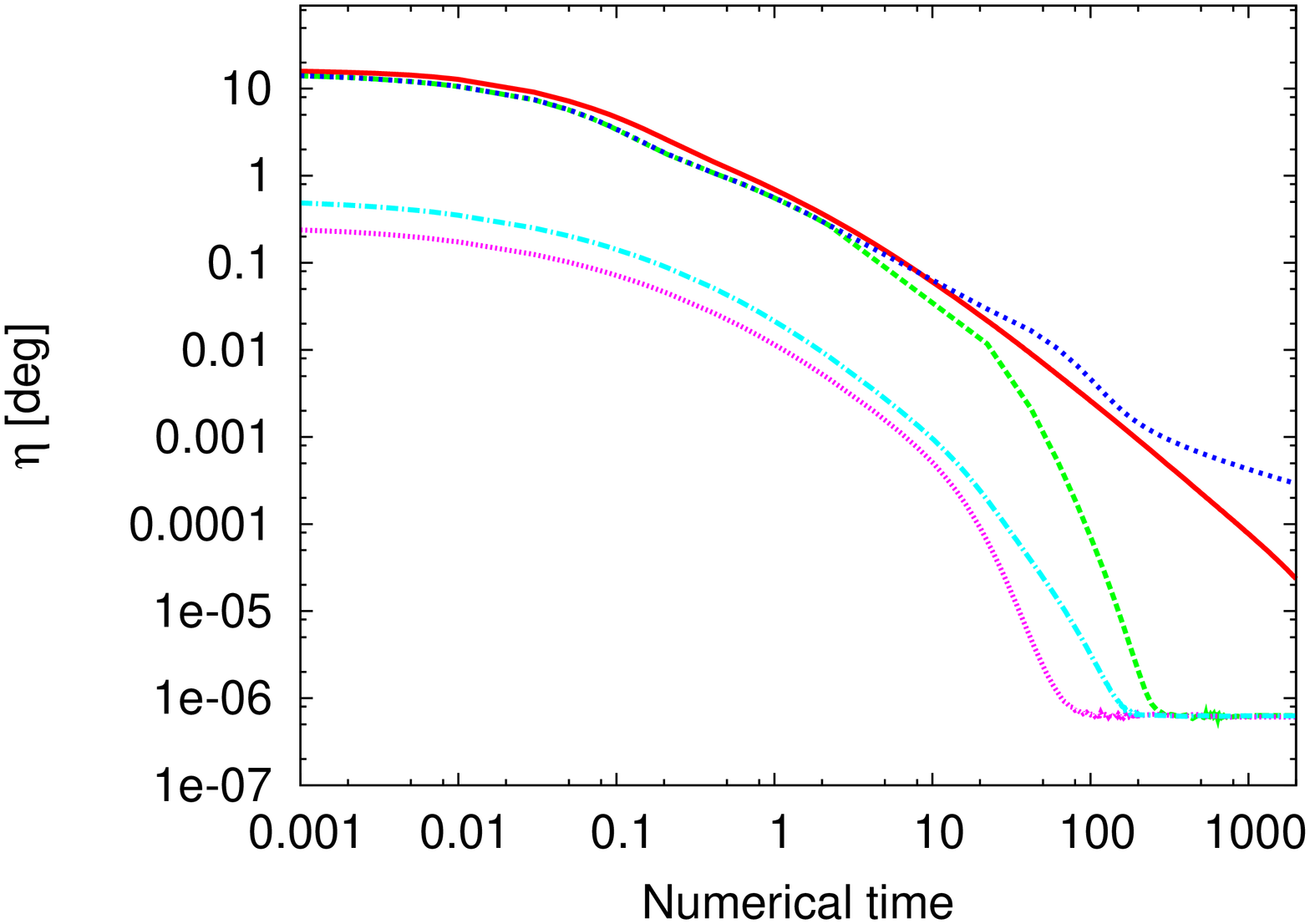}
 \caption{Influence of the outer boundary condition and the location of the external boundary on the evolution of the ratio ${\cal E}_e/{\cal E}_{b\phi}$ (left) and $\bar{\eta}$ (right).}
\label{fig_monitors}
\end{figure*}

\subsection{Boundary conditions}\label{sec_boundary}
At the polar axis, we impose the vanishing of all angular components of magnetic field and currents: $B_\theta=B_\phi=J_\theta=J_\phi=E_\theta=E_\phi=0~.$
At the surface, we have to fix the magnetic field components as provided by some interior solution. However, an arbitrary choice of  poloidal and toroidal fields may not be compatible with a force-free solution. We decided to impose $\vec{E}_f=0$, at the surface, which is equivalent to keeping the value of the radial component fixed at the surface, $B_r(1,\theta)$ and therefore to fixing the angular dependence of the magnetic flux function, $\Gamma(1,\theta)$. As a consequence, $B_\theta$ and $B_\phi$ are allowed to vary on the first radial grid point.

The external boundary is set at $r=r_{out}$. We have explored two different boundary conditions: $\vec{E}_f(r\ge r_{out})=0$ and the continuous matching to external vacuum solutions. The first choice is equivalent to fix the radial component $B_r(r_{out},\theta)$, while allowing for $B_\theta$ and $B_\phi$ to evolve.
Coupling to vacuum solutions can be done using the spectral Legendre decomposition of the radial field at the outer surface, or more precisely, of the magnetic flux function, in terms of which the vacuum boundary condition for each multipole is easily imposed \citep[e.g., as in][]{pons07}. Then we can reconstruct the angular dependence of $B_\theta(r_{out},\theta)$. The vacuum region is characterized by $B_\phi=0$ and the absence of currents or fictitious electric fields. This implies that a current sheet $J_\theta(r_{out})\ne 0$ is needed to ensure current conservation.

If we choose $\vec{E}_f=0$ as outer boundary condition, the code can actually converge to $\vec{E}_f\equiv 0$ at a round-off level, because mathematically this is the only solution compatible with $\vec{\nabla}\times \vec{E}_f=0$. The price to pay is a forced matching of the inner solution with a fixed value of $B_r(r_{out})$. In contrast, if we couple with vacuum, there is no guarantee that the final solution is $\vec{E}_f=0$ everywhere. We discuss below the influence of the different boundary conditions on the results.

\subsection{Convergence criterion and tests}\label{sec_monitors}
Since the magneto-frictional method is based on introducing an artificial, viscous electric field that drives an arbitrary initial configuration into a force-free state, we need a convergence criterion to decide when our solutions are acceptable. For that purpose, we keep track during the run of the following quantities:
\begin{itemize}
 \item Volume-integrated magnetic energy (total and contribution from the toroidal field) 
 $${\cal E}_b=\int \frac{B^2}{2}\mbox{d}V, \quad {\cal E}_{b\phi}=\int \frac{B_\phi^2}{2}\mbox{d}V .$$
 \item Volume-integrated  energy stored in the fictitious electric field 
 $${\cal E}_e=\int \frac{E_f^2}{2}\mbox{d}V .$$
 \item Total volume-integrated helicity 
 $${\cal H}\equiv \int_V B_\phi A_\phi\mbox{d}V$$ 
 (see Appendix \ref{app_helicity} for a discussion about this definition).
 \item Volume-integrated absolute value of the divergence of both, $\vec{B}$ and $\vec{J}$. These two quantities are expected to vanish at round-off level by construction.
 \item An average throughout the entire magnetosphere volume of the local angle between current and magnetic field\footnote{This average weighted with $J^2$ avoids numerical problems in regions where the numerical value of the current is
 very low and the angle $\sin^2\eta=\vec{E}\cdot\vec{J}/J^2$ is numerically ill-defined.}
 \begin{equation}\label{mean_angle}
  \sin^2\bar{\eta}=\frac{\sum  J^2\sin^2\eta}{\sum J^2}=\frac{\sum \vec{E}\cdot\vec{J}}{\sum J^2} .
 \end{equation}
  where the sum is performed over each node $(i,j)$.
 \item A last important cross-check is the consistency of the functions $I(\Gamma)$ and $\alpha(\Gamma)$. First, we check that, for each $n$-component, the three functions $\alpha_n(r,\theta)=4\pi J_n/cB_n(r,\theta)$ are the same. Second, at each radius or angle, the relation (\ref{alpha_I}) has to be satisfied.
\end{itemize}

\begin{table}
\begin{center}
 \begin{tabular}[ht!]{c c c c}
 \hline
$r_{out}$ & $n_r$ & $n_\theta$ & $t_{dis}$  \\
\hline
5        & 30	& 30	   &  $41$ \\
5        & 50	& 100	   &  $35$ \\
10       & 30	& 30	   & $183$ \\
10       & 50	& 30	   & $162$ \\
100      & 50	& 30	   & $180 \times 10^2$ \\
\hline
\end{tabular}
\caption{Time needed to dissipate the numerical currents of the vacuum dipole.}
\label{tab_dip_test}
\end{center}
\end{table}

\begin{figure*}
 \centering
 \includegraphics[width=4cm]{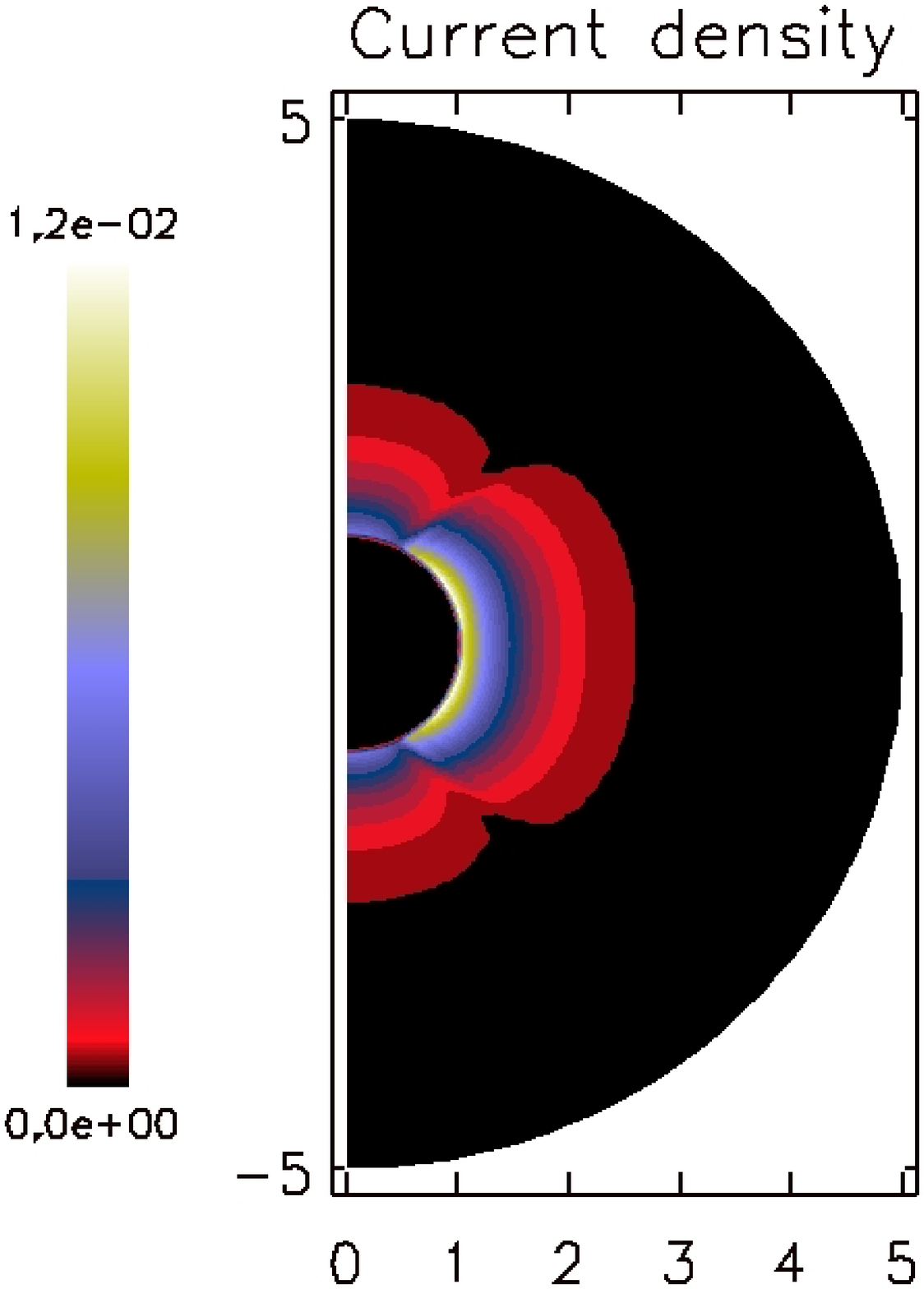}
 \includegraphics[width=4cm]{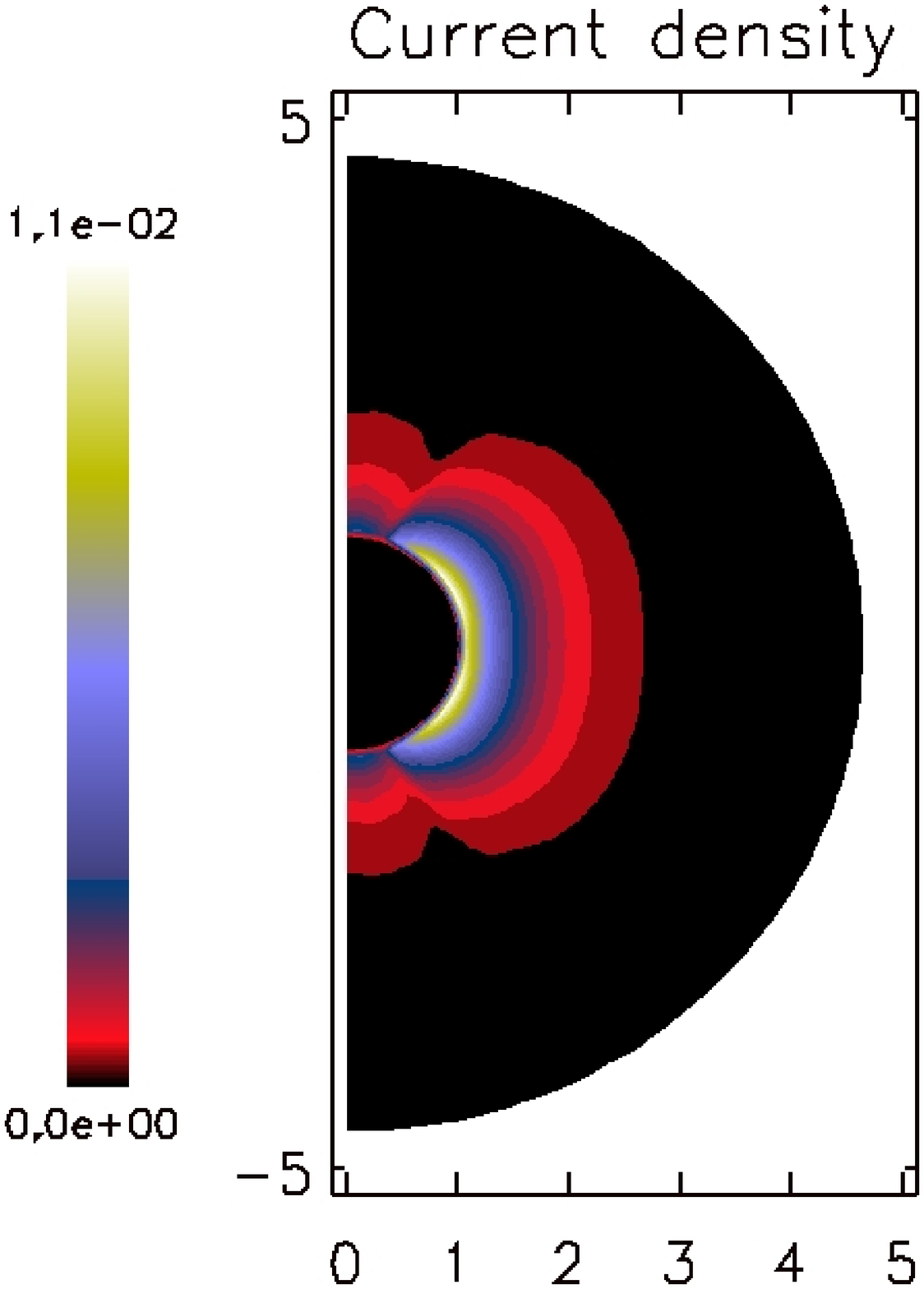}
 \includegraphics[width=4cm]{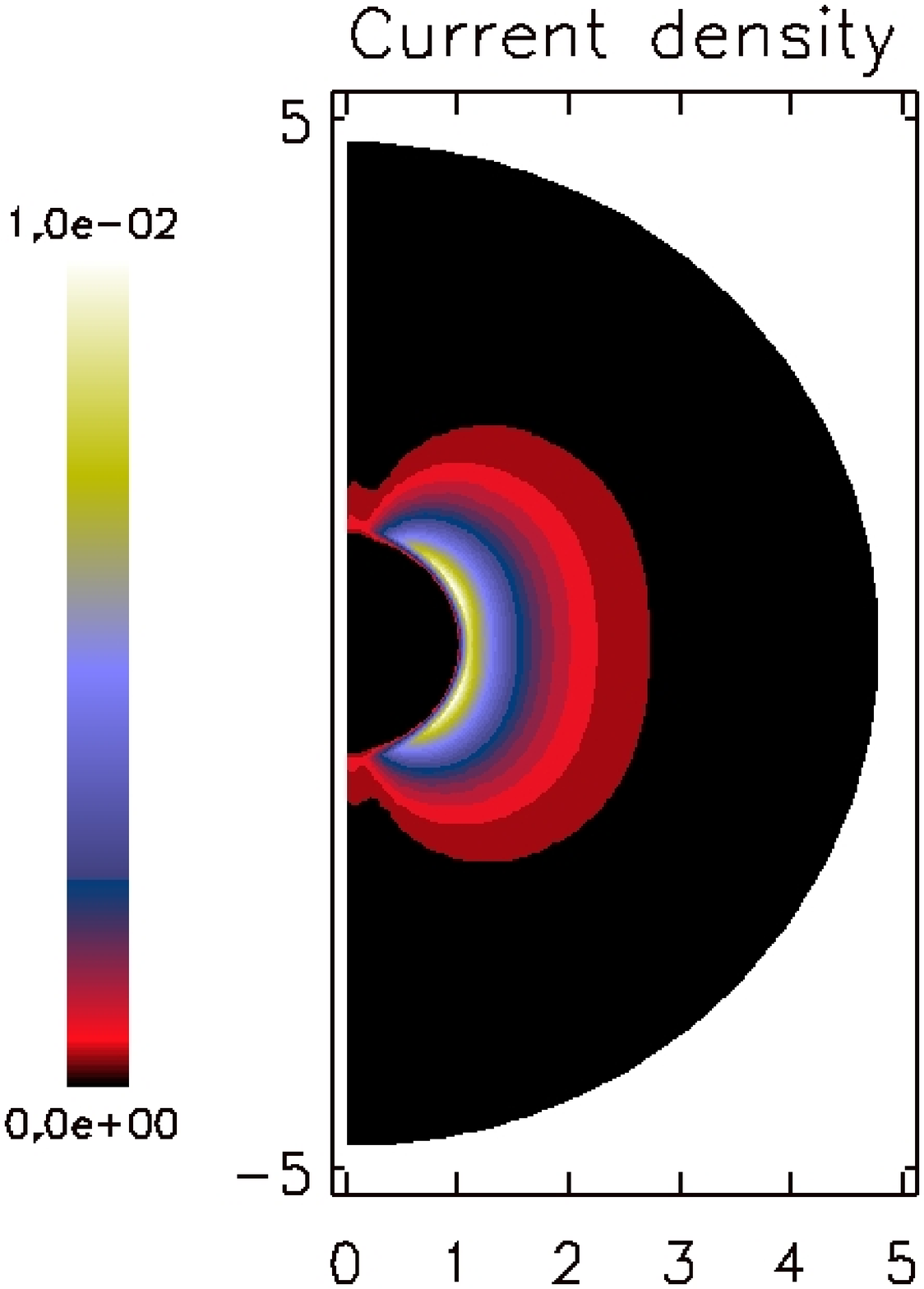}
 \includegraphics[width=4cm]{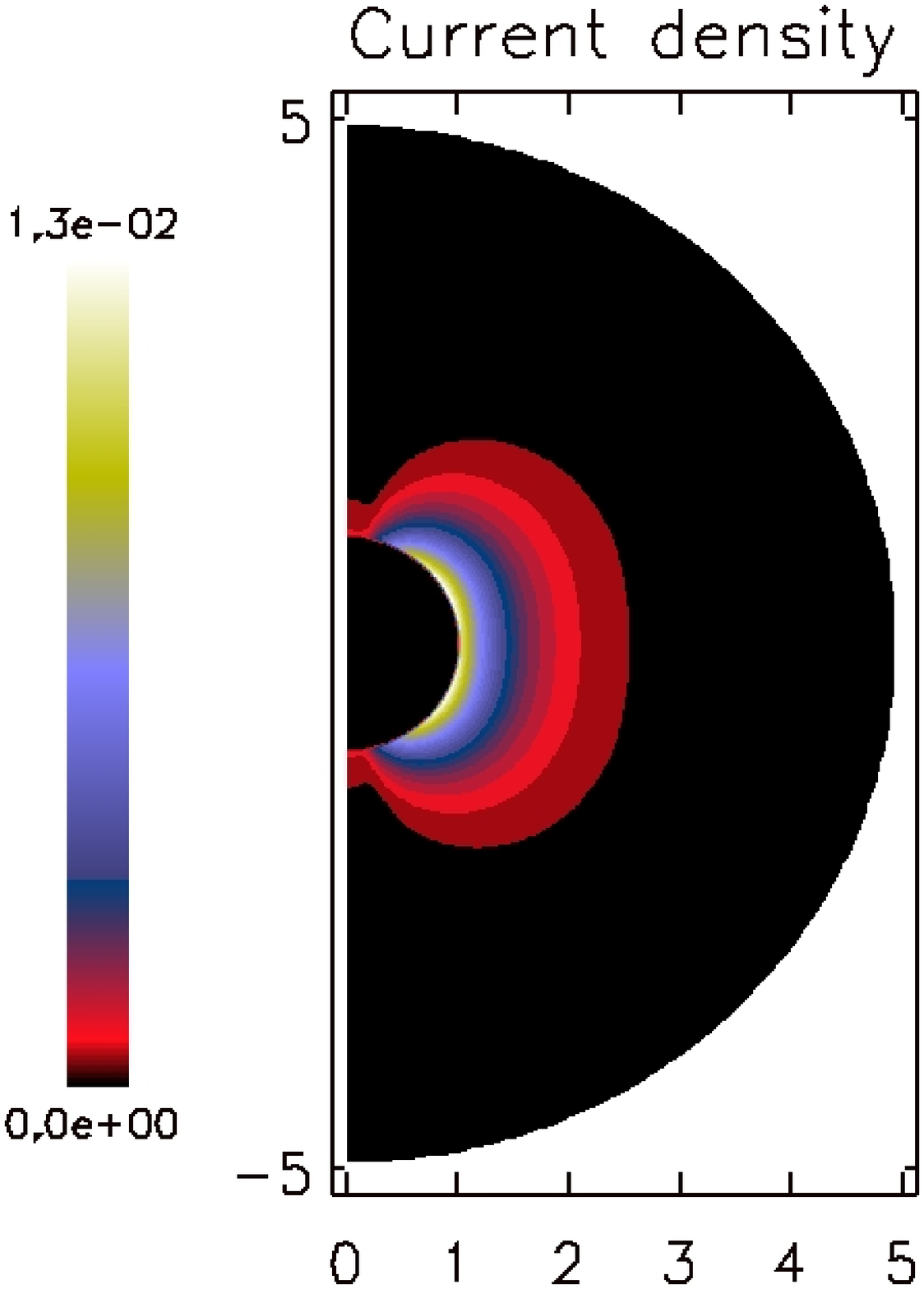}
 \caption{Current density distribution $J$ near the surface for solutions obtained with the same initial data and boundary conditions, $\vec{E}_f=0$, but varying $r_{out}=5,10,50,100$ (left to right).}
\label{fig_jrout}
\end{figure*}

\begin{figure}
 \centering
 \includegraphics[width=.45\textwidth]{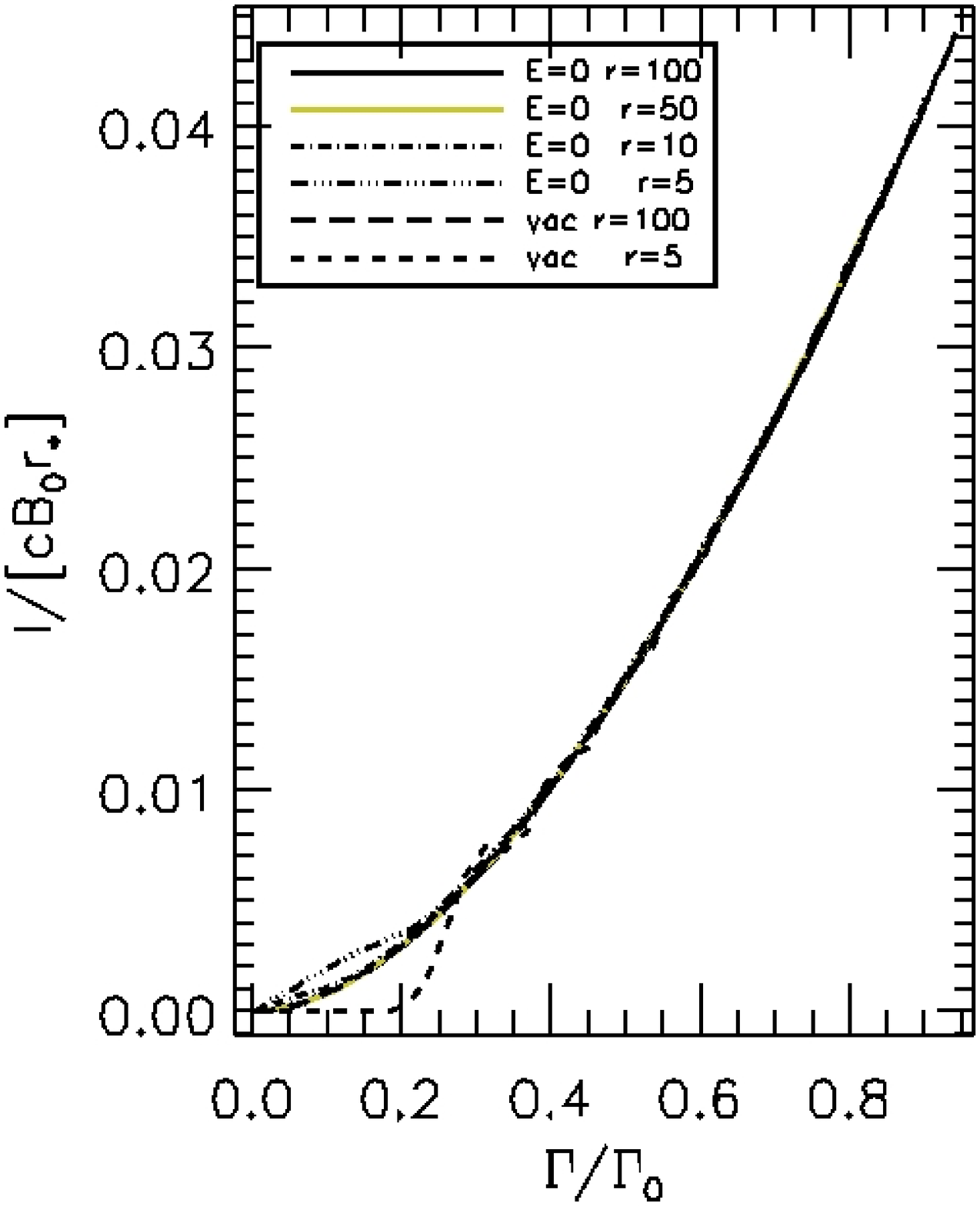}
 \caption{Enclosed current $I(\Gamma)$ for different boundary conditions ($\vec{E}_f=0$ or coupled with vacuum) at $r_{out}=5,10,50,$ or 100.}
\label{fig_igamma_bc}
\end{figure}

Hereafter, we show the electromagnetic quantities in units of $B_0$, $r_\star$, $c$, so the magnetic field $\vec{B}$ scales with $B_0$, the current density $\vec{J}$ with $cB_0/r_\star$, the enclosed current $I$ with $cB_0r_\star$, and the magnetic flux function $\Gamma$ with $\Gamma_0=B_0 r_\star^2/2$.
To test our code and to fix our convergence criteria for the realistic models, we performed a battery of tests. In the first basic test we considered the analytical vacuum dipole with $A_\phi=B_0\sin\theta/2r^2$, $B_\phi=0$ and checked the ability of the code to maintain this solution. Due to the discretization errors, a little numerical toroidal current appears, and consequently a nonvanishing toroidal fictitious electric field. These errors are small ($\Delta {\cal E}_b/{\cal E}_b\lesssim 1\%$), but it is interesting to see how long it takes to dissipate the perturbation to obtain ${\cal E}_e=0$ to machine accuracy. In Table \ref{tab_dip_test} we show the results with different resolutions and values of outer radius, always keeping the timestep close to the maximum value allowed by the Courant condition. The expected behavior $t_{dis}\sim r_{out}^2$ is obtained, with the constant of proportionality depending on the grid resolution, which affects the strength of the initial numerical current.

\par
The second, less trivial test is provided by the analytical models discussed above: the Bessel and self-similar solutions (Sect. \ref{sec_bessel}, Sect. \ref{sec_selfsimilar}). Again we began with an initial model consisting of a known solution, and let the system dissipate the currents that come from discretization errors. To obtain the initial models, the self-similar solutions require the numerical resolution of the nonlinear ODE (\ref{ode_tlk}), while the analytical Bessel solutions are directly implemented. We tried different parameters for the Bessel solutions (varying $k$) and self-similar models (varying the multipole index and the global twist). For every model tested with analytical solutions, we observe a very slight numerical readjustment of the configuration and the code rapidly reaches the relaxed state, with relative changes in ${\cal E}_b$, ${\cal E}_{b\phi}$, ${\cal H}$ less than $\sim 1\%$. We had to pay special attention to work with sufficient radial resolution in the case of highly twisted Bessel models $k\gtrsim 1$, due to their oscillatory radial dependence. For low resolution, the code may find, after a large scale reconfiguration, a completely different solution with smaller $k$, which is numerically more stable. If the resolution is high enough, all analytical solutions are found to be stable. 

We have also studied the evolution of vacuum dipolar solutions with an additional toroidal field for different values of $r_{out}$. In this case the initial currents are due not only to discretization errors, but also to an inconsistency in the initial model. Moreover, the mean angle defined in Eq. (\ref{mean_angle}) initially has a finite value $\bar{\eta}_{in}$. In Fig. \ref{fig_monitors} we show how some convergence monitors evolve as a function of time for three cases with an initial toroidal field of the form $B_\phi=0.1B_0\sin\theta/r^3$ ($\bar{\eta}_{in}=15.1^\circ$), but different external boundary conditions: matching with vacuum at $r_{out}=10$ or imposing $\vec{E}_f=0$ at  $r_{out}=10$ or 100. For comparison, we also show results for two analytical solutions: a Bessel solution and a twisted self-similar model with the same helicity.

Finally, we tested a vacuum dipole perturbed by a weak toroidal field. This is a case of physical interest for quasi periodic oscillations of magnetars, as discussed in \cite{timokhin08} and \cite{gabler11}. Given a background poloidal field described by $\Gamma$, we chose an arbitrary functional form $I(\Gamma)$ and built the toroidal field according to Eq. (\ref{Bt}). As expected, the perturbed configuration is stable and the stationary solution is rapidly reached after a small readjustment.  Typically, for $\max(B_\phi)=0.1\,B_0$, we have $\bar{\eta}_{in}\sim O(1^\circ)$ and changes $\Delta B/B \sim 1\%$.

In general, the magnetic energy is not conserved, since the system has to dissipate part of the current to reach a force-free configuration. This effect is more evident for initial configurations with high helicity. When the outer boundary condition $\vec{E}_f=0$ is imposed, the helicity is conserved within a few percent, as expected (see Appendix \ref{app_helicity} for the helicity conservation theorem), and both electric field and mean angle eventually vanish (to machine accuracy). However, when $r_{out}$ is large, or when vacuum boundary conditions are imposed, configurations with high initial helicity take a much longer time to relax (see Fig. \ref{fig_monitors}). In all cases, the relaxation process is faster near the surface, where the configuration of the magnetosphere is more important for our purposes. 

On the basis of these results, our convergence criteria for accepting that a configuration has reached a force-free state are hereafter ${\cal E}_e/{\cal E}_{b\phi}< 10^{-8}$ and $\bar{\eta}< 10^{-3~\circ}$, plus the requirement that both quantities are monotonically decreasing with time. Some short, initial relaxation phase, in which some large-scale reconfiguration occurs is possible. We chose to compare the electric energy to the magnetic energy contribution from the toroidal field, which is much more restrictive than simply the ratio of electric to magnetic energy, especially for low helicity.

\begin{table}
 \begin{center}
  \begin{tabular}{c c c c c}
Model & $k_{tor}$ & $g_{in}(r)$ & $F_{in}(\theta)$ & $\bar{\eta}_{in}$ [deg]\\
 \hline
A & 0.010  & $r^{-3}$       & $\sin\theta$ & 13.8\\
B & 0.100  & $r^{-3}$       & $\sin\theta$ & 15.1 \\
C & 0.500  & $r^{-3}$	    & $\sin\theta$ & 29.9 \\
D & 0.115  & $r^{-3}$       & $\sin^2\theta$ & 10.7 \\
E & 0.230  & $r^{-5}$       & $\sin^2\theta$ & 25.1 \\
F & 0.234  & $r^{-3}e^{-[(r-1)/0.5]^2}$ & $\sin^2\theta$ & 23.3\\
G & 0.113  & $r^{-3}$       & $\sin^2\theta$ & 34.2 \\
 \end{tabular}
\caption{Parameters defining the initial toroidal field, as indicated in Eq. (\ref{tor_r3}).}
\label{tab_initial}
\end{center}
\end{table}

\begin{table}
 \begin{center}
 \begin{tabular}{c c c c c c}
Model & ${\cal H}$ & $\Delta\phi_{max}$ & $\max(J)$  & $I_0$ & $p$\\
      & $[B_0^2r_\star^3]$ & [rad] & $[cB_0/r_\star]$ & $[cB_0r_\star]$ & \\
 \hline
S1 & 0.21 & 0.5  & $1.8\times10^{-2}$ & 0.061 & 0.97 \\
S2 & 1.11 & 1.6  & $8.1\times10^{-2}$ & 0.15 & 0.69 \\
A & 0.021 & 0.12 & $1.2\times10^{-3}$ & 0.0050 & 1.45 \\
B & 0.21  & 1.2  & $1.2\times10^{-2}$ & 0.049 & 1.40 \\
C & 1.11  & 6.5	 & $4.8\times10^{-2}$ & $\ldots$     & $\ldots$ \\
D & 0.21  & 0.7  & $1.5\times10^{-2}$ & 0.056  & 1.14 \\
E & 0.21  & 0.4  & $4.2\times10^{-2}$ & 0.106  & 0.50 \\
F & 0.21  & 0.4  & $4.3\times10^{-2}$ & $\ldots$     & $\ldots$ \\
G & 0.21  & 0.3  & $2.0\times10^{-2}$ & 0.038  & 2.22 
 \end{tabular}
\caption{Comparison between two self-similar solutions (S1-S2) and our numerical solutions (A-G).}
\label{tab_models}
\end{center}
\end{table}

\section{Results}\label{sec_results}

With the numerical code described above, we can obtain very general solutions of force-free, twisted magnetospheres.
We discuss separately the influence of the following relevant parameters:
\begin{itemize}
 \item the location of the outer radius $r_{out}$ in combination with the external boundary condition;
 \item angular and radial dependence of the initial toroidal field;
 \item initial twist and helicity, fixed by the functional form and the strength of the initial toroidal field;
 \item the geometry of the initial poloidal field.
\end{itemize}
In Fig. \ref{fig_jrout} we compare the distribution of currents in a solution obtained by imposing $\vec{E}_f=0$ at $r_{out}=5,10,50,$ or 100. In all cases the initial poloidal component is a vacuum dipole solution with $A_\phi=B_0\sin\theta/r^2$ and a toroidal field of the form $B_\phi=0.1B_0\sin\theta/r^3$. We observe that, near the axis,
the solutions are clearly affected by the location of the external boundary, if it is not far enough from the surface ($r_{out}\lesssim 10$). In such a case, the influence of the external boundary is important, and it introduces artificial features, although the current distribution in the equatorial region is similar in all cases. The final configurations become almost indistinguishable when $r_{out}=50$ or 100. Taking $r_{out} \gtrsim 100$, we ensure that the numerical noise caused by the interaction with the external boundary is negligible. By neglecting the contribution from the open field lines, for which the twist is ill-defined, the global amount of twist is similar in all models ($\sim 1.2$ rad).

\begin{figure*}[ht!]
\centering
\includegraphics[height=5cm]{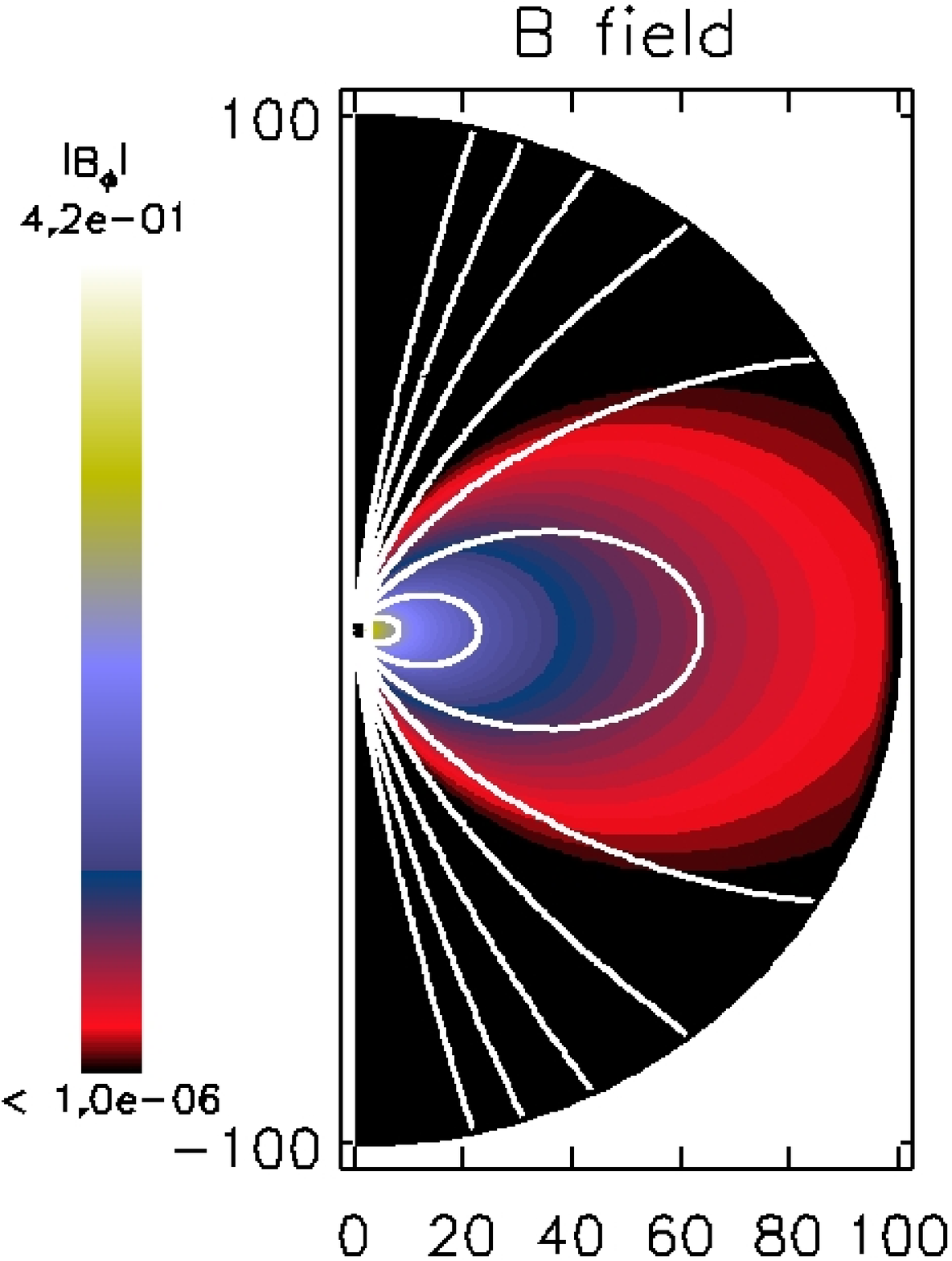}
\includegraphics[height=5cm]{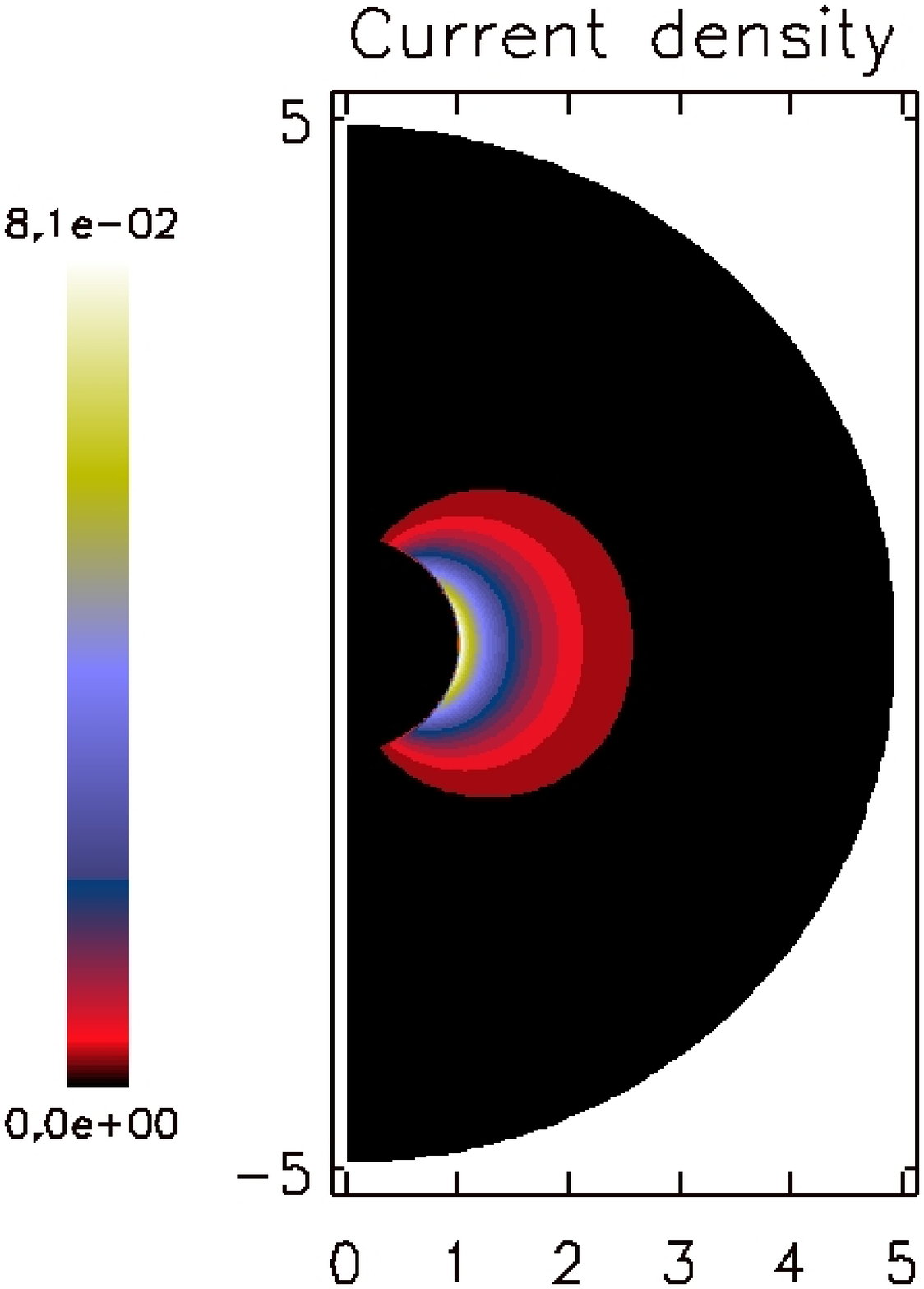}
\includegraphics[height=5cm]{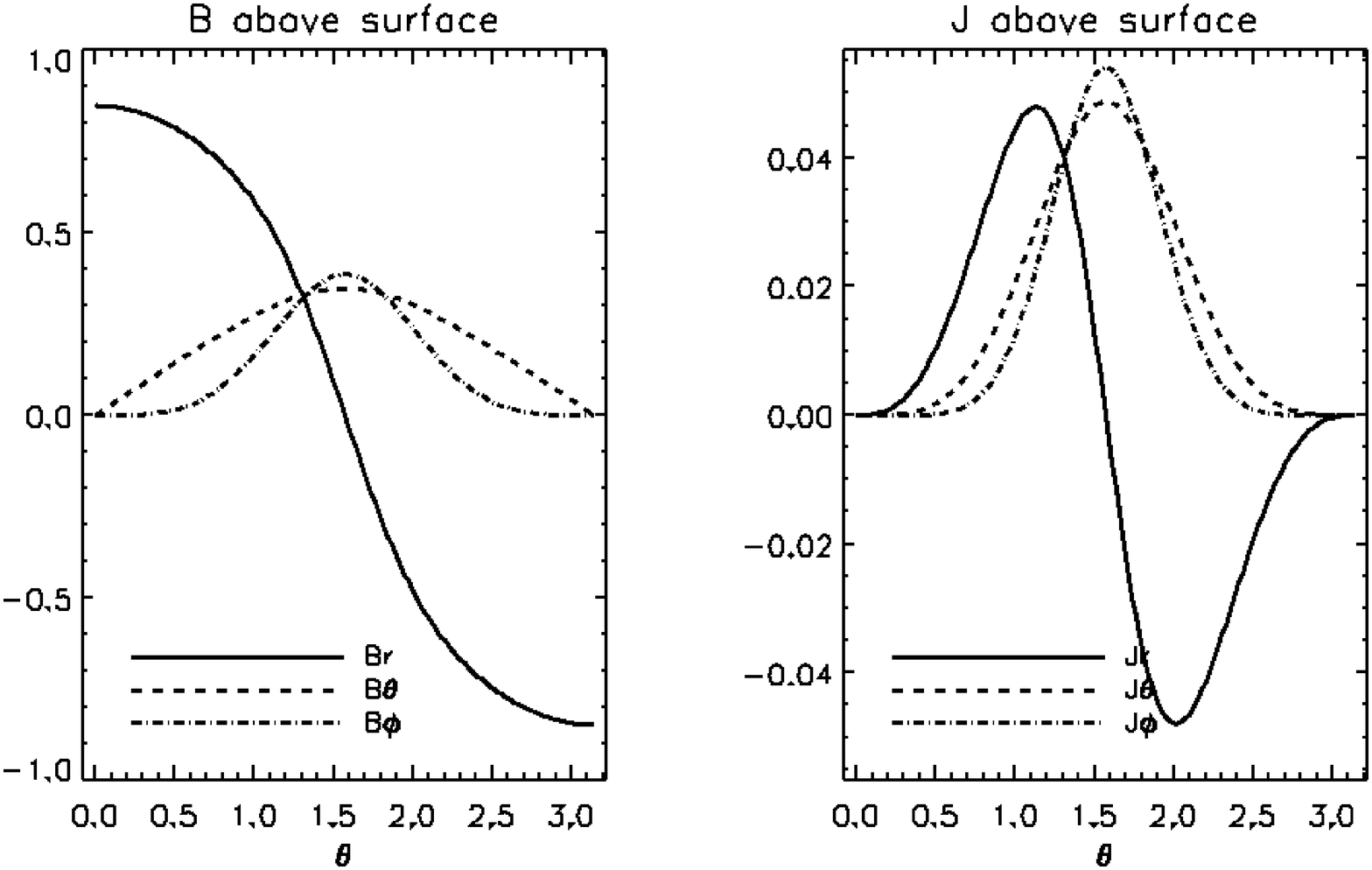}
\caption{Self-similar model S2. From left to right: poloidal magnetic field lines (white) and strength of the toroidal component (colored logarithmic scale); current density distribution $J$ in the near region ($r\le 5$) (colored linear scale); angular profiles of the magnetic field components close to the star surface, and angular profiles of the current density.}
\label{fig_s2}
\end{figure*}

\begin{figure*}[ht!]
\centering
\includegraphics[height=5cm]{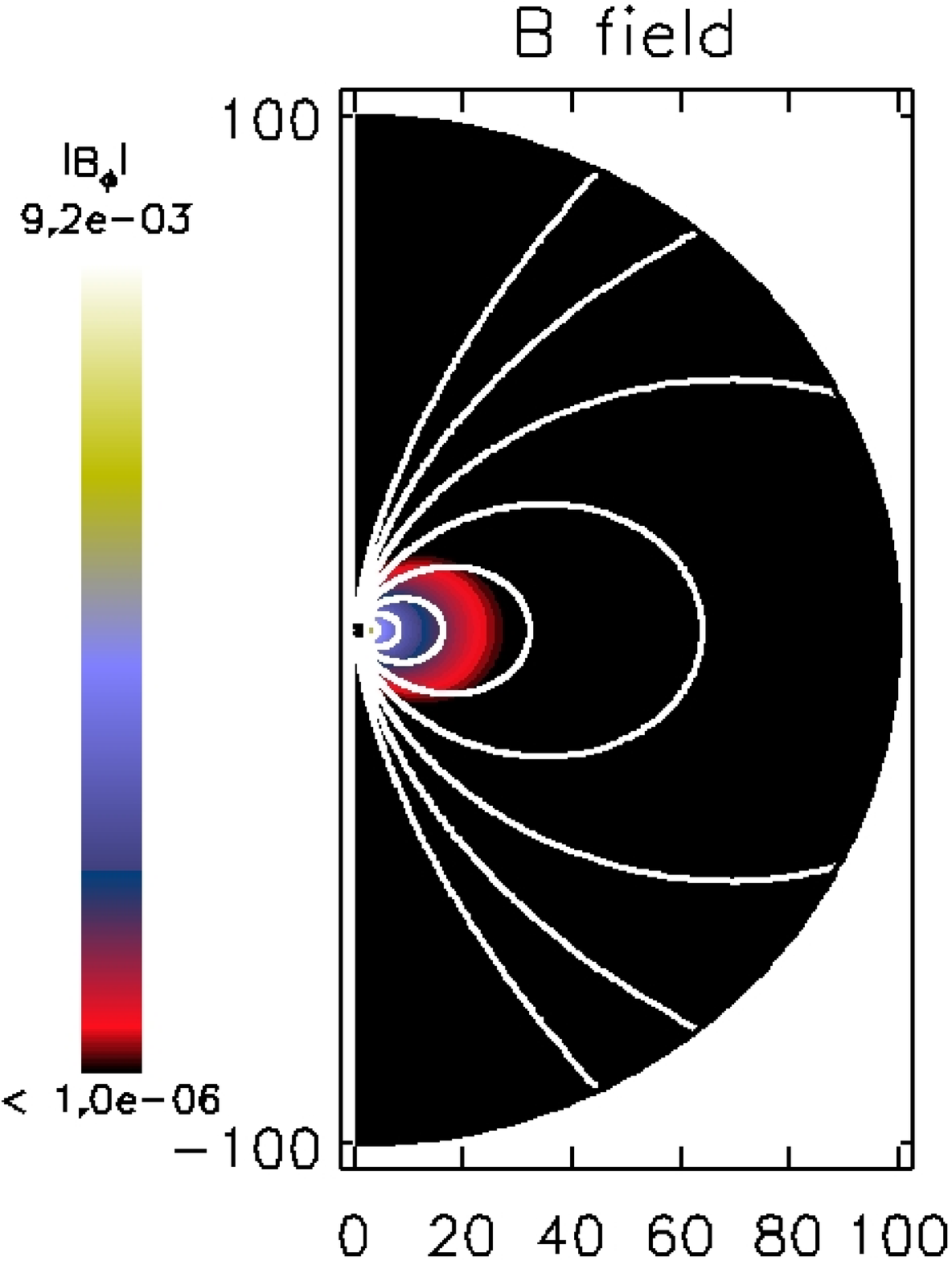}
\includegraphics[height=5cm]{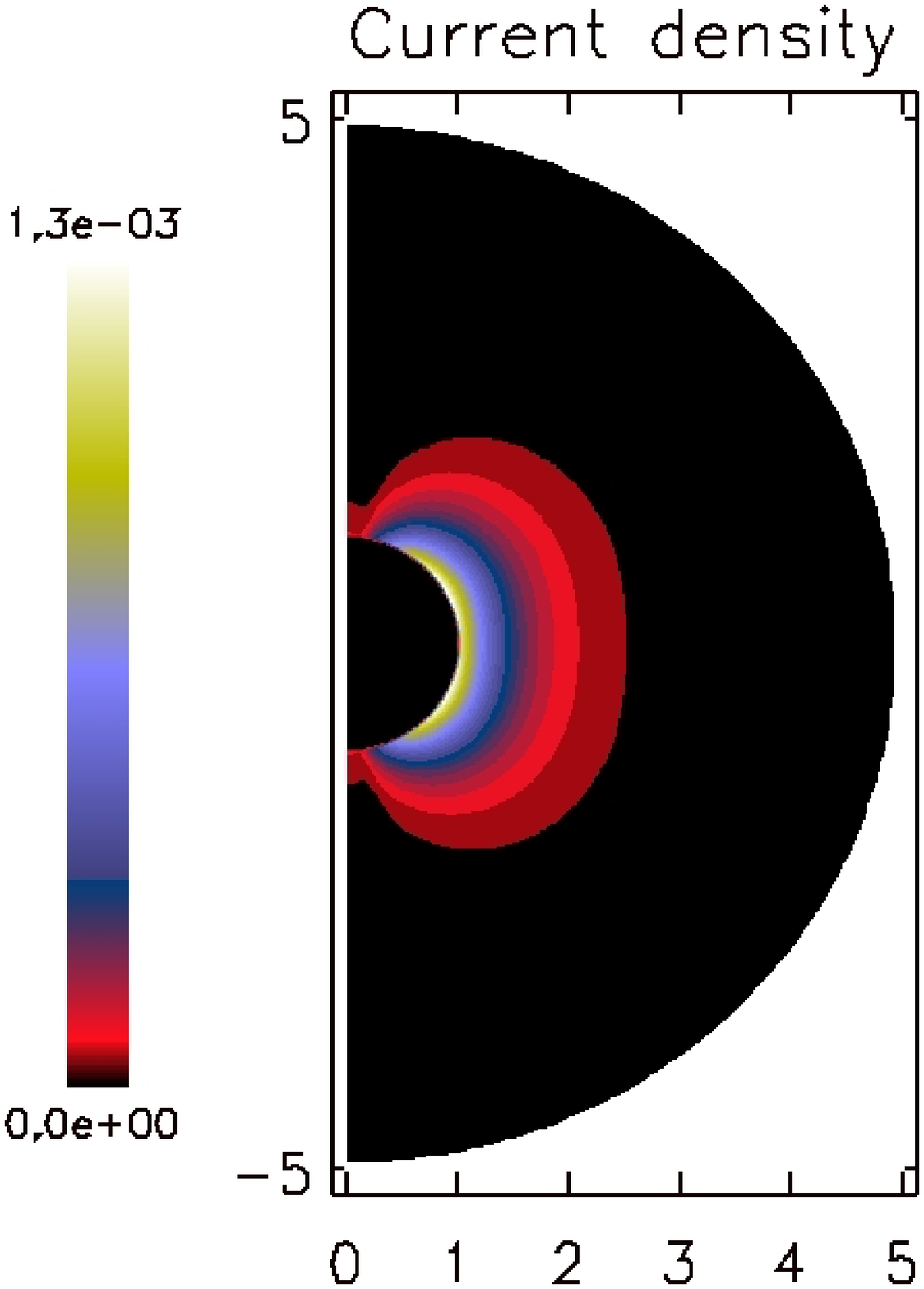}
\includegraphics[height=5cm]{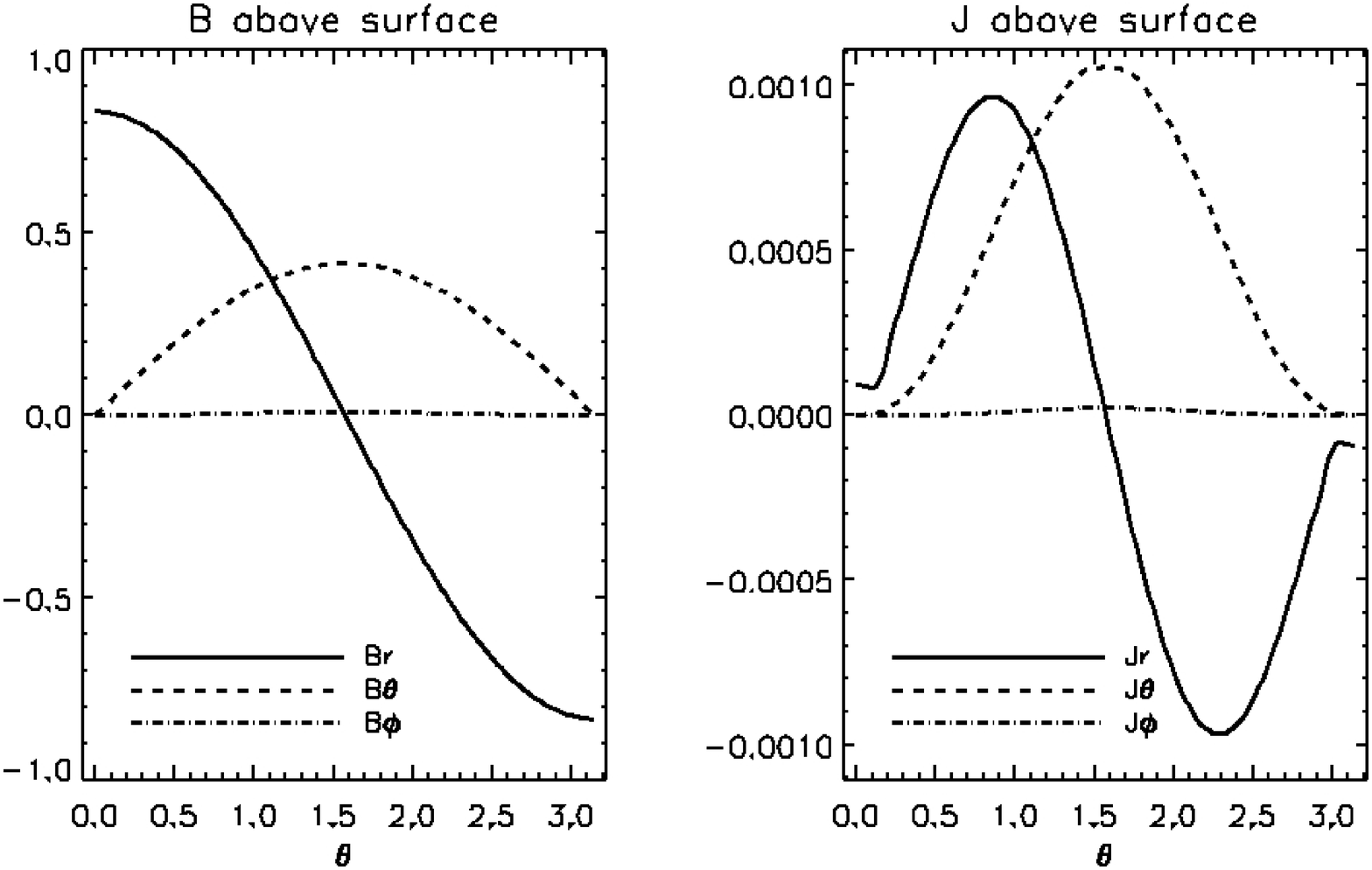}
\caption{Same as Fig. \ref{fig_s2} for model A.} 
\label{fig_moda}
\end{figure*}

\begin{figure*}[ht!]
\centering
\includegraphics[height=5cm]{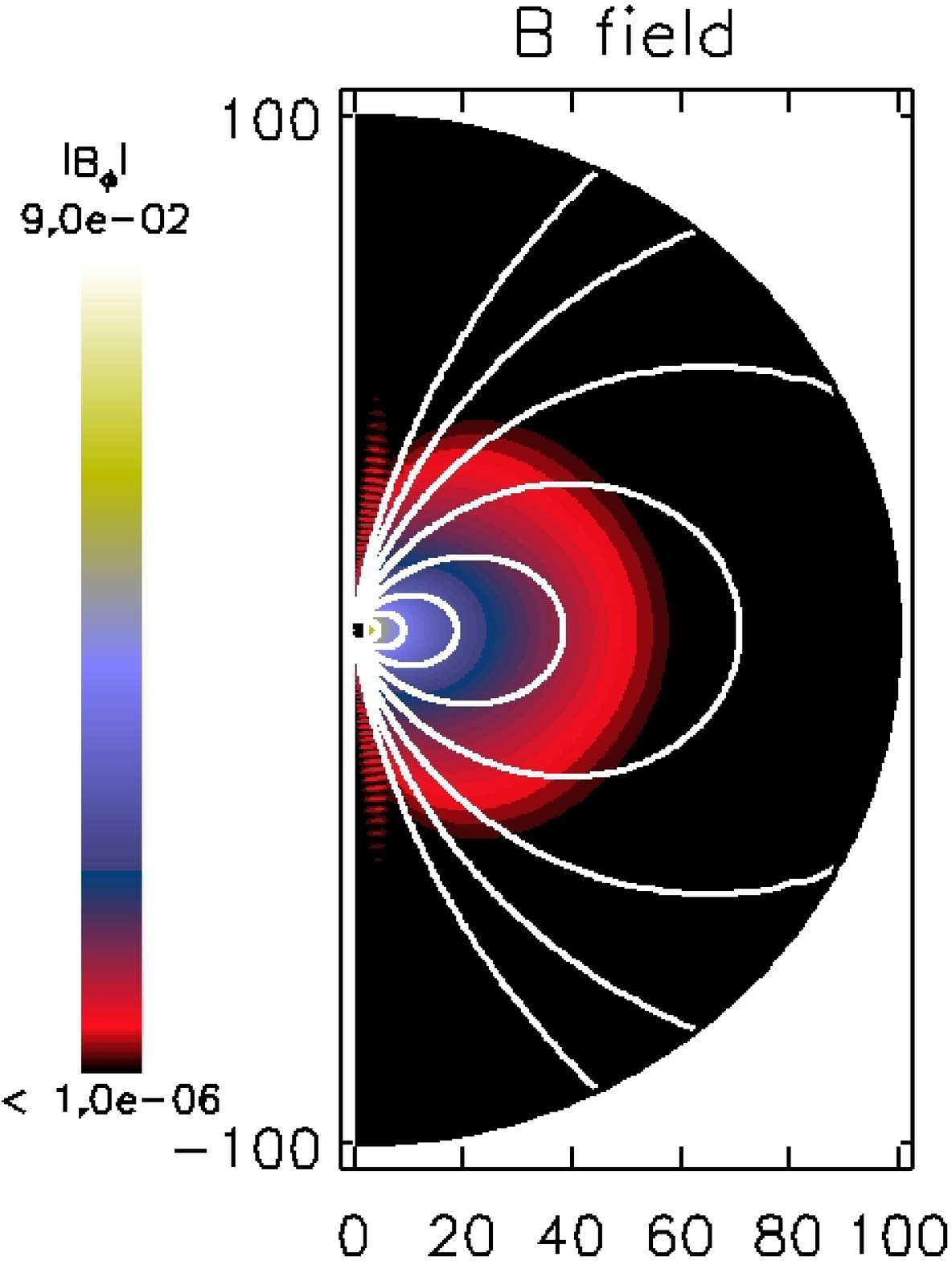}
\includegraphics[height=5cm]{images/jb.eps}
\includegraphics[height=5cm]{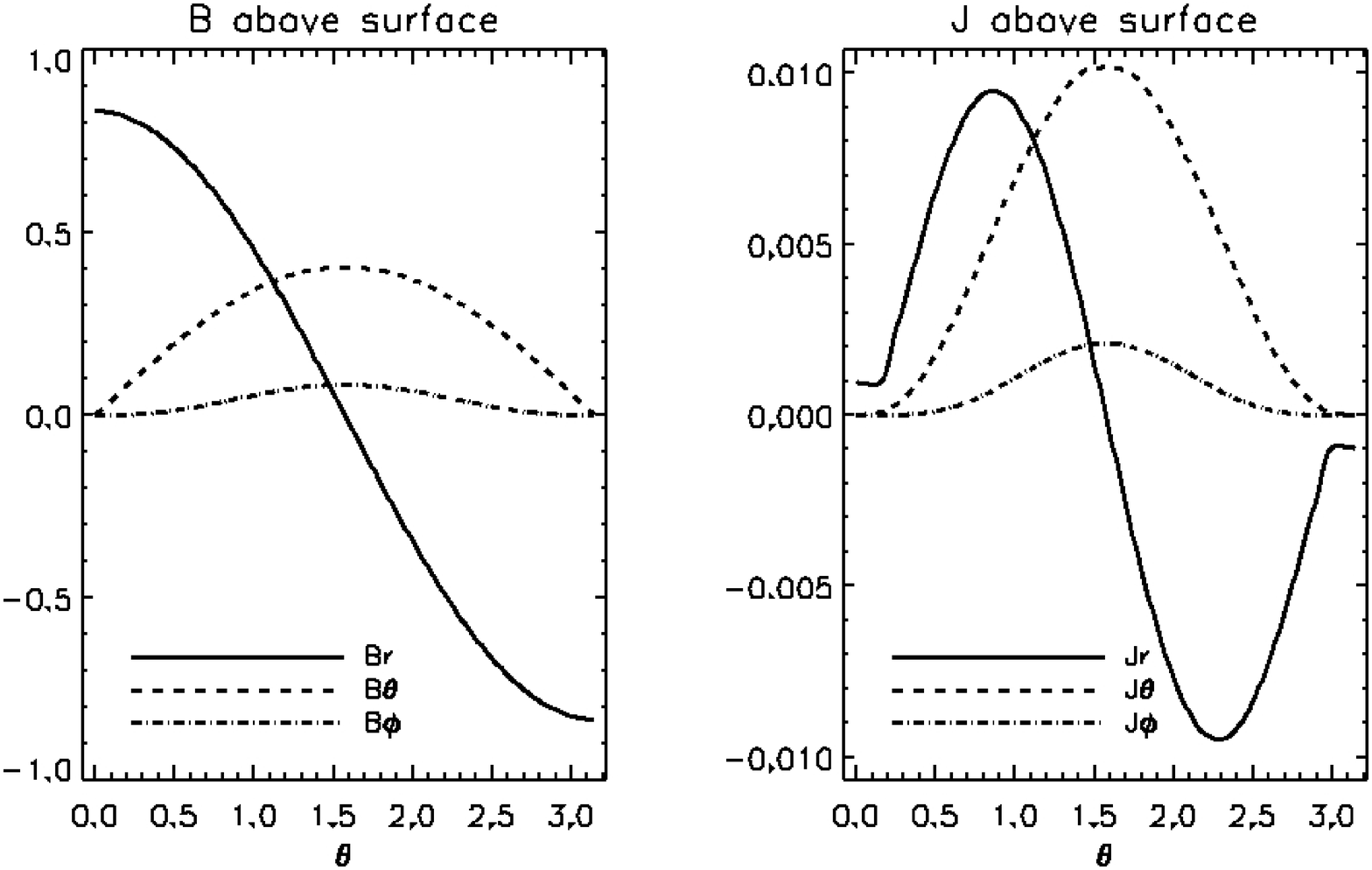}
\caption{Same as Fig. \ref{fig_s2} for model B. } 
\label{fig_modb}
\end{figure*}

\begin{figure*}[ht!]
\centering
\includegraphics[height=5cm]{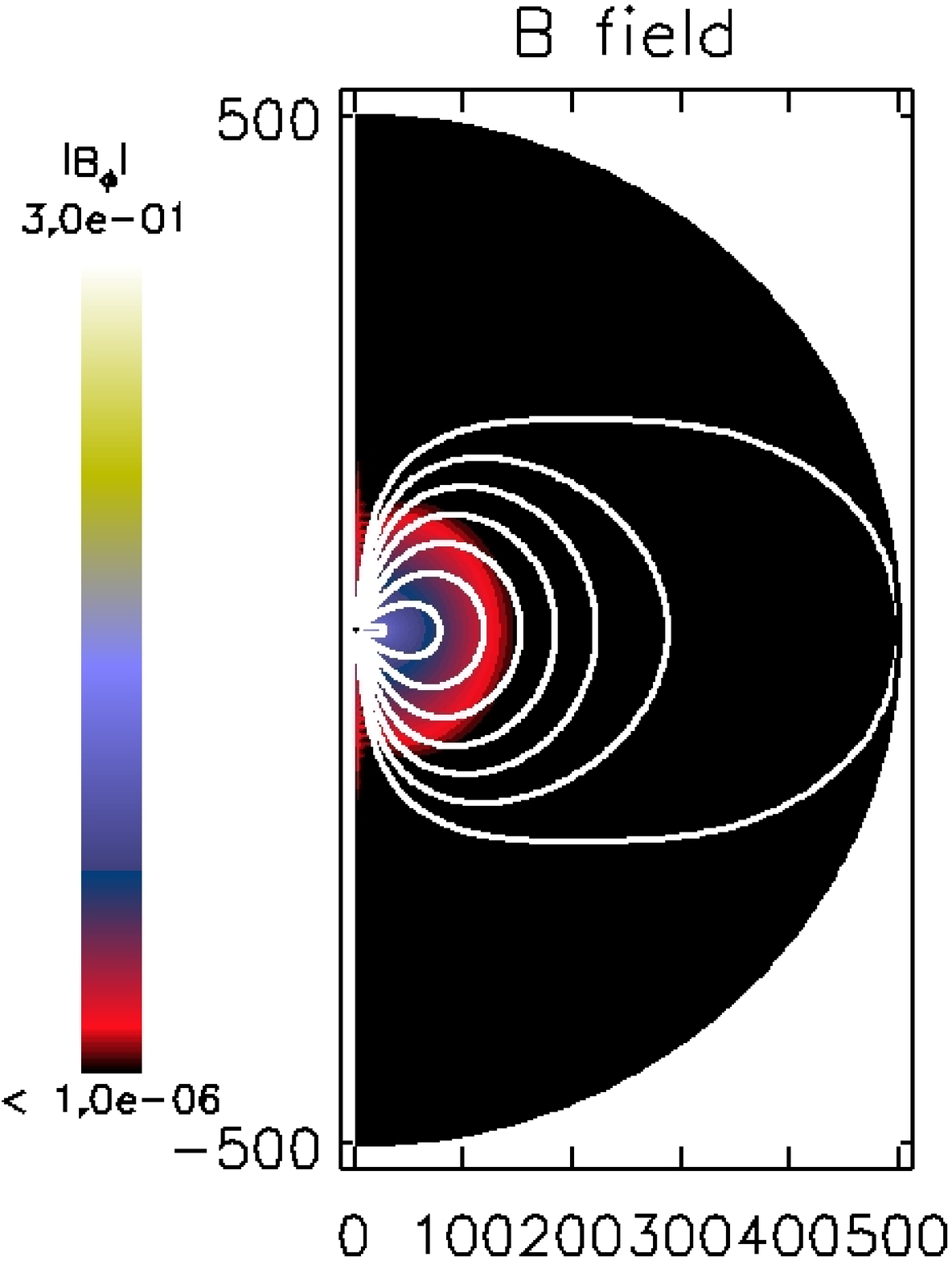}
\includegraphics[height=5cm]{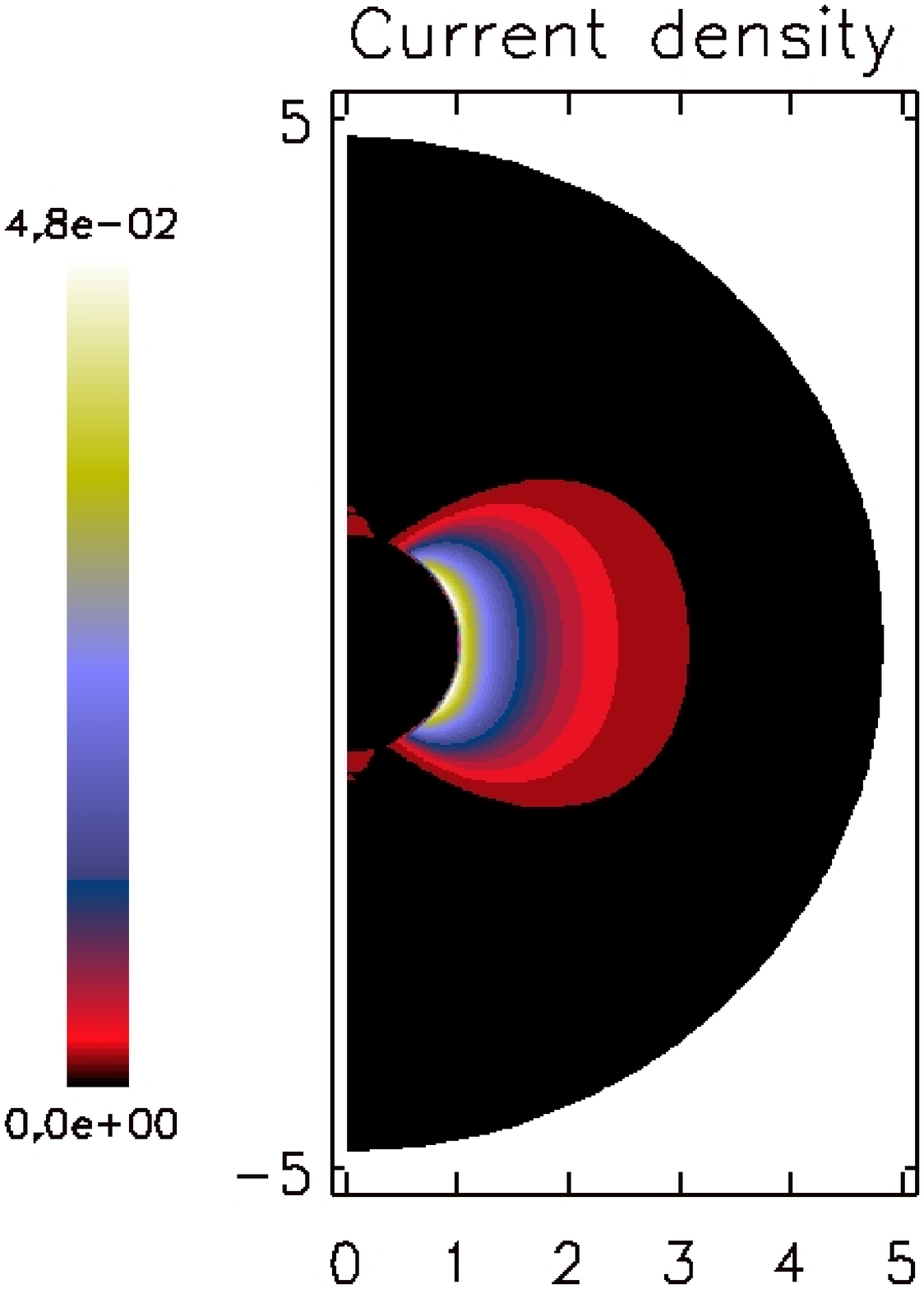}
\includegraphics[height=5cm]{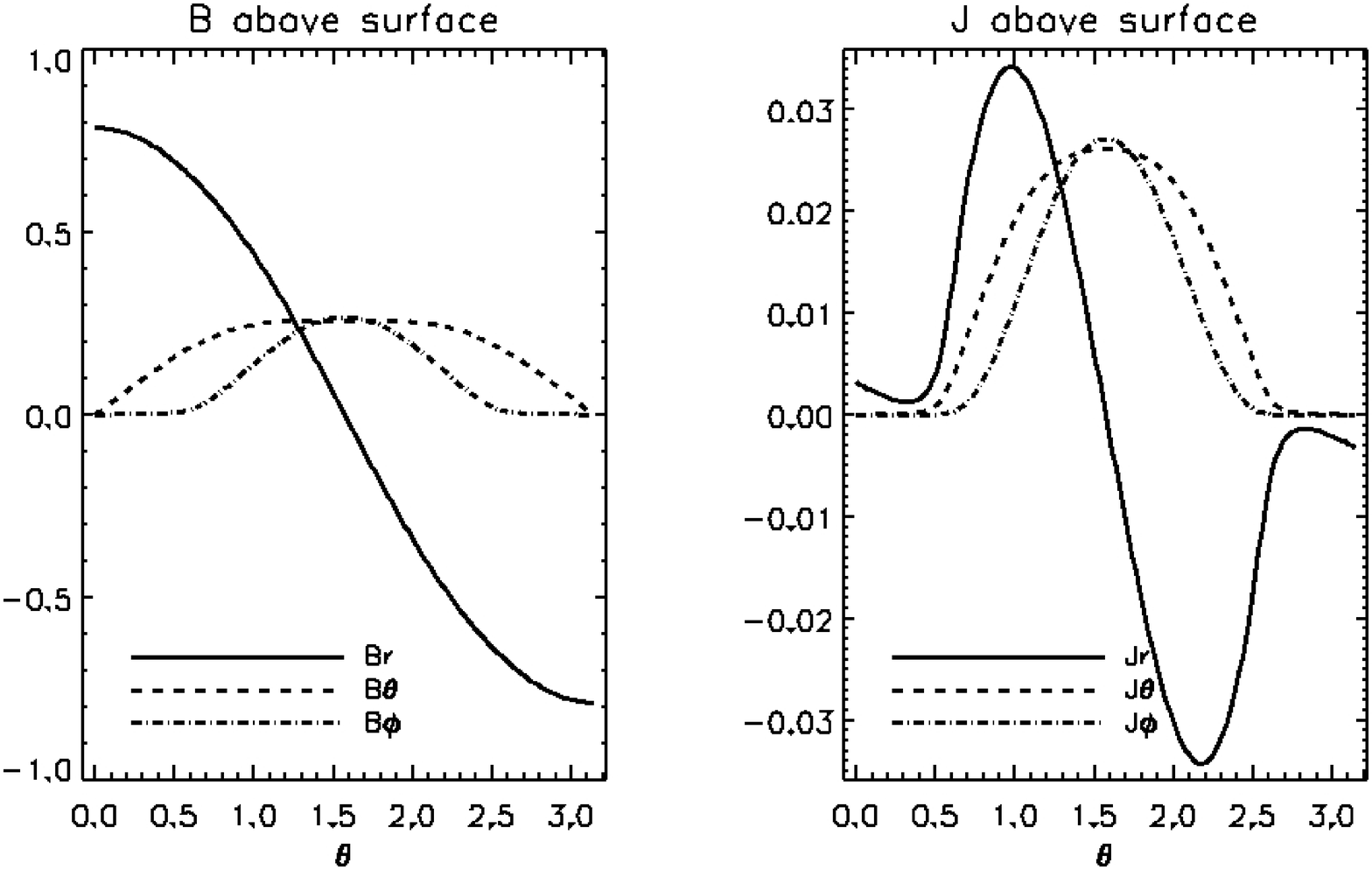}
\caption{Same as Fig. \ref{fig_s2} for model C. } 
\label{fig_modc}
\end{figure*}

\begin{figure*}[ht!]
\centering
\includegraphics[height=5cm]{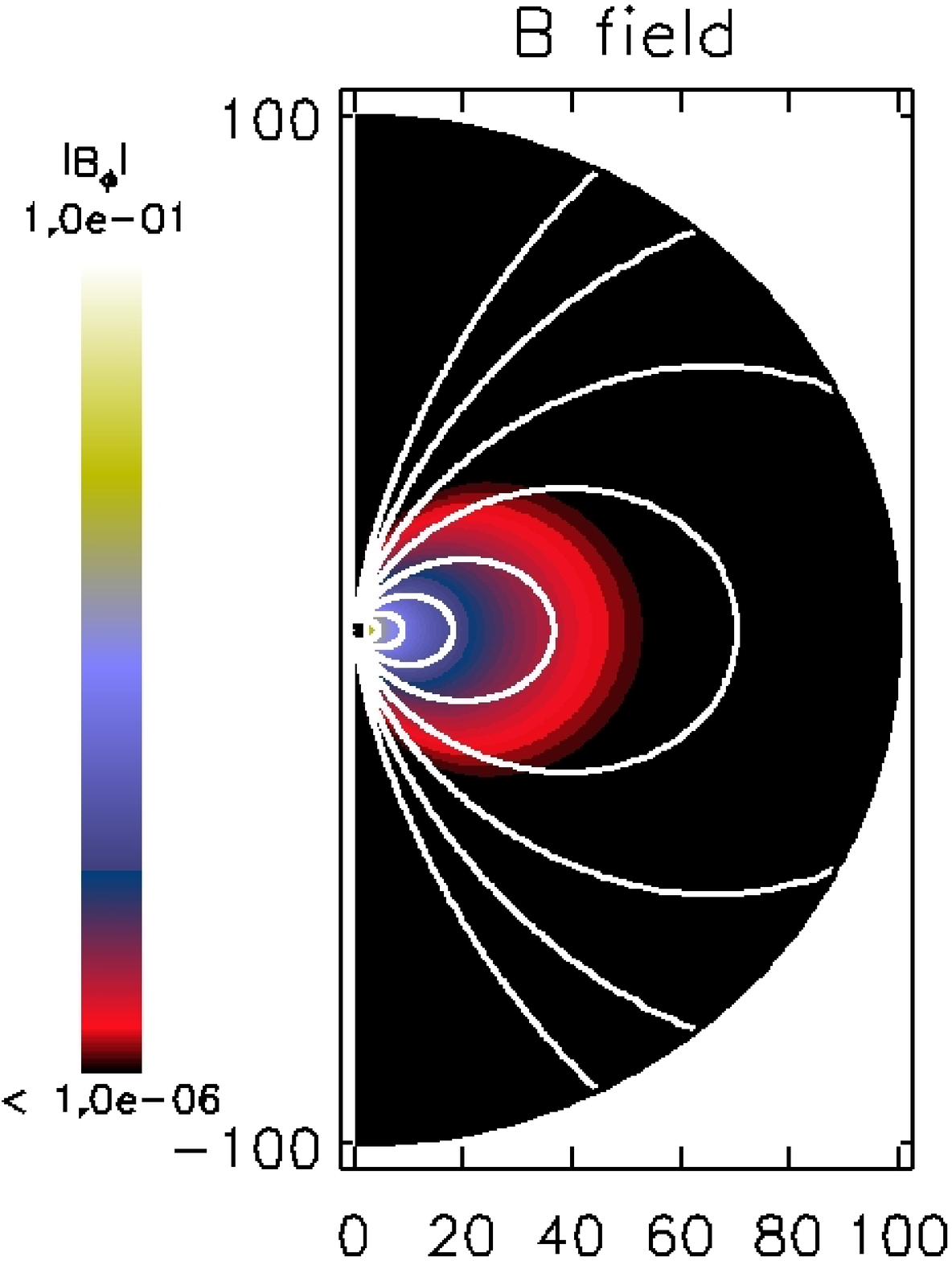}
\includegraphics[height=5cm]{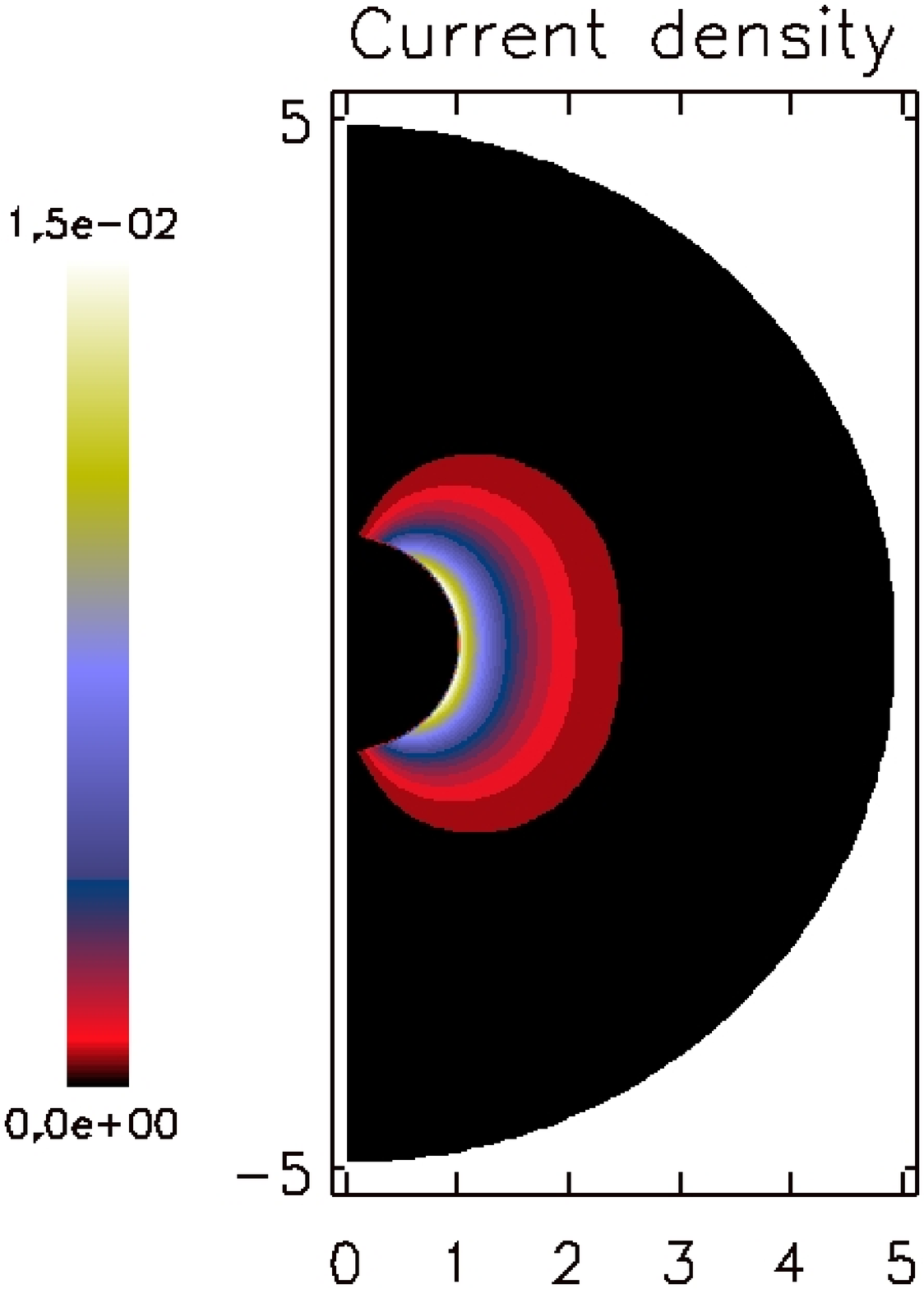}
\includegraphics[height=5cm]{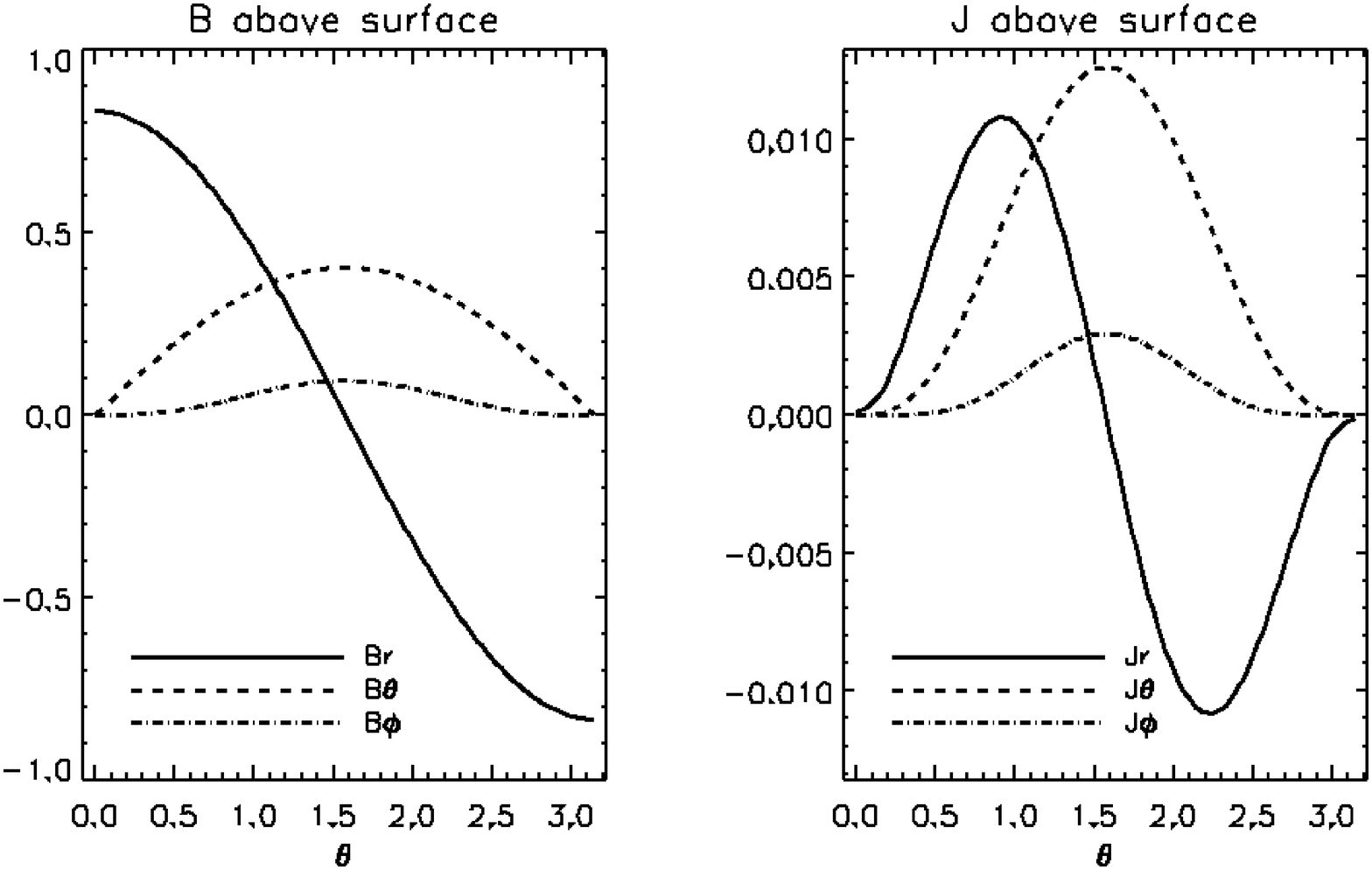}
\caption{Same as Fig. \ref{fig_s2} for model D. } 
\label{fig_modd}
\end{figure*}

\begin{figure*}[ht!]
\centering
\includegraphics[height=5cm]{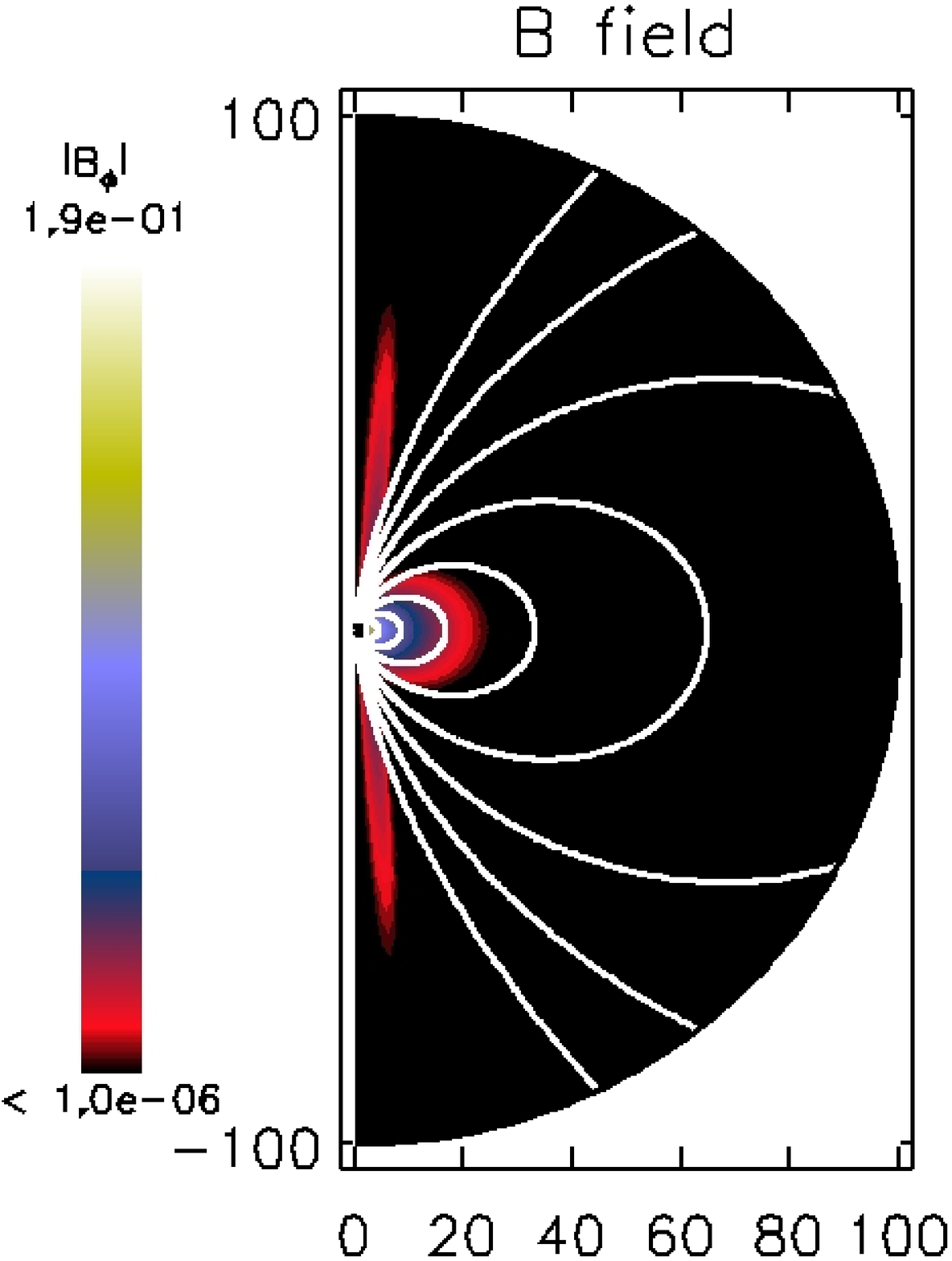}
\includegraphics[height=5cm]{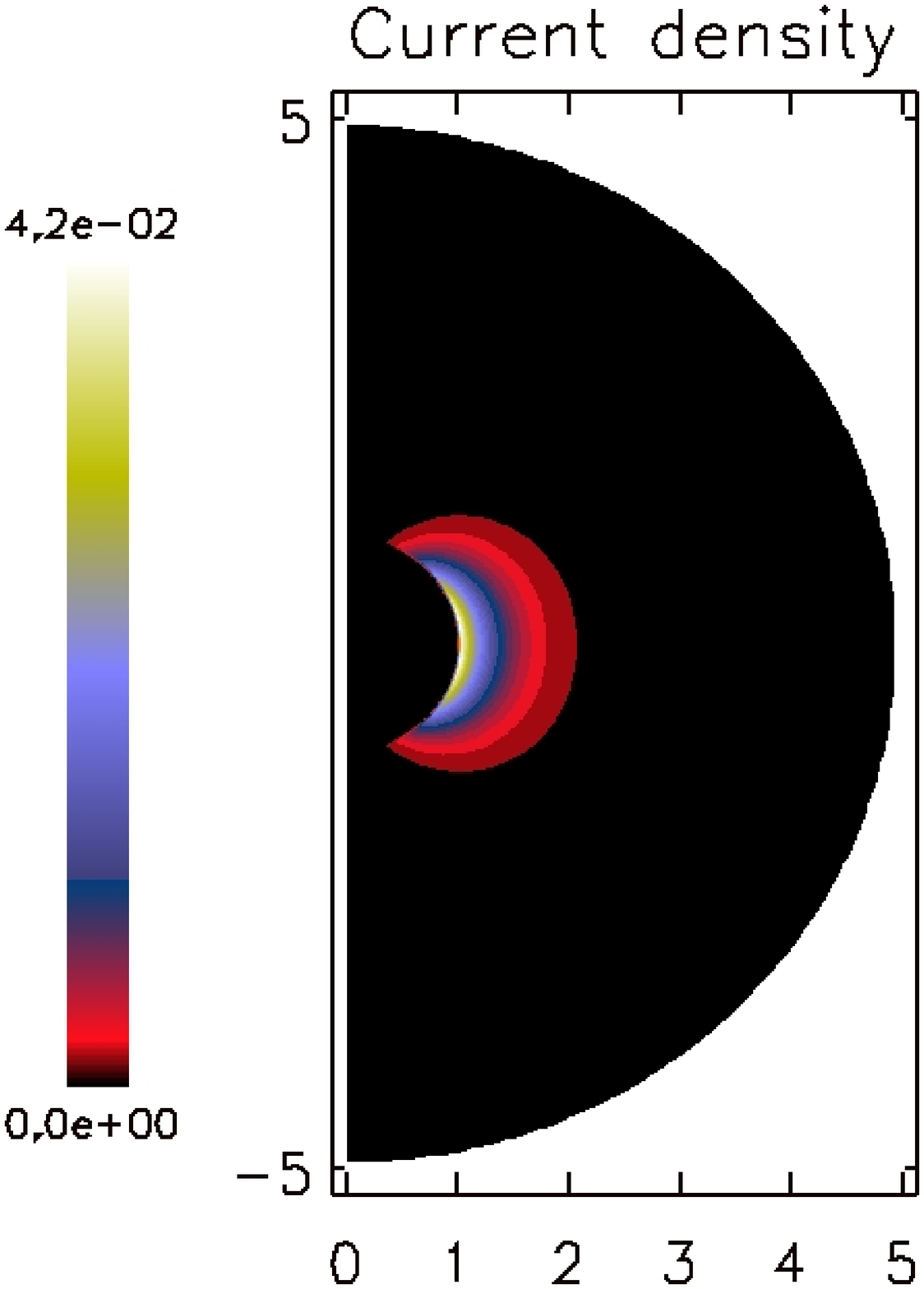}
\includegraphics[height=5cm]{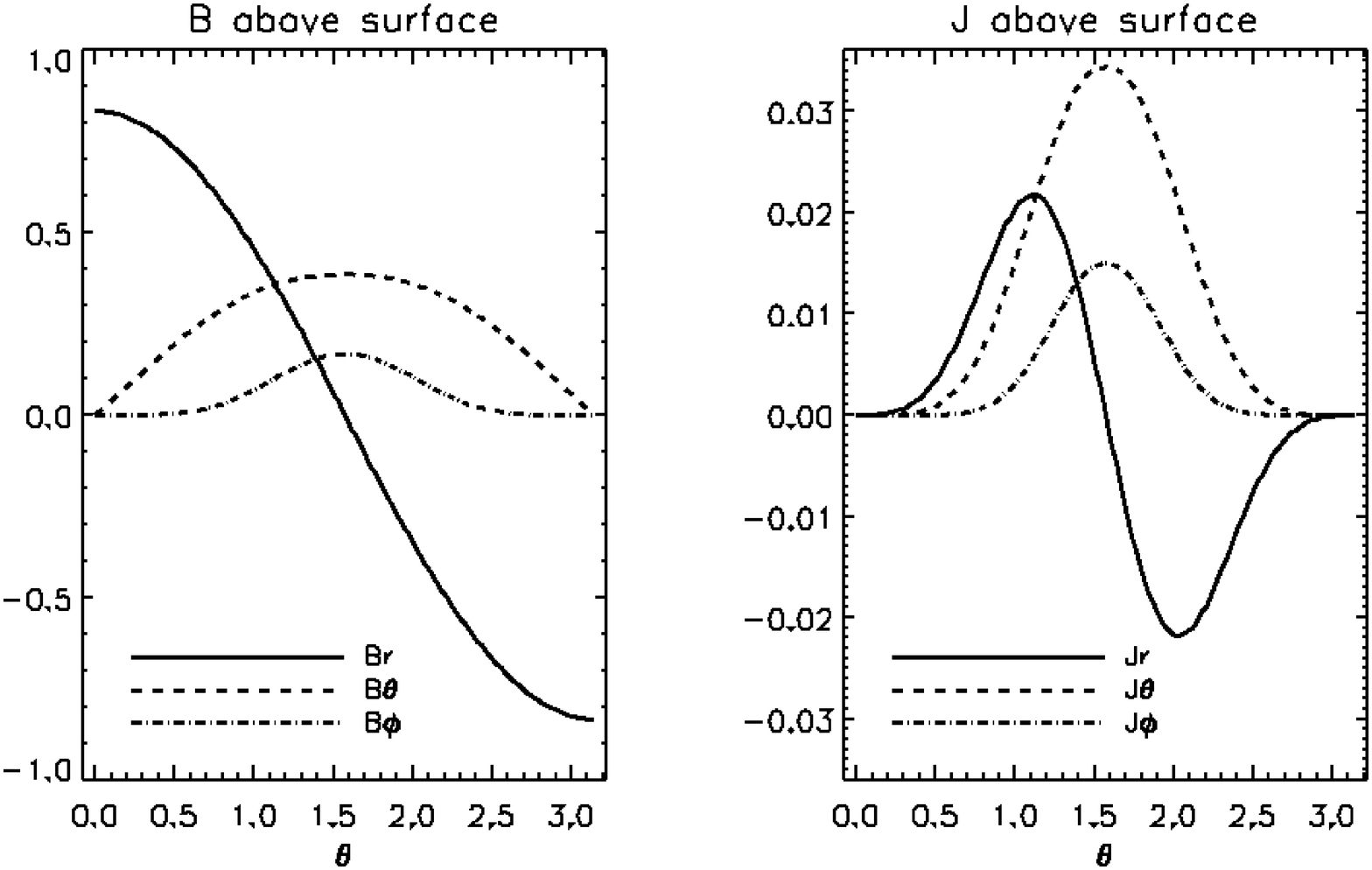}
\caption{Same as Fig. \ref{fig_s2} for model E. }
\label{fig_mode}
\end{figure*}

\begin{figure*}[ht!]
\centering
\includegraphics[height=5cm]{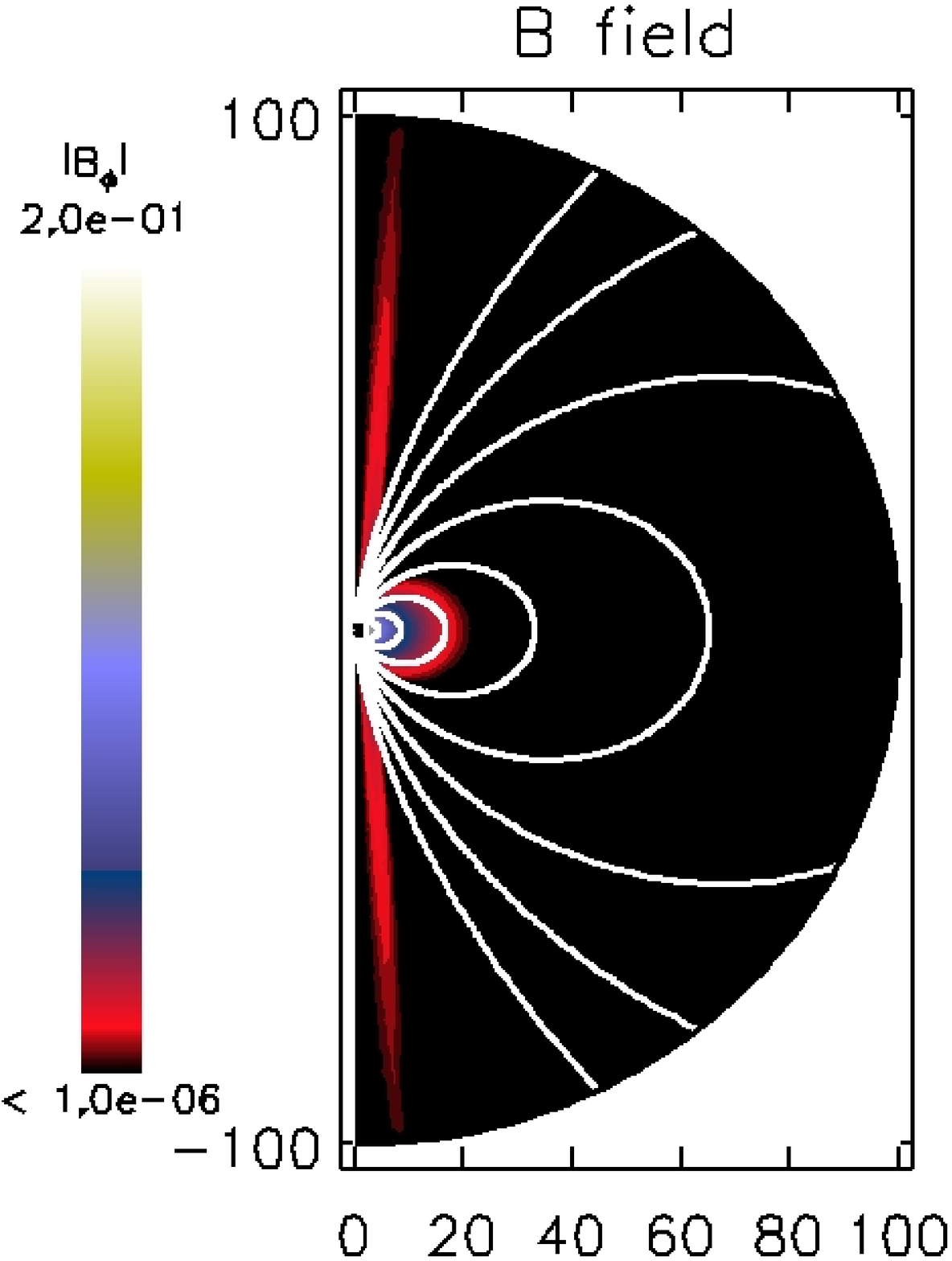}
\includegraphics[height=5cm]{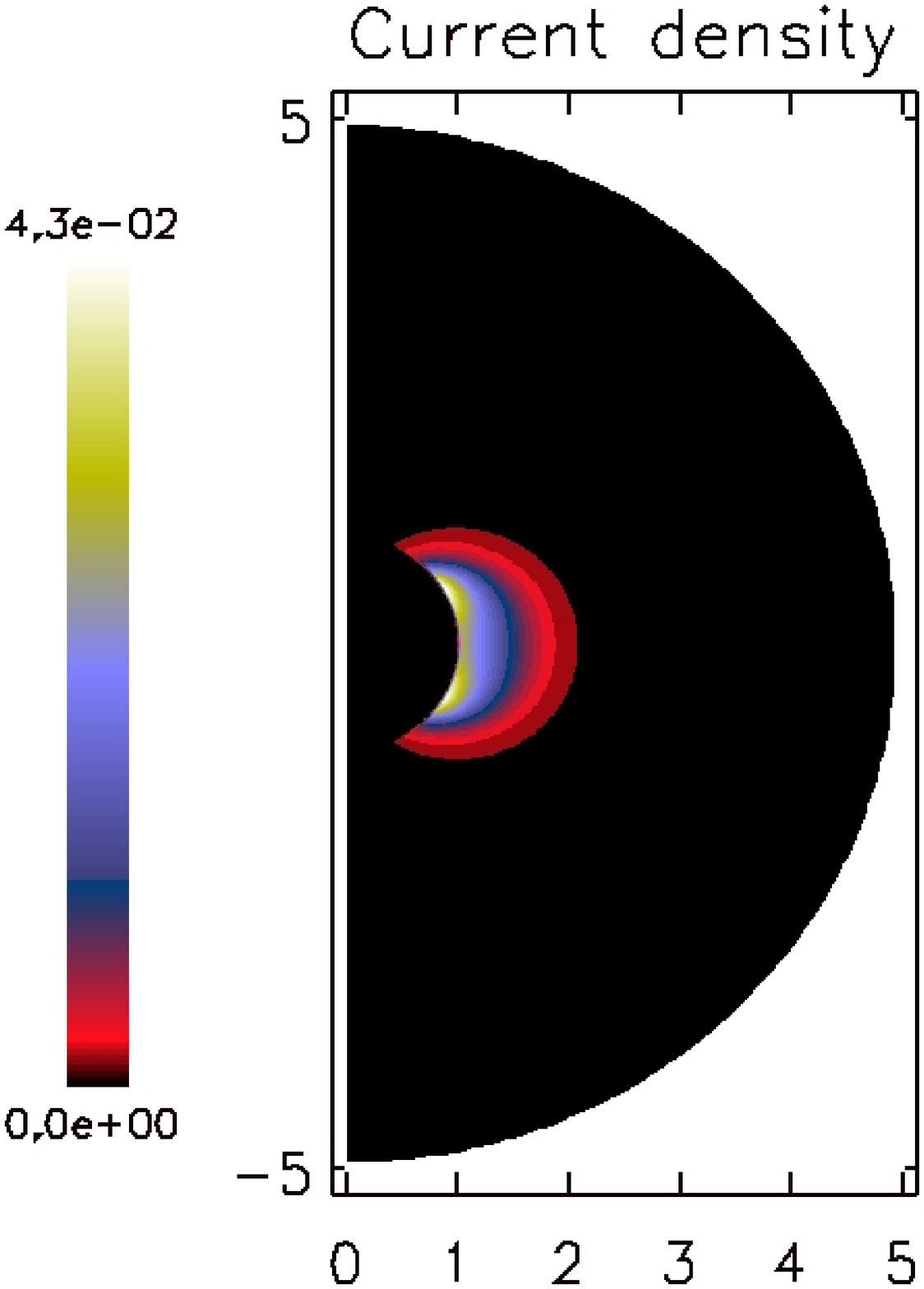}
\includegraphics[height=5cm]{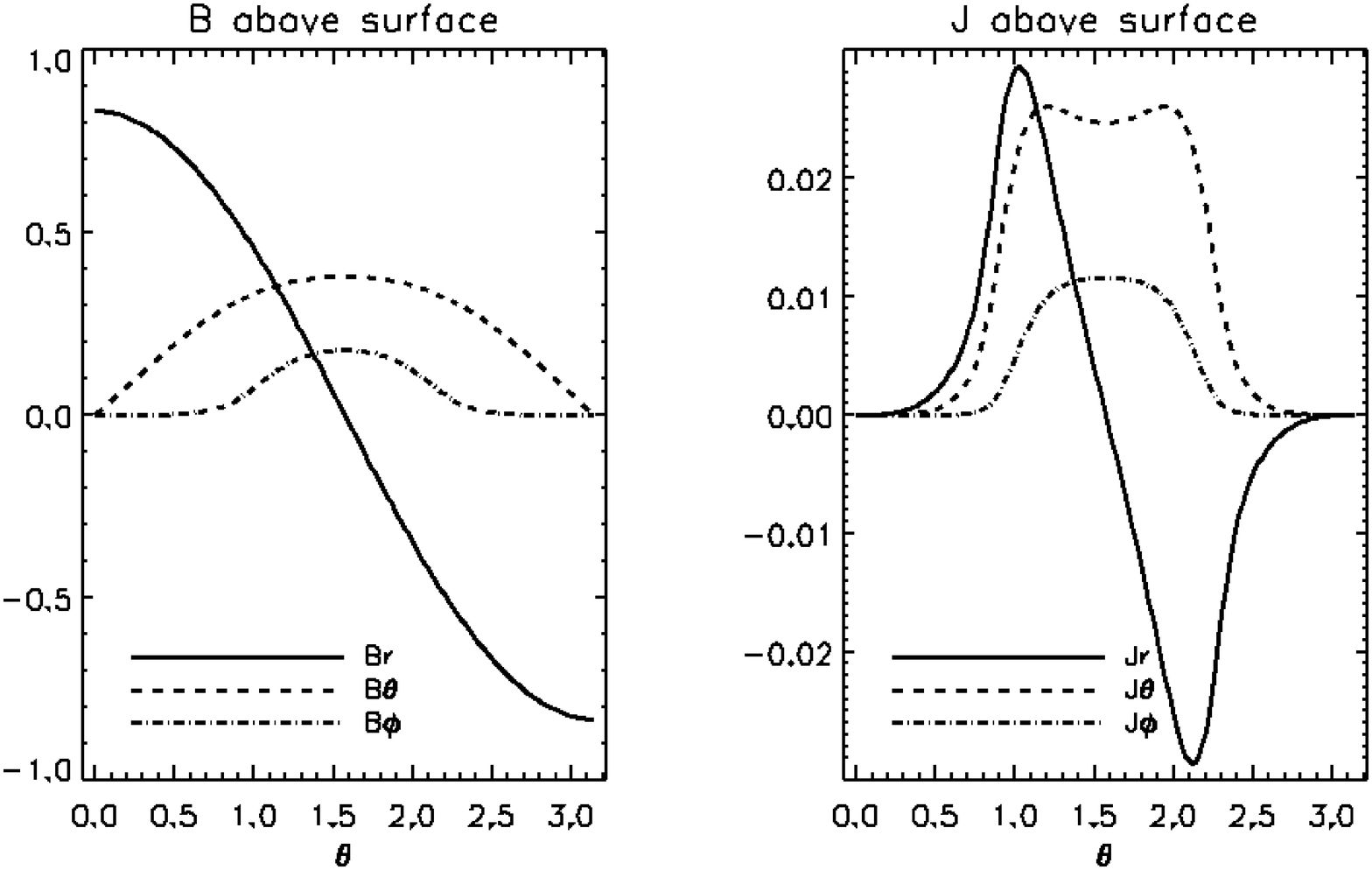}
\caption{Same as Fig. \ref{fig_s2} for model F. }
\label{fig_modf}
\end{figure*}

\begin{figure*}[ht!]
\centering
\includegraphics[height=5cm]{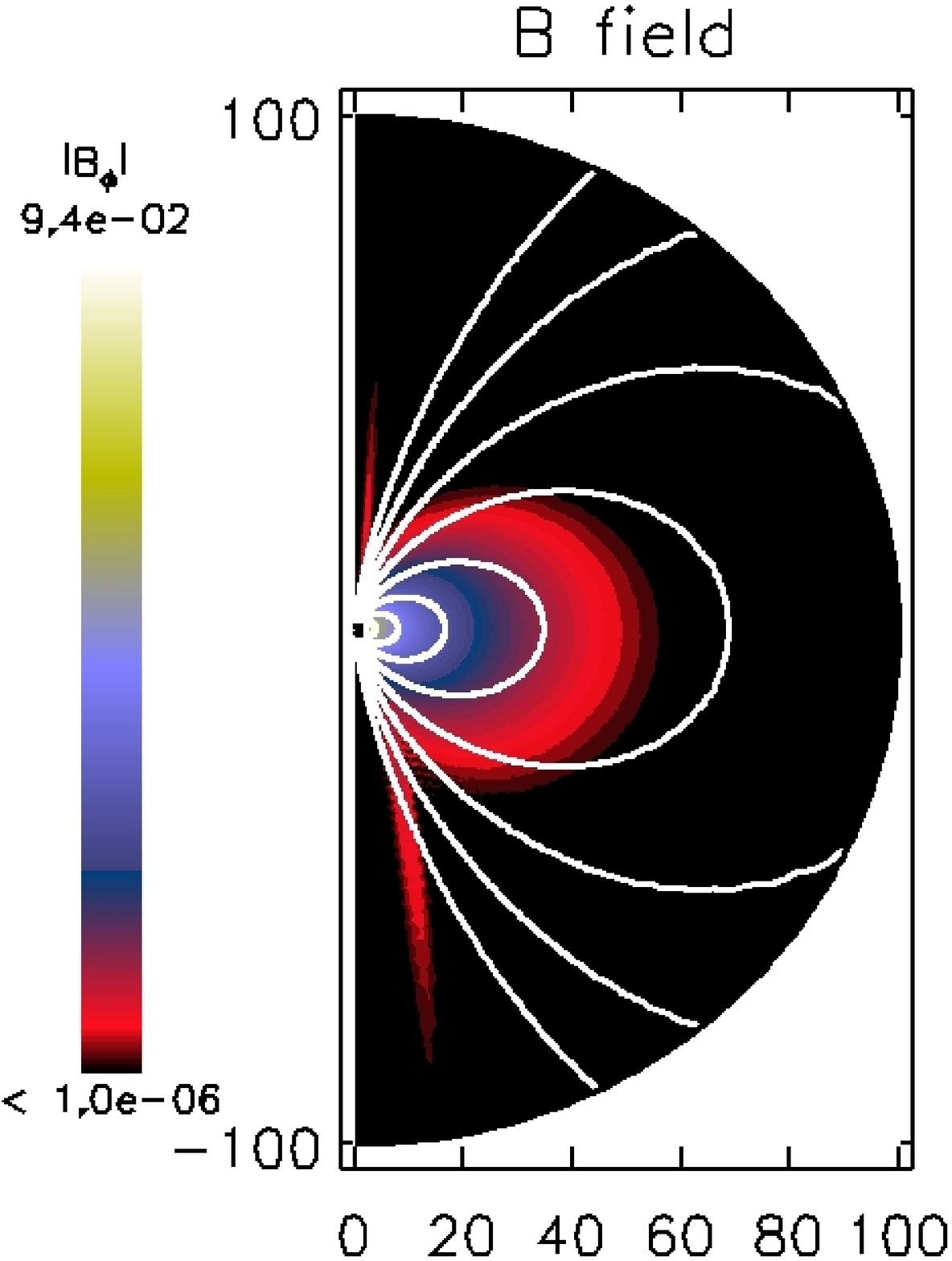}
\includegraphics[height=5cm]{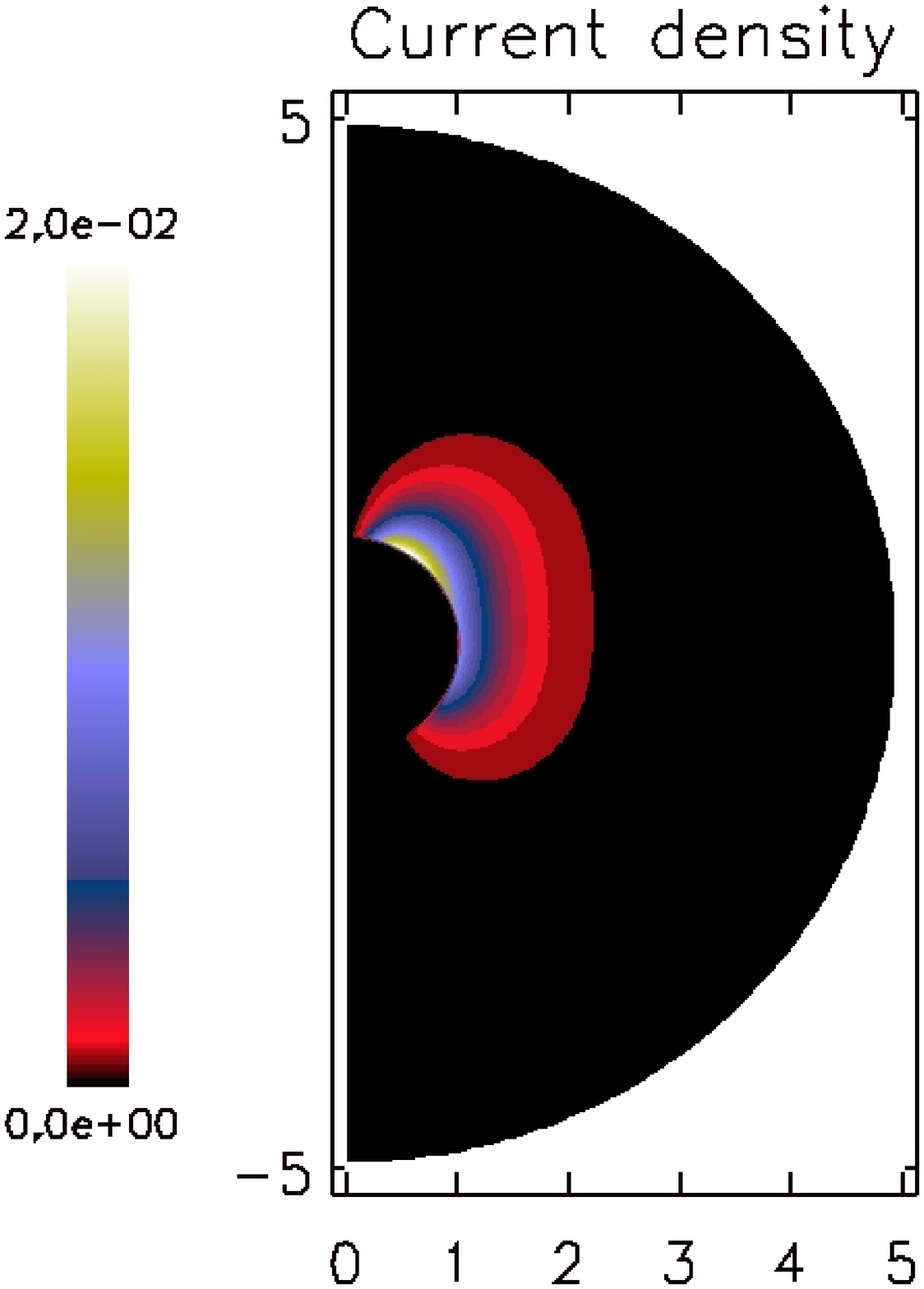}
\includegraphics[height=5cm]{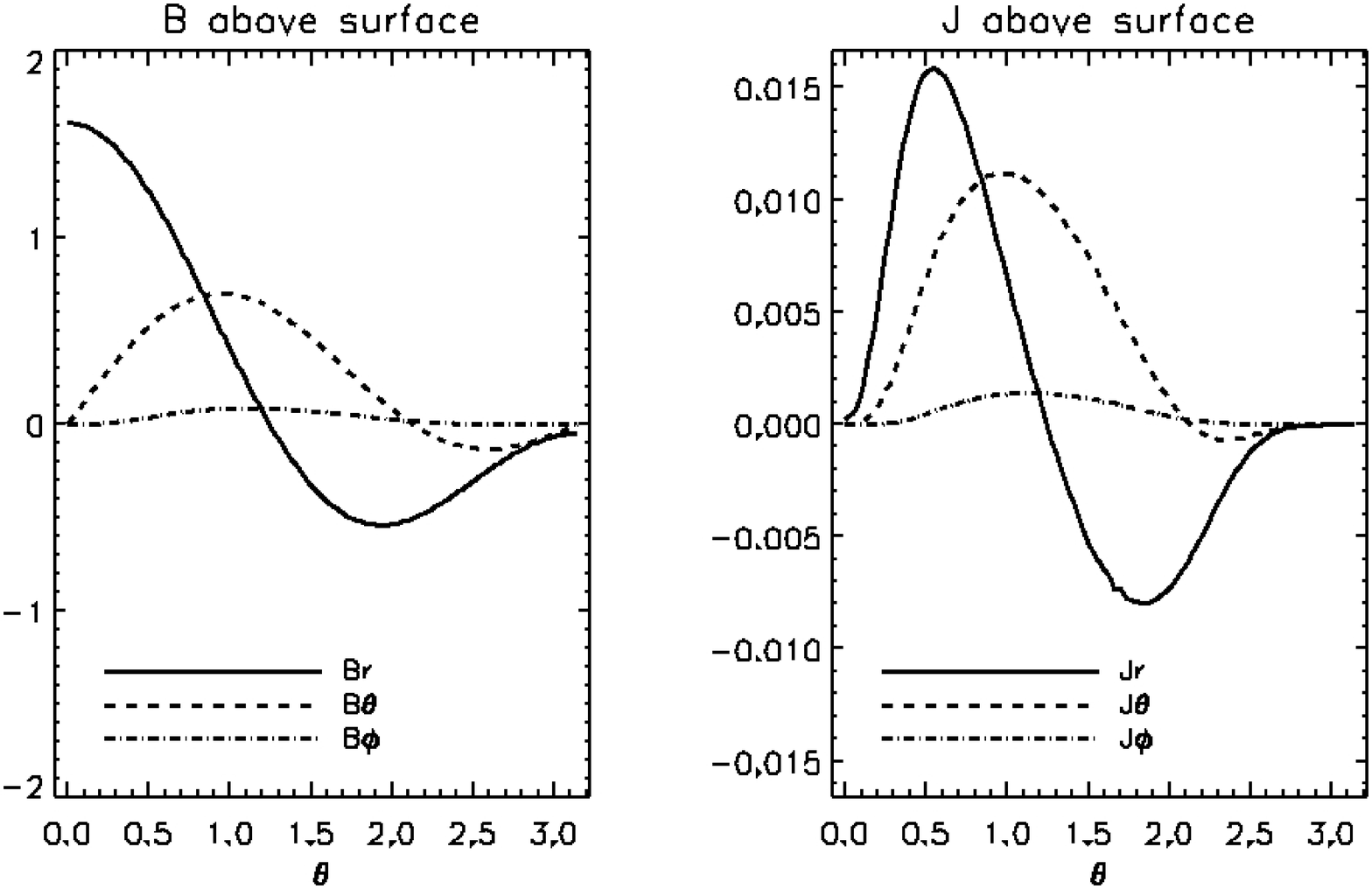}
\caption{Same as Fig. \ref{fig_s2} for model G. } 
\label{fig_modg}
\end{figure*}


The function $I(\Gamma)$ for the same four cases is shown in Fig. \ref{fig_igamma_bc}, together with two more cases corresponding to $r_{out}=5$ and $r_{out}=100$ but replacing the external boundary condition $\vec{E}_f=0$ by a smooth matching with vacuum solutions. The use of a different boundary condition affects the final solution only for low values of $r_{out}$. Matching with external vacuum implies that no currents can flow through the boundary, so that $I(\Gamma)=\alpha(\Gamma)=0$ along every field line crossing the outer boundary. As a consequence, a force-free configuration, coupled with a vacuum, will be characterized by a plateaux $I(\Gamma)=0$ for $\Gamma<\Gamma_c$, with $\Gamma_c$ labeling the first closed field line. This means that a bundle of open field lines is potential. The fraction of open field lines (the length of the plateaux for low $\Gamma$) is large only for low values of $r_{out}$. 
For $\Gamma>0.2$ (equatorial region), all curves coincide. 
Increasing $r_{out}>100$ further has no visible effect on the models with this helicity. For models with higher helicity, the interaction with the boundary becomes more important, and $r_{out}$ needs to be accordingly increased to minimize the
boundary effects.

The next step is to explore the influence of the strength and form of the initial toroidal field. Table \ref{tab_initial} summarizes the parameters of the initial models employed and the initial mean angle between $\vec{B}$ and $\vec{J}$, Eq. (\ref{mean_angle}). Models A-F are obtained starting from a dipolar poloidal component with strength at pole $B_0$ and a toroidal component given by the general form
\begin{equation}\label{tor_r3}
 B_\phi^{in}=k_{tor}B_0~g_{in}(r)f_{in}(\theta)~.
 \end{equation}
The angular part is chosen to be of the form $f_{in}(\theta)=\sin^d\theta$, with $d$ being a positive integer. The radial dependence of models A to E is a power law, $g_{in}(r)=r^{-s}$, while in model F we use a rapidly decaying function:
\begin{equation}\label{torF}
 g_{in}(r)=r^{-3}e^{-[(r-1)/0.5]^2} ~.
\end{equation}
The initial configuration of model G is asymmetric with respect to the equator: it is a superposition of dipolar and quadrupolar poloidal components plus a toroidal component, as follows:
\begin{eqnarray}\label{eq_asy}
 && A_\phi=\frac{B_0}{2}\frac{\sin\theta}{r^2}+\frac{B_0}{2}\frac{\sin\theta\cos\theta}{r^3}~,\\
 && B_\phi=0.1B_0\frac{\sin\theta}{r^3}~.\nonumber
\end{eqnarray}
We fix $r_{out}=100$ and the external boundary condition to $\vec{E}_f(r_{out})=0$ for all models, except for model C (highest helicity), where $r_{out}=500$ to reduce the influence of boundary, as discussed before.

Table \ref{tab_models} shows the features of our final configurations: helicity, maximum line twist, maximum value of current density, and parameters of $I(\Gamma)=I_0(\Gamma/\Gamma_0)^{1+1/p}$.  We also include (model S1 and S2) two self-similar solutions (Sect. \ref{sec_selfsimilar}) of similar helicity, where all components have the same radial dependence $B_i\sim r^{-(p+2)}$. In this case the current function is analytical, $I=I_0(\Gamma/\Gamma_0)^{1+1/p}$ with $(p,I_0)$ belonging to the family of solutions in Fig. \ref{fig_par_ss}. In the other cases $I_0$ and $p$ are obtained with fits to the numerical function.

The final geometry of the magnetic field and currents for all these models is shown in Figs. \ref{fig_s2} to \ref{fig_modg}. For $k_{tor}\lesssim 0.1$ (models A and B), the initial poloidal field remains almost unaltered, and the behavior of the solutions is nearly linear. Toroidal field strength, helicity, current density $J$, enclosed current function $I(\Gamma)$, and global twist scale linearly with $k_{tor}$. In contrast, for $k_{tor}\gtrsim 0.1$, the high initial helicity results in a larger twist angle, up to several radians, which in turn corresponds to a highly deformed poloidal field. A direct comparison of models A, B and C, which differ from each other only in the strength of the initial toroidal component, illustrates this effect: models A and B have the same shape with just a different scale factor, but model C is qualitatively different. 

The general features in the low $k_{tor}$ models do not differ much from self-similar solutions, because the initial conditions were close to a slightly twisted dipole. In self-similar models, two features are the absence of radial currents on the axis and a higher concentration of currents around the equatorial plane. Conversely, in our models, currents are spread more over the angular direction, and we allow for the existence of radial current on the axis. As a consequence, comparing numerical solutions A and B with self-similar models with comparable helicity, the former reaches lower maximum values of current density with a higher global twist. We also note that in the most extreme case (model C) the angular dependence of the toroidal field and radial currents becomes steeper.\footnote{In some cases it approaches the formation of a current sheet near the equatorial plane, and this introduces numerical noise that does not allow reaching a smooth solution or calculating the twist angle accurately.} It is also interesting to compare models C and S2 (Figs. \ref{fig_modc} and \ref{fig_s2}). Both have a similar helicity, but the global twist is higher in model C, while the maximum current density is higher in model S2.

Comparing models B and D, which only differ in the angular dependence of the initial data and the normalization (fixed to obtain the same helicity), we find that the converged solutions are very similar, except near the axis where model D has no radial currents. The effect of varying the initial radial dependence can be estimated by comparing model D to model E. It seems that in this case the final solution keeps memory of the initial model: the converged solution shows a toroidal field that decreases with distance faster in the model E than in model D.

In model E a tiny toroidal field (note the color scale in the figures) appears near the axis. This is likely a numerical artifact that can be partly ascribed to the (narrow) bundle of lines that depart from polar region and interact with the outer boundary. As a matter of fact, these structures are stronger for low values of $r_{out}$, as already shown in  Fig. \ref{fig_jrout}. Moreover, the numerical dissipation of the current is slower near the axis and longer runs are needed to reach more restrictive convergence criteria.

\begin{figure}
\centering
\includegraphics[width=.24\textwidth]{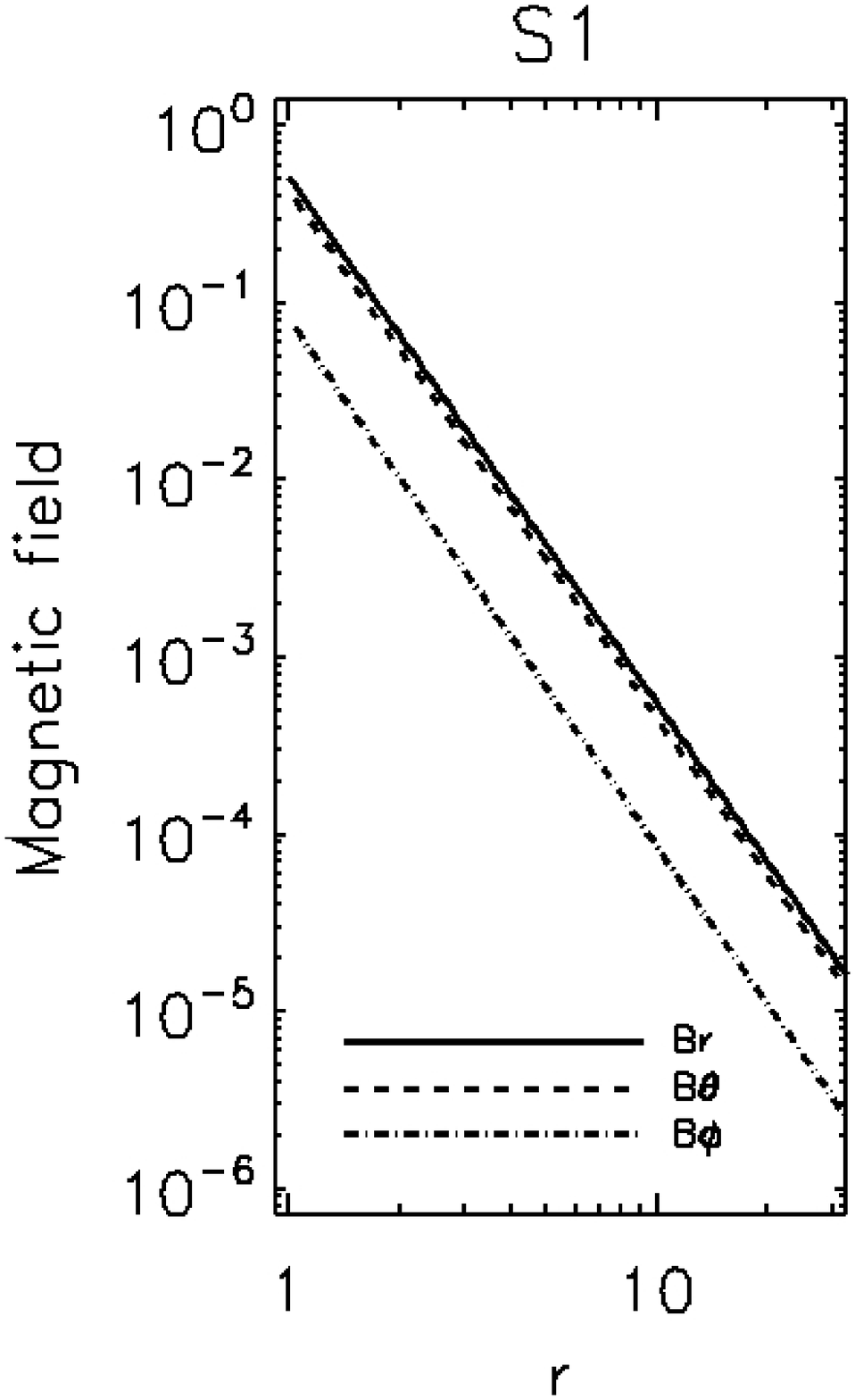}
\includegraphics[width=.24\textwidth]{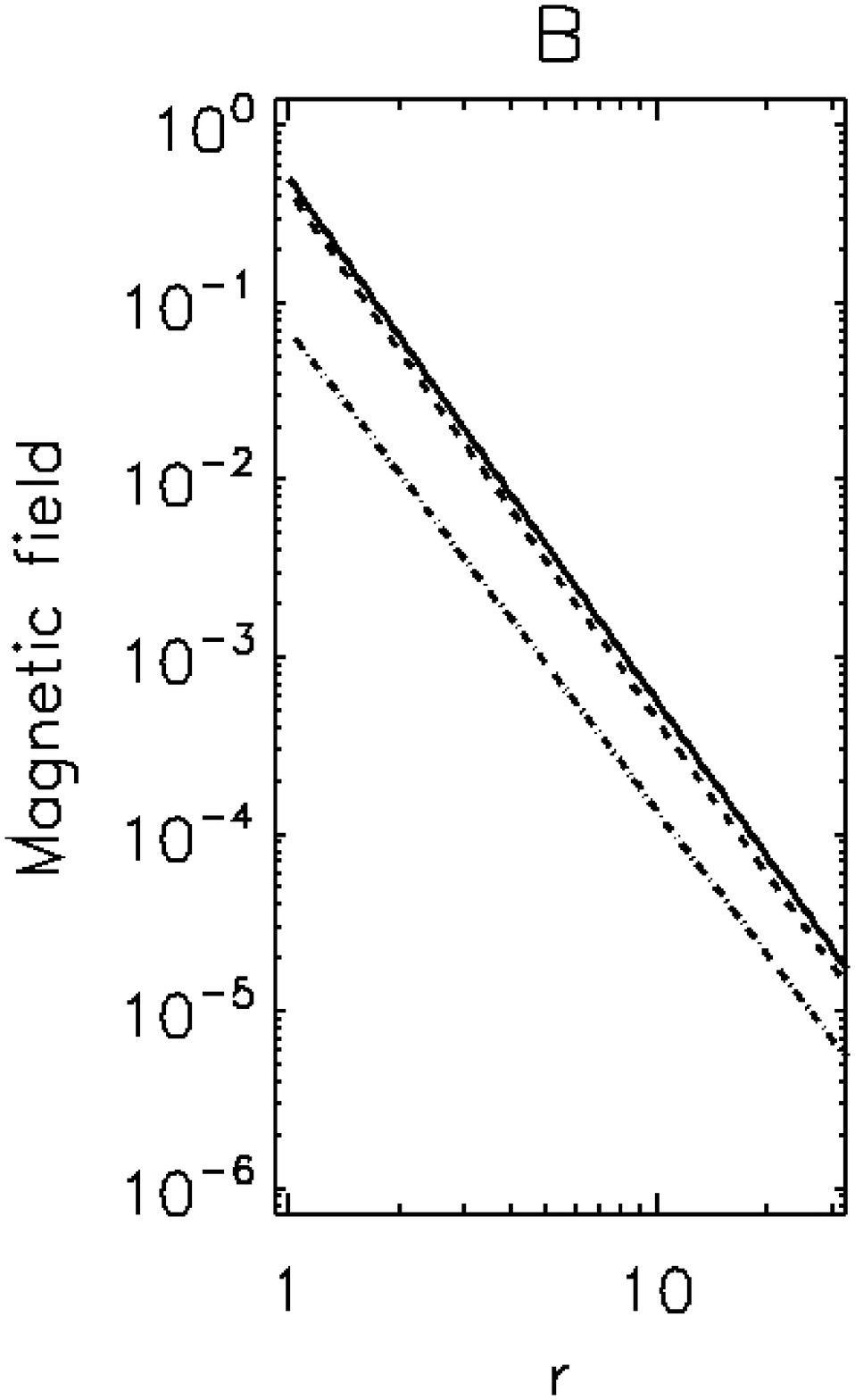}
\includegraphics[width=.24\textwidth]{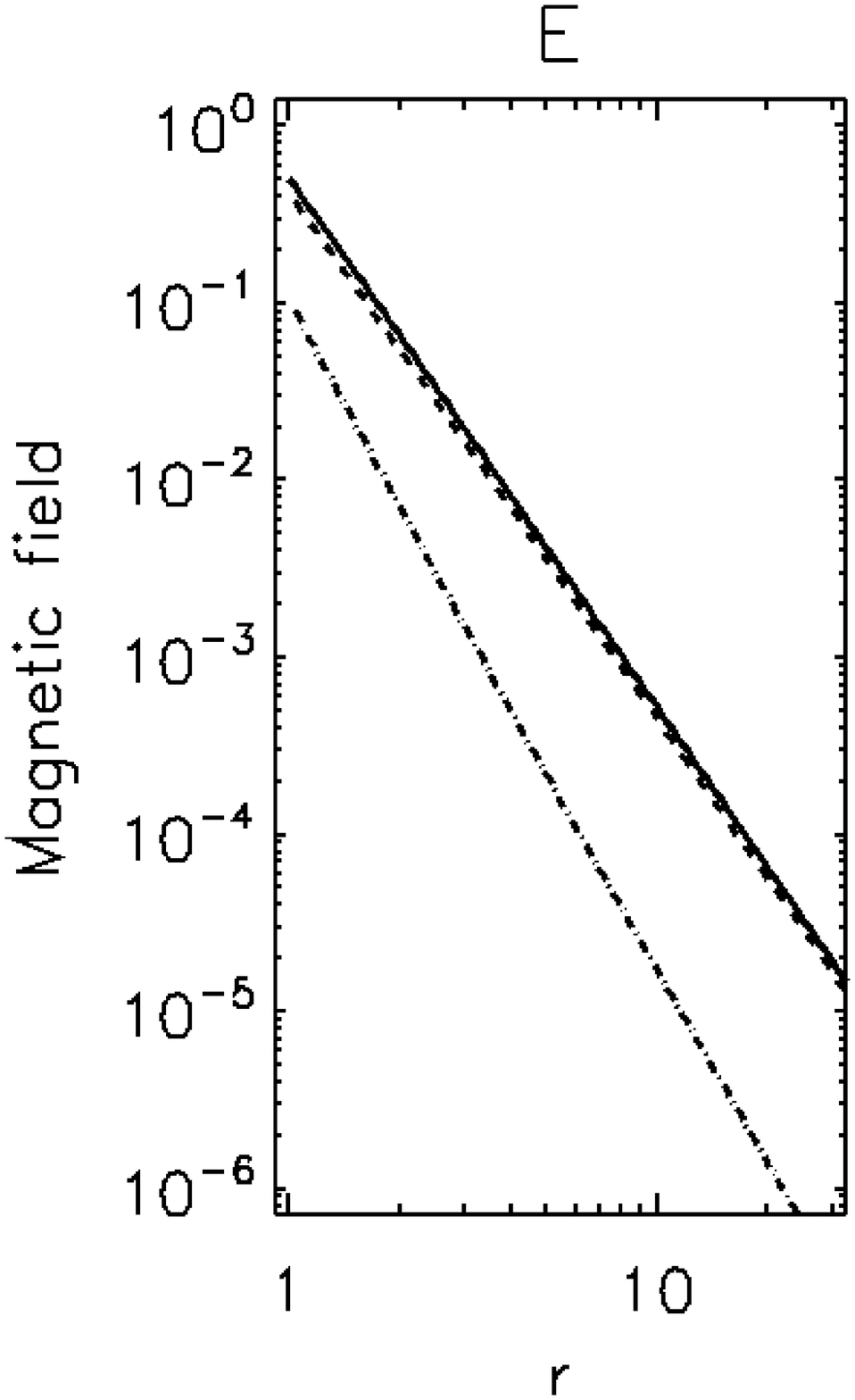}
\includegraphics[width=.24\textwidth]{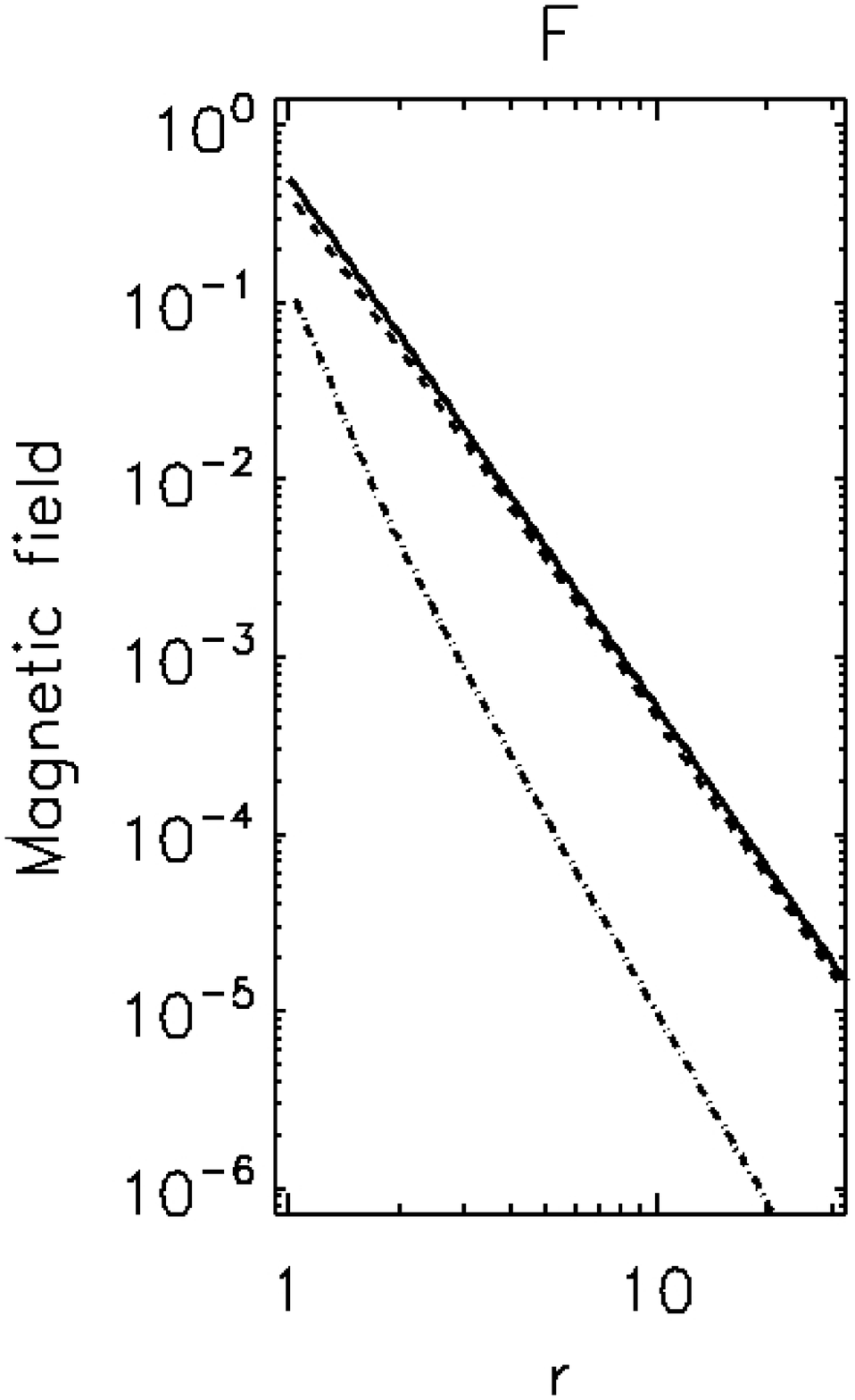}

\caption{Radial profiles of the magnetic field components for models S1, B (top), E and F (bottom). We show $B_r$ on the axis and $B_\theta,B_\phi$ at the equator for $r\in[1,30]$.}
\label{fig_profr}
\end{figure}

The different radial dependences of the magnetic field components in models B, E and F, together with S1, are shown in Fig. \ref{fig_profr}. The self-similar solution has the same radial dependence $r^{-(p+2)}$ for the three components, but in the numerical solutions the toroidal field can decrease faster (model E and F) or slower (model B) than the poloidal components. Furthermore, the radial behavior may depend on the magnetic colatitude $\theta$, too.
In addition, the radial dependence of the poloidal components is close to the power law $r^{-3}$ when the twist is low, but it may significantly deviate from a power law for the more complex models. The departure from self-similar solutions is likely to have a visible effect on the observed spectrum, as discussed in more detail in the next section.

A different solution that can only be found numerically is model G, the asymmetric configuration. Most of the current density is concentrated at a high latitude $\theta_m$.

An interesting case (not shown) consists of an initial model with toroidal and poloidal fields of opposite parity (e.g. a dipolar poloidal field plus a quadrupolar toroidal field). In this case, the total initial helicity is zero and, since this is a conserved quantity, the current is dissipated and a potential solution is found by the numerical code. This is consistent with the fact that the numerical evolution converges towards the most trivial solution with the same helicity.

\begin{figure}
\centering
\includegraphics[width=.5\textwidth]{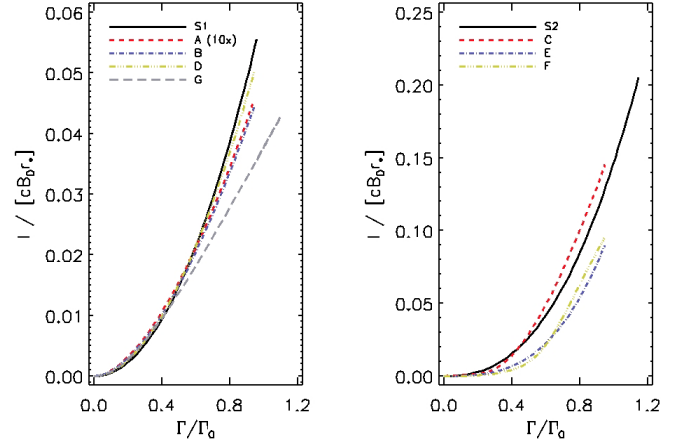}
\caption{$I(\Gamma)$ for different models. The curve of model A is magnified by a factor of 10.}
\label{cfr_igamma}
\end{figure}

Finally, to better understand the differences among models, a comparison between the enclosed current function $I(\Gamma)$ is very helpful, as shown in Fig. \ref{cfr_igamma}. In principle, if we know this function or an approximation fits the results of numerical simulations, the pulsar equation (Eq. \ref{pulsar_eq}) can be solved and the magnetospheric structure can be determined. In our models A, B, D, E, G, the current $I(\Gamma)$ is monotonic and can be well fitted by a single power law $I_0(\Gamma/\Gamma_0)^{1+1/p}$, but the values obtained for the parameter $p$ are not consistent with the value of $I_0$ describing the self-similar family of solutions (Fig. \ref{fig_par_ss}), as expected. As a matter of fact, for models A, B, D, the values of $p$ are greater than 1 (the self-similar dipolar family is described by $p\in[0,1]$). In contrast, model E lies quite close to the self-similar solution. The enclosed current of models C and E varies more rapidly, describing the concentration of currents in a smaller angular region. For some models, a power-law fit to $I(\Gamma)$ simply does not work. In model G, even if the symmetry with respect to the equatorial plane is broken, the resulting enclosed current $I(\Gamma)$ is not very different from models A, B, D, and S1 (after rescaling accordingly to the factor $k_{tor}$). The difference is the mapping between $\Gamma$ and the surface footprints. In all symmetric models, the maximum magnetic flux (proportional to $\Gamma$) is located at the equator, while it corresponds to $\theta_m\neq \pi/2$ for model G. Depending on the viewing angle, this has a strong imprint on the processed spectra.

\subsection{Resonant optical depth}\label{sec_rcs}

A prime candidate mechanism for generating the hard tail component observed in magnetar X-ray spectra is the resonant Compton upscattering of photons in an external magnetic field. We denote the cyclotron frequency by $\omega_B=ZeB/mc$, where $Ze$ and $m$ are the charge and mass of the particle. For electrons, this corresponds to an energy $\hbar\omega_B=11.6 (B/10^{12}~{\rm G})$ keV. In the X-ray band, the plasma frequency is much lower than the photon frequency, thus taking the polarization vector orthogonal to the propagation direction $\hat{\vec{k}}$ (semitransverse approximation) works well (see \cite{canuto71} and \cite{ventura79}). The normal modes of propagation are in general elliptically polarized and the degree of ellipticity depends on the ratio $\omega/\omega_B$ and $\theta_{kB}$, defined as the the angle between $\hat{\vec{k}}$ and $\vec{B}$.
It is common to introduce the ordinary (O) and extraordinary (X) modes. In the limit of propagation perpendicular to the magnetic field ($\theta_{kB}=\pi/2$), the modes are linearly polarized, with the polarization vector parallel (O-mode) or orthogonal (X-mode) to $\vec{B}$. For parallel propagation ($\theta_{kB}=0$), we recover the circularly polarized modes.

The cross section of O and E modes depends in a nontrivial way on $\theta_{kB}$, and on the ratio $\omega/\omega_B$. The cross section of the O-mode is close to the Thomson cross section $\sigma_T$ for $\theta_{kB}$ near to $\pi/2$, it scales with $\sin^2\theta_{kB}$ if $\omega\ll\omega_B$, and it does not depend dramatically on the frequency. Conversely, for low photon energies $\hbar \omega \ll \hbar\omega_B$, the cross section of the X-mode is strongly suppressed due to the reduced mobility of charged particles across magnetic field lines. Moreover, at photon frequency $\omega=\omega_B$, the X-mode becomes resonant \citep{ventura79}. If we neglect the thermal velocity of the particles (cold plasma approximation) and the natural width of resonance due to the radiation-damping force, the resonant cross section can be approximated by a delta function:
\begin{equation}\label{sigma_res}
 \sigma_{res}=\pi^2(1+\cos^2\theta_{kB}) \frac{(Ze)^2}{mc}\delta(\omega-\omega_B)
\end{equation}
where the factor $(1+\cos^2\theta_{kB})$ arises from the assumption of unpolarized light. The first attempts to consider this process quantitatively in magnetars are presented by \cite{lyutikov06}. They study a simplified, semi-analytical 1-D model by assuming that seed photons are radially emitted from the NS surface with a blackbody spectrum (BB) and the resonant Compton scattering occurs in a thin, plane parallel magnetospheric slab. By assuming a bulk velocity of electrons, but neglecting all effects of their recoil (Thomson limit $\hbar\omega_B\ll m_ec^2/\gamma$, where $\gamma$ is the Lorentz factor of electrons), it can be estimated how a BB spectra is affected. \cite{lyutikov06} find an average upscattering for the transmitted radiation (forward scattered photons), while the mean energy of the reflected radiation remains the same. Later, \cite{nobili08a,nobili08b} predicted spectra obtained via a new 3-D Monte Carlo code, accounting for polarization, QED effects, and relativistic cross sections. Although including our magnetospheric models in such sophisticated Monte Carlo simulations is beyond the scope of this paper, we can estimate some of the possible effects by looking at the behavior of relevant quantities.

In particular, we can estimate the resonant optical depth, given by the integration of Eq. (\ref{sigma_res}) along a given line of sight of constant $\theta$ \citep{lyutikov06}:
\begin{equation}\label{tau_res}
 \tau_{res}(\theta)=\pi^2 n_ZZe(1+\cos^2\theta_{kB})\left|\frac{dB}{dr}\right|^{-1}
\end{equation}
where $n_Z$ is the density of scatterer particles, and all the quantities are evaluated at the resonant radius $r_{res}$. As the latter depends on the photon energy, the energy dependence of the optical depth is given basically by the local ratio $(1+\cos^2\theta_{kB})n_Z/|dB/dr|$ (provided that $r_{res}$ lies above the star surface). If we assume charge separation, the particle density is proportional to the current density, $J=\kappa cn_ZZe$. The dimensionless factor $\kappa$ depends locally on the plasma composition and on the (bulk and/or thermal) velocity of the particles. Following \cite{thompson02} and \cite{nobili08a}, we can (very roughly) estimate the optical depth along a line of sight (assuming $\kappa$ constant along it), and considering only radially directed photons:
\begin{equation}\label{tau_simple}
 \kappa\tau_{res}(\theta)=\pi^2 \frac{J}{c}\left(1+\frac{B_r^2}{B^2}\right)\left|\frac{dB}{dr}\right|^{-1}~.
\end{equation}

If we consider self-similar magnetosphere models under these naive assumptions, the optical depth becomes independent of $r_{res}$, hence, independent of where the scattering happens. This is because the local ratio $(1+\cos^2\theta_{kB})J/|dB/dr|$ is the same for each angle, since for every component $i$, $B_i\propto r^{-p+2}$, and $J_i\propto r^{-p+3}$. Furthermore, the optical depth does not depend on the normalization of the magnetic field, $B_0$.

\begin{figure*}[ht]
\centering
\includegraphics[width=.24\textwidth]{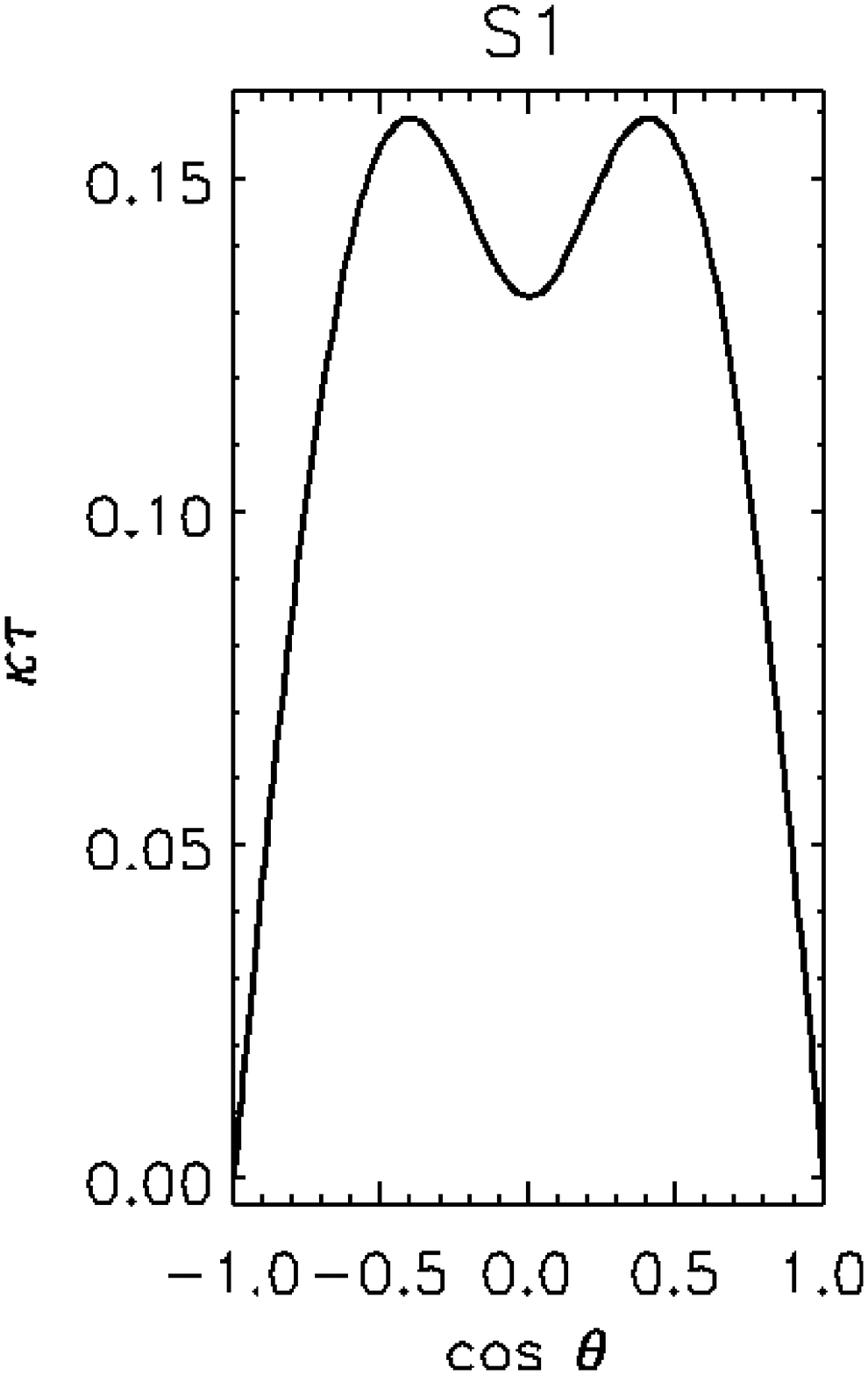}
\includegraphics[width=.24\textwidth]{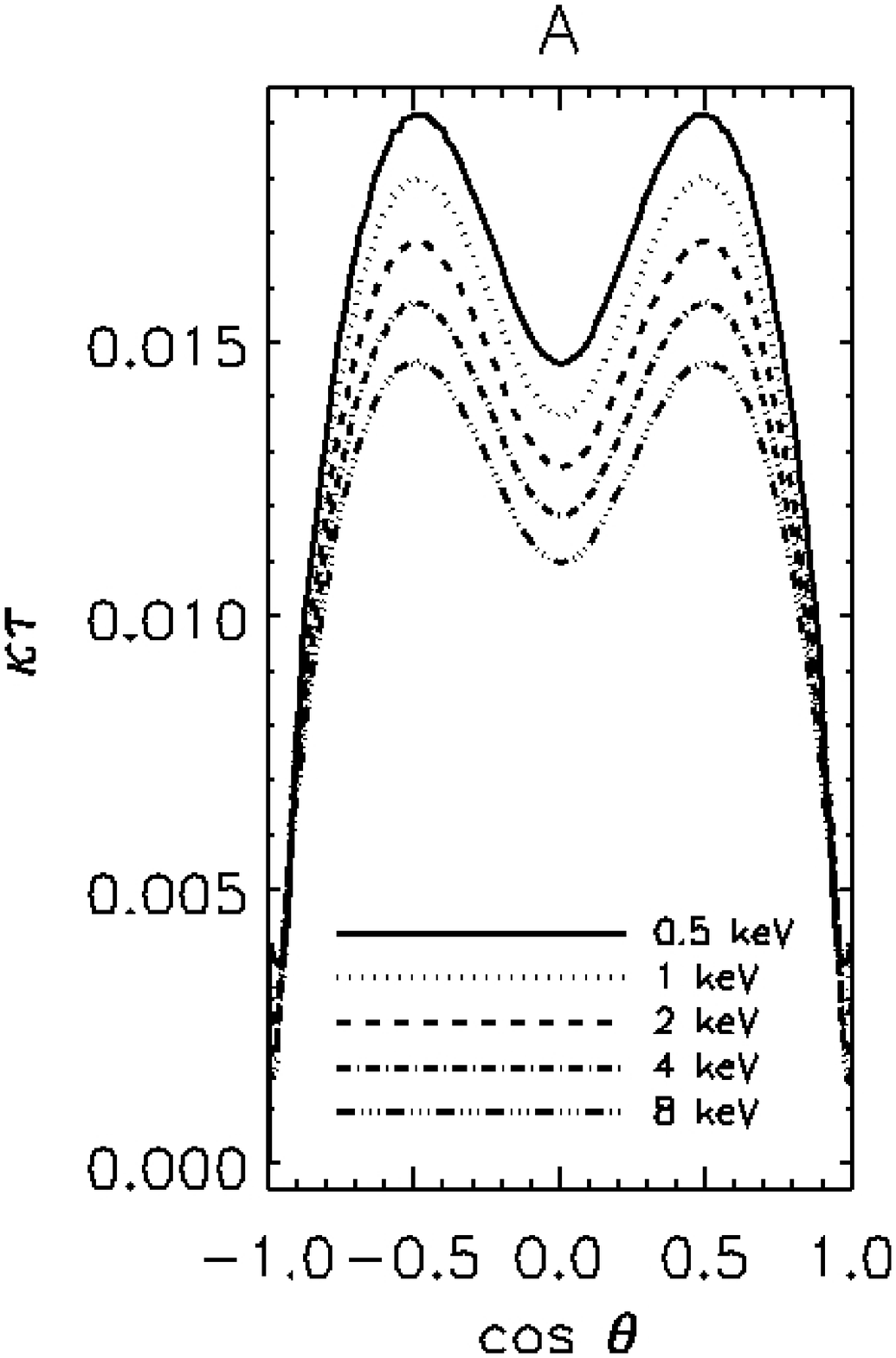}
\includegraphics[width=.24\textwidth]{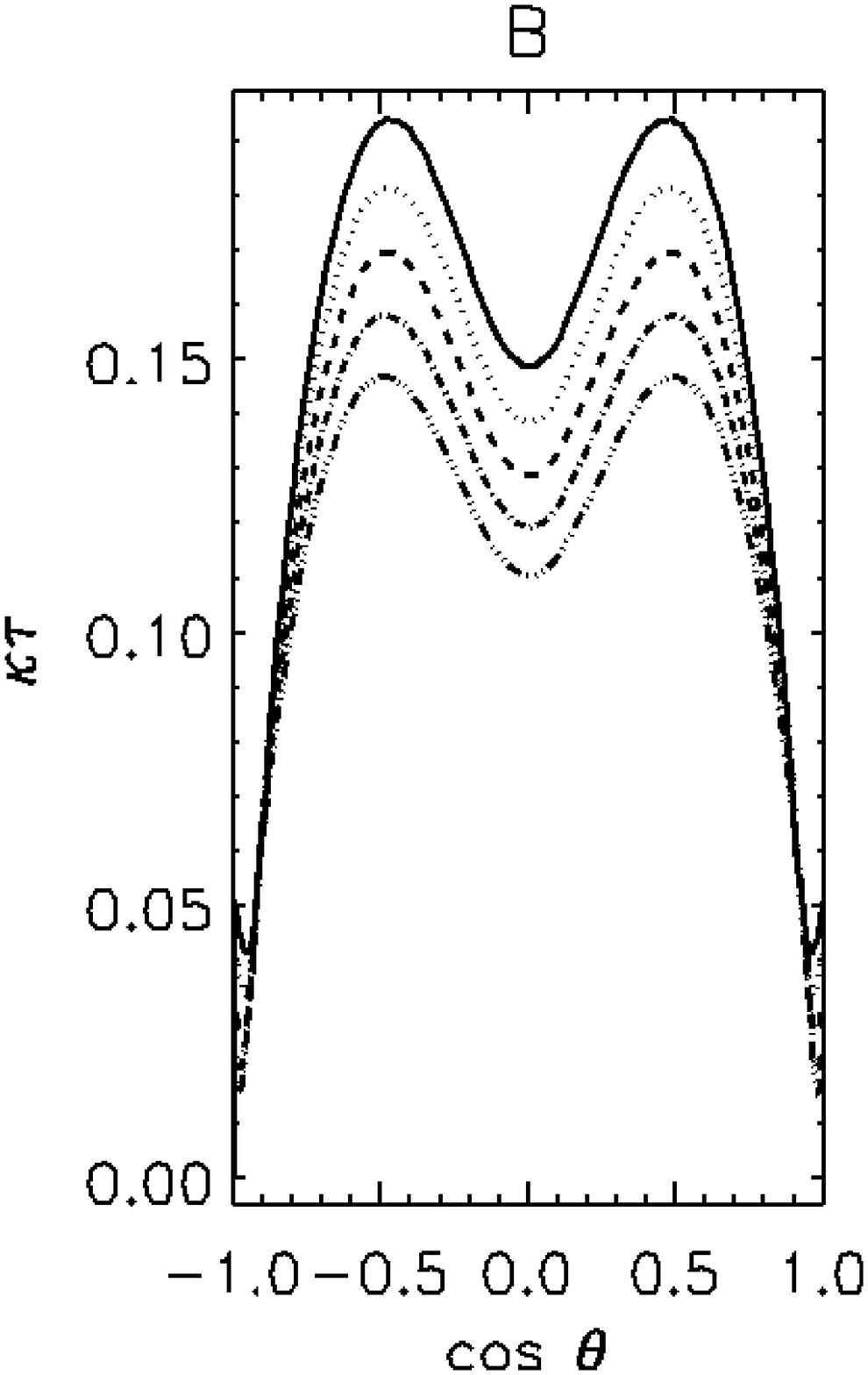}
\includegraphics[width=.24\textwidth]{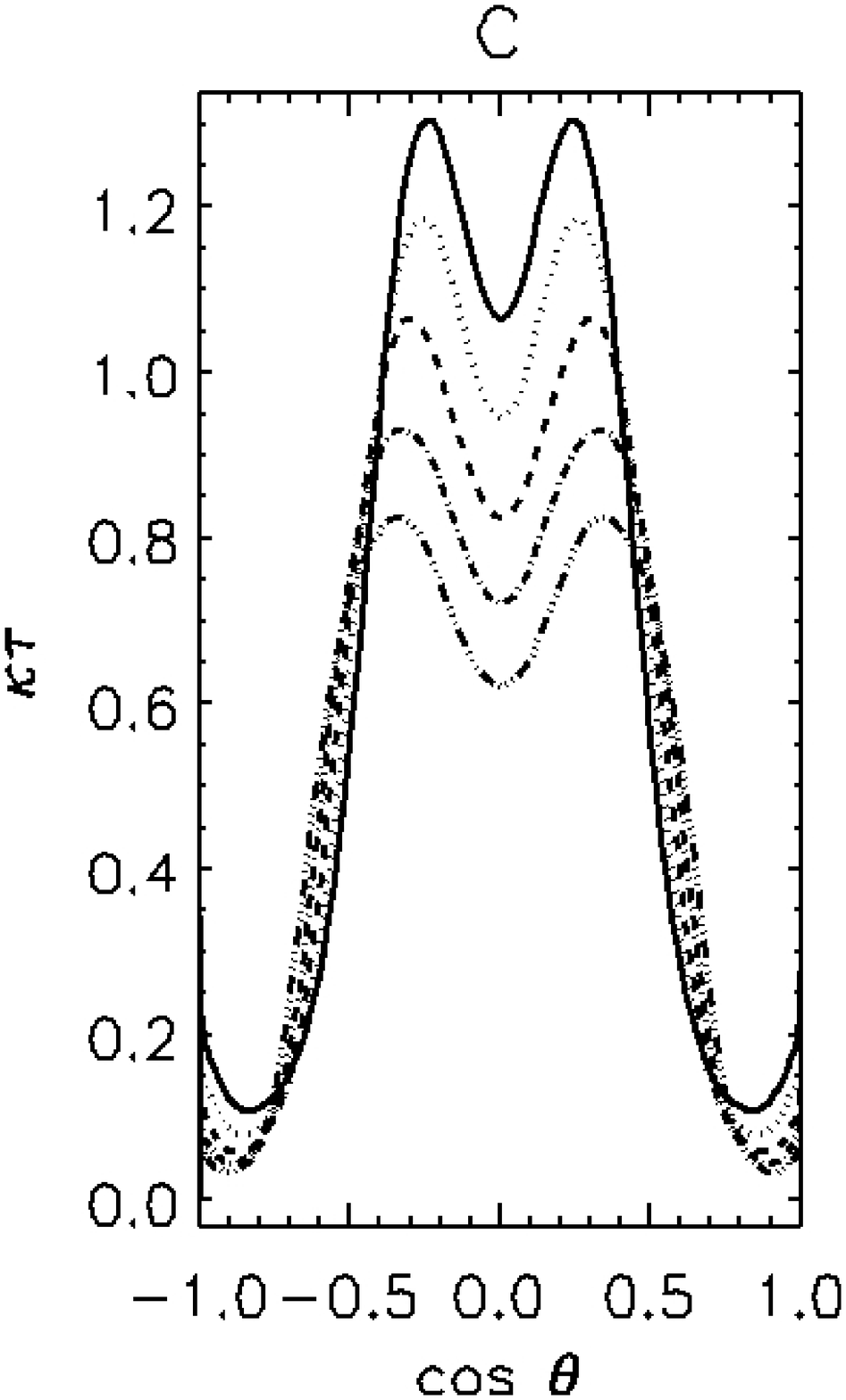}
\includegraphics[width=.24\textwidth]{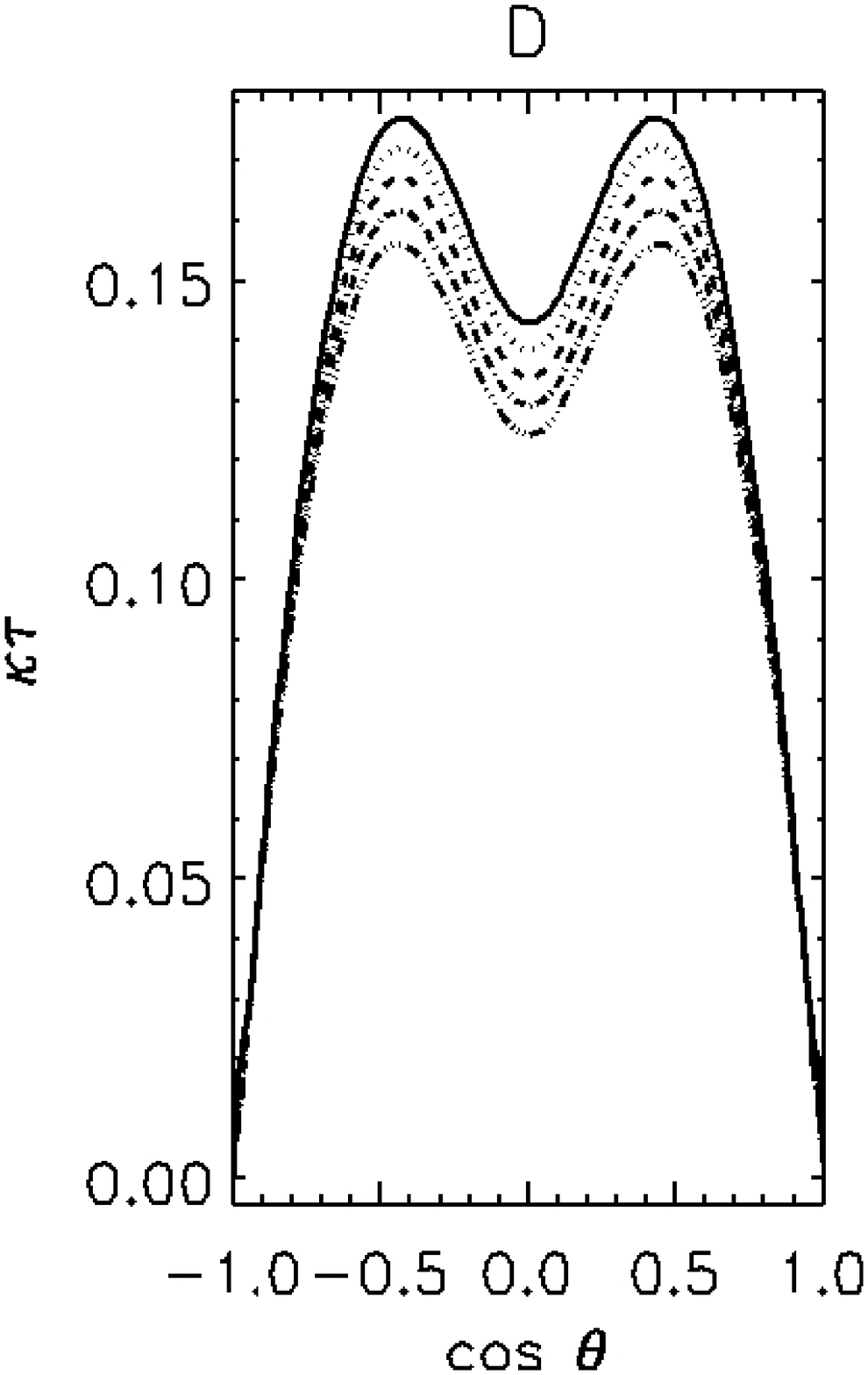}
\includegraphics[width=.24\textwidth]{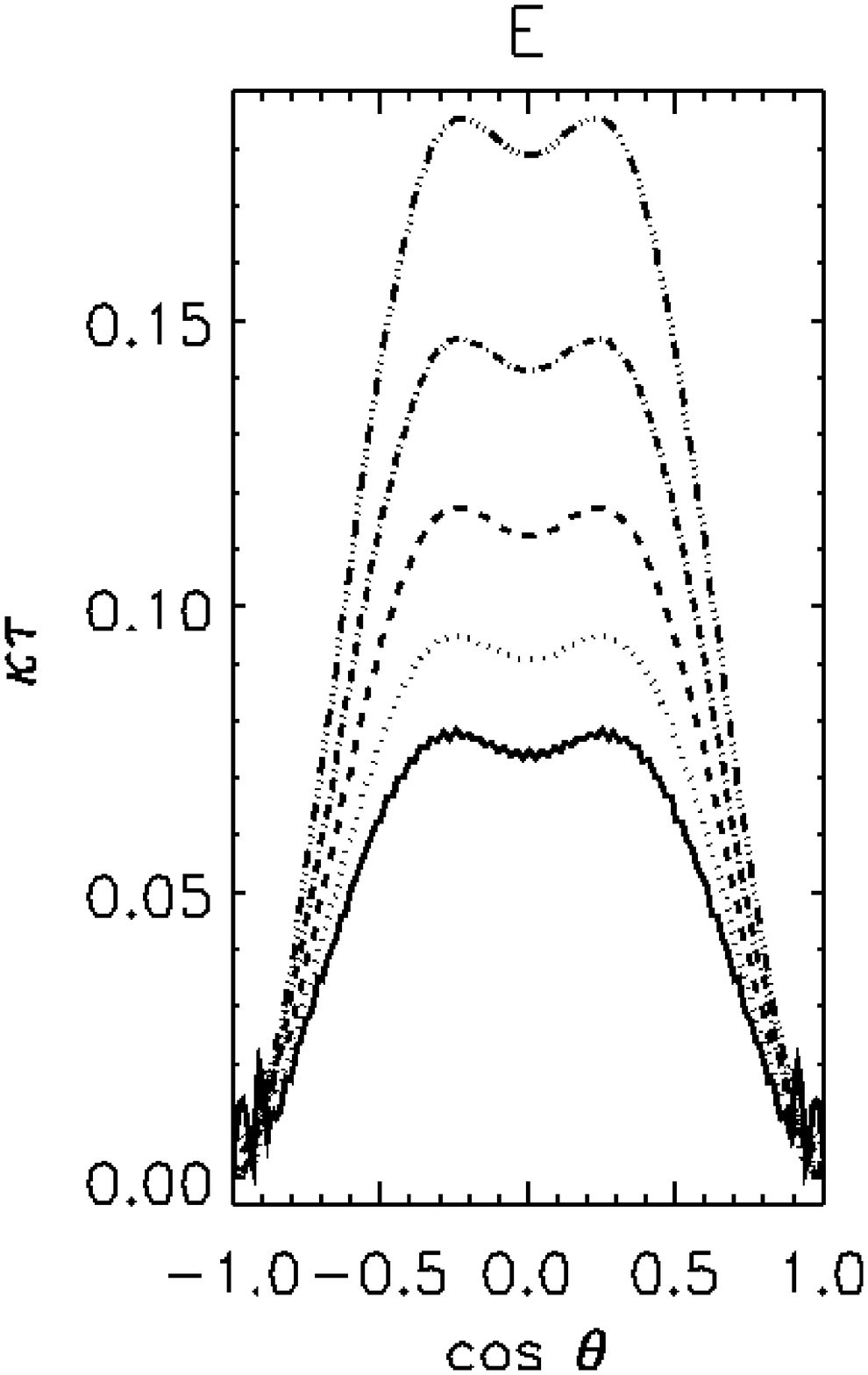}
\includegraphics[width=.24\textwidth]{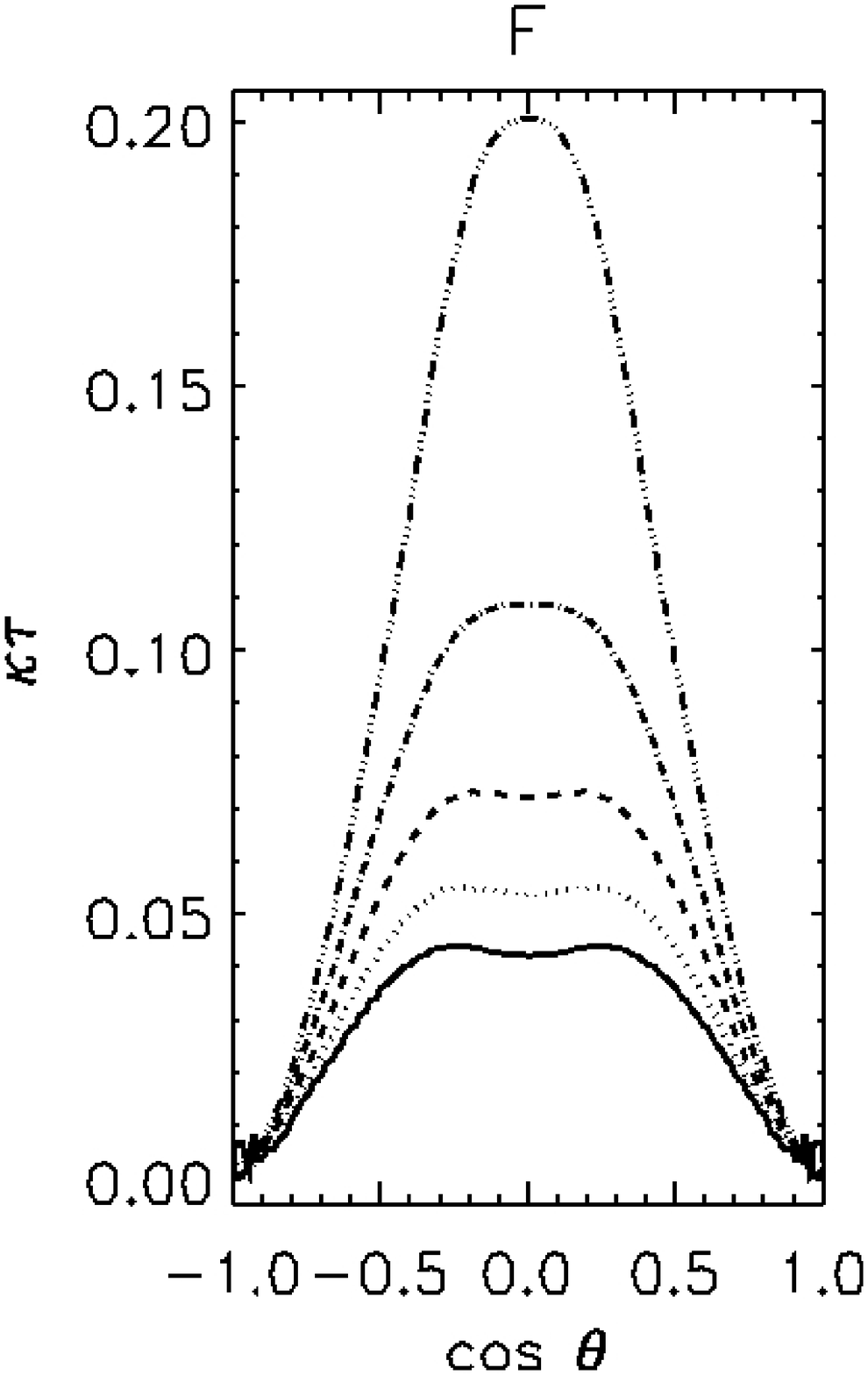}
\includegraphics[width=.24\textwidth]{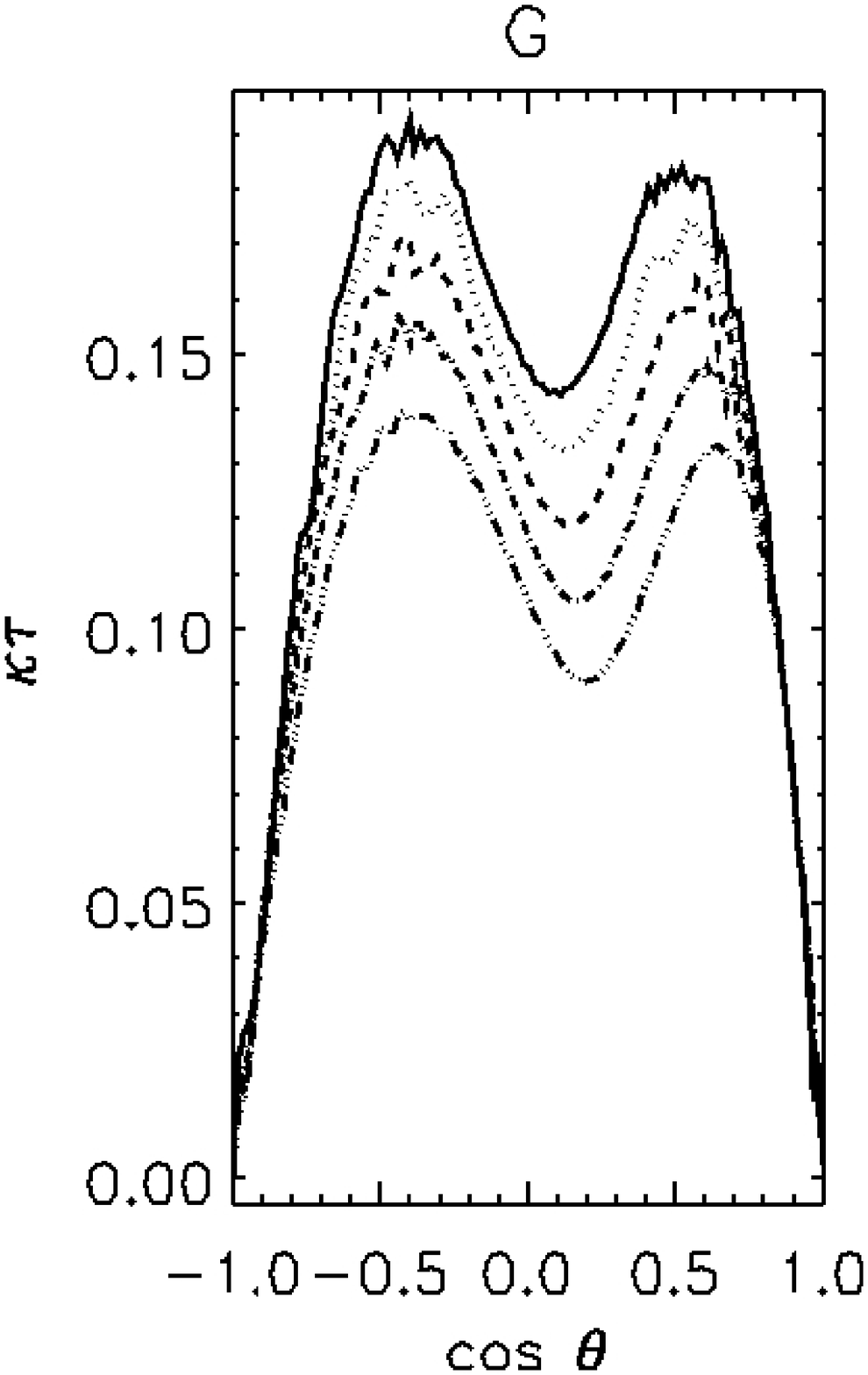}

\caption{Estimated resonant optical depth (multiplied by the microphysical factor $\kappa$) for radial photons as a function of the polar angle $\theta$. We show results for models S1, A, B, C (first row), D, E, F, G (second row) for different photon energies as indicated in the legend. We set $B_0=10^{13}$ G.}
\label{tau}
\end{figure*}

In contrast, the optical depth in our numerical models depends on the photon energy, because the components of $\vec{J}$ and $\vec{B}$ cannot be described by the same power law. In Fig. \ref{tau} we show the estimated resonant optical depth, Eq. (\ref{tau_simple}), for our models, compared with the self-similar model S1. We plot $\kappa\tau_{res}$ as a function of the polar angle $\theta$, for different energies of the seed photons, $\hbar\omega=0.5,1,2,4,8$ keV, taking $B_0=10^{13}$ G. If the magnetic field is predominantly poloidal (low helicity), the optical depth is roughly given by the ratio between toroidal field (which provides poloidal current) and poloidal field at  the resonant radius. Increasing the photon energy, the resonant radius will be closer to the surface, modifying the estimate of $J$ and $dB/dr$ which are involved in Eq. (\ref{tau_simple}). Thus, if the toroidal field decays slower than the poloidal component, a higher photon energy implies a lower ratio $J/|dB/dr|$ at the resonant radius. Looking at the radial profiles (Fig. \ref{fig_profr}), we can understand why the optical depth increases with the photon energy for models E and F, while for other models the optical depth is greater for soft photons. A more precise prediction of how the spectrum is modified when using one or another magnetosphere model requires more elaborated calculations than those presented here. We also point out that, due to the linear relation between $B$ and $\omega_B$, increasing the energy is equivalent to decreasing the normalization $B_0$ by the same factor.

\section{Conclusions}\label{conclusions}

The study of force-free magnetosphere models from an analytical or semi-analytical point of view have led to remarkable results in a few cases. In the literature, the twisted configurations usually implemented in the modeling of synthetic spectra are often based on the self-similar solutions by \cite{thompson02}, which are restricted to a twisted dipole or to a single higher multipole \citep{pavan09}. Since semi-analytical studies of a general combination of multipoles are more difficult due to the nonlinear character of the equations, we faced the problem by constructing and thoroughly testing a numerical code that relaxes an arbitrary initial model to a force-free solution with given boundary conditions. With our numerical simulations we can find general solutions of twisted magnetospheres with complex geometries. Our only input is the radial field at the surface, which can be completely general (i.e. provided by numerical simulations of the internal evolution).

Our main conclusion is that the possible family of relaxed configurations depends on the initial data. As already pointed out by \cite{contopoulos11} in a different context, the solution is not unique, given the surface radial magnetic field and the outer boundary conditions. This reflects the freedom we have in choosing the enclosed current function $I(\Gamma)$, for a given $\Gamma(\theta)$ at the surface. Starting with two different configurations with the same value of volume-integrated helicity does not necessarily lead to the same final configuration. Thus we were able to find a variety of solutions, which can be qualitatively different from the self-similar models.
 
Some of the new magnetosphere models that we found are characterized by a high degree of nonlinearity, twists up to a few radians, and a nontrivial functional form of the threading current $I(\Gamma)$. These models establish a new basis for generalizing the study of radiation transfer in neutron star magnetospheres, as already extensively done with the self-similar models \citep{rea08, fernandez07, zane09}. In particular, our approach provides a more general framework for studying how the geometry of the magnetic field and currents affects the output X-ray spectrum of a magnetar. If resonant Compton scattering is a crucial ingredient in forming the spectra, relaxing the self-similarity constraint in the models may have a significant effect, as our preliminary estimates of optical depths have shown. This opens the possibility of implementing alternative models in simulations of radiative transfer. In the future we aim at providing spectra to be compared with observations.

\begin{acknowledgements}

We are grateful to P.~Cerd\'a-Dur\'an, M.~Gabler, and T.~Akg\"un for interesting discussions and comments. This work was partly supported by CompStar, a Research Networking Program of the European Science Foundation and the Spanish grant AYA 2010-21097-C03-02. D.~Vigan\`o is supported by a fellowship from the {Prometeo} program for research groups of excellence of the Generalitat Valenciana.

\end{acknowledgements}

\bibliography{ff}

\begin{thebibliography}{38}
\expandafter\ifx\csname natexlab\endcsname\relax\def\natexlab#1{#1}\fi

\bibitem[{{Beloborodov} \& {Thompson}(2007)}]{beloborodov07}
{Beloborodov}, A.~M. \& {Thompson}, C. 2007, \apj, 657, 967

\bibitem[{{Canuto} {et~al.}(1971){Canuto}, {Lodenquai}, \&
  {Ruderman}}]{canuto71}
{Canuto}, V., {Lodenquai}, J., \& {Ruderman}, M. 1971, \prd, 3, 2303

\bibitem[{{Chiu} \& {Hilton}(1977)}]{chiu77}
{Chiu}, Y.~T. \& {Hilton}, H.~H. 1977, \apj, 212, 873

\bibitem[{{Contopoulos} {et~al.}(2011){Contopoulos}, {Kalapotharakos}, \&
  {Georgoulis}}]{contopoulos11}
{Contopoulos}, I., {Kalapotharakos}, C., \& {Georgoulis}, M.~K. 2011, \solphys,
  269, 351

\bibitem[{{Contopoulos} {et~al.}(1999){Contopoulos}, {Kazanas}, \&
  {Fendt}}]{contopoulos99}
{Contopoulos}, I., {Kazanas}, D., \& {Fendt}, C. 1999, \apj, 511, 351

\bibitem[{{Duncan} \& {Thompson}(1992)}]{duncan92}
{Duncan}, R.~C. \& {Thompson}, C. 1992, \apjl, 392, L9

\bibitem[{{Fern{\'a}ndez} \& {Davis}(2011)}]{fernandez11}
{Fern{\'a}ndez}, R. \& {Davis}, S.~W. 2011, \apj, 730, 131

\bibitem[{{Fern{\'a}ndez} \& {Thompson}(2007)}]{fernandez07}
{Fern{\'a}ndez}, R. \& {Thompson}, C. 2007, \apj, 660, 615

\bibitem[{{Gabler} {et~al.}(2011){Gabler}, {Cerd{\'a} Dur{\'a}n}, {Font},
  {M{\"u}ller}, \& {Stergioulas}}]{gabler11}
{Gabler}, M., {Cerd{\'a} Dur{\'a}n}, P., {Font}, J.~A., {M{\"u}ller}, E., \&
  {Stergioulas}, N. 2011, \mnras, 410, L37

\bibitem[{{Gaunt}(1930)}]{gaunt30}
{Gaunt}, J.~A. 1930, Royal Society of London Philosophical Transactions Series
  A, 229, 163

\bibitem[{{Goldreich} \& {Julian}(1969)}]{goldreich69}
{Goldreich}, P. \& {Julian}, W.~H. 1969, \apj, 157, 869

\bibitem[{{Gruzinov}(2005)}]{gruzinov05}
{Gruzinov}, A. 2005, Physical Review Letters, 94, 021101

\bibitem[{{Kalapotharakos} \& {Contopoulos}(2009)}]{kalapotharakos09}
{Kalapotharakos}, C. \& {Contopoulos}, I. 2009, \aap, 496, 495

\bibitem[{{Low} \& {Lou}(1990)}]{low90}
{Low}, B.~C. \& {Lou}, Y.~Q. 1990, \apj, 352, 343

\bibitem[{{Lyutikov} \& {Gavriil}(2006)}]{lyutikov06}
{Lyutikov}, M. \& {Gavriil}, F.~P. 2006, \mnras, 368, 690

\bibitem[{{Mereghetti}(2008)}]{mereghetti08}
{Mereghetti}, S. 2008, \aapr, 15, 225

\bibitem[{{Michel}(1973{\natexlab{a}})}]{michel73dis}
{Michel}, F.~C. 1973{\natexlab{a}}, \apj, 180, 207

\bibitem[{{Michel}(1973{\natexlab{b}})}]{michel73}
{Michel}, F.~C. 1973{\natexlab{b}}, \apjl, 180, L133+

\bibitem[{{Michel}(1980)}]{michel80}
{Michel}, F.~C. 1980, \apss, 72, 175

\bibitem[{{Nobili} {et~al.}(2008{\natexlab{a}}){Nobili}, {Turolla}, \&
  {Zane}}]{nobili08a}
{Nobili}, L., {Turolla}, R., \& {Zane}, S. 2008{\natexlab{a}}, \mnras, 386,
  1527

\bibitem[{{Nobili} {et~al.}(2008{\natexlab{b}}){Nobili}, {Turolla}, \&
  {Zane}}]{nobili08b}
{Nobili}, L., {Turolla}, R., \& {Zane}, S. 2008{\natexlab{b}}, \mnras, 389, 989

\bibitem[{{Pavan} {et~al.}(2009){Pavan}, {Turolla}, {Zane}, \&
  {Nobili}}]{pavan09}
{Pavan}, L., {Turolla}, R., {Zane}, S., \& {Nobili}, L. 2009, \mnras, 395, 753

\bibitem[{{Pons} \& {Geppert}(2007)}]{pons07}
{Pons}, J.~A. \& {Geppert}, U. 2007, \aap, 470, 303

\bibitem[{{Rea} \& {Esposito}(2011)}]{rea11}
{Rea}, N. \& {Esposito}, P. 2011, in High-Energy Emission from Pulsars and
  their Systems, ed. {D.~F.~Torres \& N.~Rea}, 247--+

\bibitem[{{Rea} {et~al.}(2008){Rea}, {Zane}, {Turolla}, {Lyutikov}, \&
  {G{\"o}tz}}]{rea08}
{Rea}, N., {Zane}, S., {Turolla}, R., {Lyutikov}, M., \& {G{\"o}tz}, D. 2008,
  \apj, 686, 1245

\bibitem[{{Roumeliotis} {et~al.}(1994){Roumeliotis}, {Sturrock}, \&
  {Antiochos}}]{roumeliotis94}
{Roumeliotis}, G., {Sturrock}, P.~A., \& {Antiochos}, S.~K. 1994, \apj, 423,
  847

\bibitem[{{Scharlemann} \& {Wagoner}(1973)}]{scharlemann73}
{Scharlemann}, E.~T. \& {Wagoner}, R.~V. 1973, \apj, 182, 951

\bibitem[{{Seehafer}(1978)}]{seehafer78}
{Seehafer}, N. 1978, \solphys, 58, 215

\bibitem[{{Spitkovsky}(2006)}]{spitkovsky06}
{Spitkovsky}, A. 2006, \apjl, 648, L51

\bibitem[{{Taflove} \& {Brodwin}(1975)}]{taflove75}
{Taflove}, A. \& {Brodwin}, M.~E. 1975, IEEE Trans. Microwave Theory and
  Techniques, 23, 623

\bibitem[{{Thompson} \& {Duncan}(1996)}]{thompson96}
{Thompson}, C. \& {Duncan}, R.~C. 1996, \apj, 473, 322

\bibitem[{{Thompson} {et~al.}(2002){Thompson}, {Lyutikov}, \&
  {Kulkarni}}]{thompson02}
{Thompson}, C., {Lyutikov}, M., \& {Kulkarni}, S.~R. 2002, \apj, 574, 332

\bibitem[{{Timokhin} {et~al.}(2008){Timokhin}, {Eichler}, \&
  {Lyubarsky}}]{timokhin08}
{Timokhin}, A.~N., {Eichler}, D., \& {Lyubarsky}, Y. 2008, \apj, 680, 1398

\bibitem[{{Ventura}(1979)}]{ventura79}
{Ventura}, J. 1979, \prd, 19, 1684

\bibitem[{{Wolfson}(1995)}]{wolfson95}
{Wolfson}, R. 1995, \apj, 443, 810

\bibitem[{{Yang} {et~al.}(1986){Yang}, {Sturrock}, \& {Antiochos}}]{yang86}
{Yang}, W.~H., {Sturrock}, P.~A., \& {Antiochos}, S.~K. 1986, \apj, 309, 383

\bibitem[{{Yee}(1966)}]{yee66}
{Yee}, K.~S. 1966, IEEE Trans. on Antennas and Propagat., 302

\bibitem[{{Zane} {et~al.}(2009){Zane}, {Rea}, {Turolla}, \& {Nobili}}]{zane09}
{Zane}, S., {Rea}, N., {Turolla}, R., \& {Nobili}, L. 2009, \mnras, 398, 1403

\end{thebibliography}

\appendix

\section{Magnetic helicity}\label{app_helicity}
For a given magnetic field configuration, the usual definition of magnetic helicity 
\begin{equation}
 {\cal H}\equiv\int (\vec{A}\cdot \vec{B}) \mbox{d}V
\end{equation}
is not unique due to the gauge freedom in the
vector potential, $\vec{A}\rightarrow \vec{A}+\vec{\nabla}\Psi$, where $\Psi$ is a scalar function. 
However, in axial symmetry
(no $\phi$-dependence), the following definition of helicity
\begin{equation}\label{helicity_tor}
 {\cal H}\equiv\int A_\phi B_\phi\mbox{d}V
\end{equation}
is gauge-independent because ${\cal H}\rightarrow {\cal H} + \int B_\phi\nabla_\phi\Psi={\cal H}$. We have employed this definition of helicity in numerical code. It can also be shown that this quantity is conserved during the evolution. Taking the time derivative (denoted by $\partial_t$) and using the induction equation, we have
\begin{eqnarray}\label{helicity_cons_tor}
 \partial_t {\cal H} &=& \int A_\phi (\partial_t B_\phi)\mbox{d}V + \int (\partial_t A_\phi) B_\phi\mbox{d}V = \nonumber\\
 &=& -\int A_\phi \hat{\phi}\cdot(\vec{\nabla}\times \vec{E})\mbox{d}V - \int E_\phi B_\phi\mbox{d}V = \nonumber\\
 &=& \int \left[ \vec{\nabla} \cdot (A_\phi \hat{\phi} \times \vec{E} ) - \vec{E} \cdot \vec{\nabla}\times (A_\phi \hat{\phi}) \right]\mbox{d}V - \int E_\phi B_\phi\mbox{d}V = \nonumber\\
	      &=&  \oint_{\partial V} (A_\phi\hat{\phi}\times \vec{E})\cdot \mbox{d}\vec{S} - \int (\vec{E}\cdot\vec{B})\mbox{d}V ~.
\end{eqnarray}
If the poloidal electric field vanishes at the boundaries, the total helicity is conserved.
\par
With the definition (\ref{helicity_tor}), the explicit expression of the magnetic helicity for self-similar models (Sect. \ref{sec_selfsimilar}) is
\begin{equation}\label{hel_tlk}
 {\cal H} =\frac{\pi}{4}B_0^2\frac{k}{p+1}\int\frac{F(\mu)^{2+1/p}}{1-\mu^2}\mbox{d}\mu,
\end{equation}
and for a dipolar Bessel model (Sect. \ref{sec_bessel}) we have
\begin{equation}\label{gauge_bes}
 {\cal H}=\int_V kB_0^2Y_1(x)^2\sin^2\theta\mbox{d}V=\frac{4}{3}kB_0^2\int_{k}^{\infty}Y_1(x)^2\mbox{d}x~.
\end{equation}

\section{A nontrivial semi-analytical configuration}\label{app_legendre}
Here we show the derivation of Eq. (\ref{ode_coup}) in Sect. \ref{sec_legendre}. Substituting $\alpha=\frac{k}{r_\star}|\Gamma/\Gamma_0|^{1/2}$ in Eq. (\ref{ode_leg}), with $\Gamma$ given by Eq. (\ref{potential_gen}), leads to
\begin{eqnarray}\label{ode_alpha3}
  && \frac{l(l+1)}{2l+1}\left[\frac{d^2 (ra_l(r))}{dr^2}-l(l+1)\frac{a_l(r)}{r}\right]=\nonumber\\
  && -\frac{k^2}{3}\frac{1}{r_\star}\int_{-1}^1\frac{d P_l(\mu)}{d\mu}\left(\frac{\Gamma}{\Gamma_0}\right)^2\mbox{sgn}(\Gamma)\mbox{d}\mu ~.
\end{eqnarray}
The integral at the righthand side can be written as
\begin{eqnarray}\label{alpha3_int_part}
 {\cal I} &=& \frac{1}{\Gamma_0^2}\int_{-1}^1\frac{d P_l(\mu)}{d\mu}\Gamma^2\mbox{sgn}(\Gamma)\mbox{d}\mu=\frac{1}{\Gamma_0^2}[\Gamma^2 \mbox{sgn}(\Gamma)P_l(\mu)]^1_{-1} + \nonumber\\
   && - \frac{2}{\Gamma_0^2}\left[\int_{-1}^1\Gamma\frac{d \Gamma}{d\mu}\mbox{sgn}(\Gamma)P_l(\mu)\mbox{d}\mu  + \int_{-1}^1\Gamma^2\delta(\Gamma)\frac{d \Gamma}{d\mu}P_l(\mu)\mbox{d}\mu \right]\nonumber ~,
\end{eqnarray}
where the third term comes from $\mbox{sgn}'(x)=2\delta(x)$. The first and third terms in this equation are zero. Next, we assume by simplicity that $\Gamma \ge 0$ in the whole domain, so that $\mbox{sgn}(\Gamma)\equiv 1$, and we define the dimensionless radial functions $f_m(r)=a_m(r)r/r_\star$. Hereafter we drop the dependences on $\mu$, $r$, and the integration limits for conciseness. Using the Legendre equation we obtain from Eq. (\ref{potential_gen})
\begin{eqnarray}\label{alpha3_int_part2}
 \frac{d \Gamma}{d\mu} &=& -\Gamma_0\sum_n n(n+1)f_nP_n~,
 \end{eqnarray}
and we can express the integral ${\cal I}$ in a more compact form
\begin{eqnarray}
 {\cal I} &=& 2 \sum_{n=1}^{\infty} n(n+1) f_n\sum_{m=1}^\infty f_m \int (1-\mu^2) P_l \frac{d P_m}{d\mu} P_n=\\
   &=& \sum_{m,n=1}^{\infty}\frac{2m(m+1)}{2m+1}n(n+1)f_m f_n \left[\int P_lP_{m-1}P_n - \int P_lP_{m+1}P_n\right]\nonumber ~,
\end{eqnarray}
where we have also used the recurrence relations between Legendre polynomials. Finally, the ODE (\ref{ode_leg}) for each $f_l(r)$ is
\begin{equation}\label{ode_sol3}
  \frac{d^2 f_l}{d r^2}-l(l+1)\frac{f_l}{r^2}=-\left(\frac{k}{r_\star}\right)^2 \sum_{m,n=1}^\infty f_mf_n g_{lmn}~,
\end{equation}
where
\begin{eqnarray} 
g_{lmn}=\frac{2}{3}\frac{2l+1}{2m+1}\frac{m(m+1)}{l(l+1)}n(n+1) \left[\int P_lP_{m-1}P_n - \int P_lP_{m+1}P_n\right]~.\nonumber
\end{eqnarray}
The integral of the product of three Legendre polynomials can be expressed by Wigner $3-j$ symbols:

\begin{equation}\label{wigner}
 \int P_aP_bP_c=\left(
  \begin{array}{ccc}
  a & b & c\\
  0 & 0 & 0
  \end{array}\right)^2~.
\end{equation}
Alternatively, we can express the factors above in terms of associated Legendre polynomials:
\begin{equation}\label{glmn}
  g_{lmn}=\frac{2}{3}\frac{n(n+1)}{l(l+1)}\left[\int P_{l+1}^1P_m^1P_n^0 - \int P_{l-1}^1P_m^1P_n^0\right]~.
\end{equation}
These integrals can be evaluated analytically using \textit{Gaunt's formula} \citep{gaunt30}
\begin{eqnarray}\label{gaunt}
  && {\cal I}_3=\int P_a^u P_b^v P_c^w=\nonumber\\
  && 2\,(-)^{(s-b-w)}\frac{(b+v)!(c+w)!(2s-2c)!s!}{(b-v)!(s-a)!(s-b)!(s-c)!(2s+1)!}\times\nonumber\\
  && \times\sum_{t=p}^q (-)^t \frac{(a+u+t)!(b+c-u-t)!}{t!(a-u-t)!(b-c+u+t)!(c-w-t)!}\\
  && s=\frac{a+b+c}{2}\nonumber\\
  && p=max[0,c-b-u]\nonumber\\
  && q=min[b+c-u,a-u,c-w]~.\nonumber
\end{eqnarray}
The integral ${\cal I}_3$ is non-zero if and only if both the following relations are satisfied:
\begin{itemize}
 \item $b-c\le a\le b+c$, a triangular condition;
 \item $a+b+c$ is even.
\end{itemize}
For each fixed $l$ in Eq. (\ref{glmn}), the conditions above are satisfied for an infinite number of pairs $(m,n)$, with $g_{lmn}$ having numerical values of $\sim O(1)$. Thus Eq. (\ref{ode_sol3}) couples each $l$-multipole with an infinite number of other multipoles.

It should be noted that this is valid only if $\Gamma \ge 0$ in its whole domain $(r,\mu)\in[1,r_{max}]\times[-1,1]$, otherwise we cannot use the formulas above (Eqs. (\ref{wigner}), (\ref{gaunt})). A non positive $\Gamma$ would make the calculation of these factors much harder.

\end{document}